\documentclass[iop,apj]{emulateapj}
\usepackage{natbib}
\usepackage{multirow}
\usepackage{graphicx}
\usepackage{epstopdf}
\usepackage{amsmath}
\shorttitle{Hot ISM of NGC 4490}
\shortauthors{Richings, Fabbiano, Wang \& Roberts}
\begin{document}
\title{The Hot Interstellar Medium of the Interacting Galaxy NGC 4490}
\author{A. J. Richings, G. Fabbiano, Junfeng Wang}
\affil{Harvard-Smithsonian Center for Astrophysics, 60 Garden Street, Cambridge, MA, 02138}
\and
\author{T. P. Roberts}
\affil{Department of Physics, University of Durham, South Road, Durham, DH1 3LE}

\begin{abstract}
We present an analysis of the hot interstellar medium (ISM) in the spiral galaxy NGC 4490, which is interacting with the irregular galaxy NGC 4485, using $\sim 100$ks of {\it Chandra} ACIS-S observations. The high angular resolution of {\it Chandra} enables us to remove discrete sources and perform spatially resolved spectroscopy for the star forming regions and associated outflows, allowing us to look at how the physical properties of the hot ISM such as temperature, hydrogen column density and metal abundances vary throughout these galaxies. We find temperatures of $>0.41$ keV and $0.85_{-0.12}^{+0.59}$ keV, electron densities of $>1.87 \eta^{-1/2} \times 10^{-3} \mathrm{cm}^{-3}$ and $0.21_{-0.04}^{+0.03} \eta^{-1/2} \times 10^{-3} \mathrm{cm}^{-3}$, and hot gas masses of $>1.1 \eta^{1/2} \times 10^{7} \mathrm{M}_{\odot}$ and $\sim 3.7 \eta^{1/2} \times 10^{7} \mathrm{M}_{\odot}$ in the plane and halo of NGC 4490 respectively, where $\eta$ is the filling factor of the hot gas. The abundance ratios of Ne, Mg and Si with respect to Fe are found to be consistent with those predicted by theoretical models of type II supernovae. The thermal energy in the hot ISM is $\sim 5\%$ of the total mechanical energy input from supernovae, so it is likely that the hot ISM has been enriched and heated by type II supernovae. The X-ray emission is anticorrelated with the H$\alpha$ and mid-infrared emission, suggesting that the hot gas is bounded by filaments of cooler ionized hydrogen mixed with warm dust.
\end{abstract}


\section{Introduction}

NGC 4490/85 are a pair of closely interacting galaxies that are relatively nearby, at a distance of $7.8 \: \mathrm{Mpc}$ \citep{tully}. NGC 4490 is a type SB(s)d spiral galaxy, and its companion, NGC 4485, is a type IB(s)m irregular galaxy \citep{devaucouleurs}. The optical isophotes of NGC 4490 indicate that its disk has a high inclination of $60^{\circ}$ \citep{bertola}, which allows us to separate emission from the hot gaseous medium in the plane and the halo of the galaxy.  

These galaxies are embedded in an unusually large envelope of HI gas \citep{viallefond}, which \citet{clemens98} suggest may have been formed by star formation driven outflows. \citet{clemens99} studied the thermal and synchrotron components of the radio continuum emission from these galaxies and concluded that active star formation has been progressing throughout the disk of NGC 4490 at an approximately constant rate of $4.7 \; \mathrm{M}_{\odot} \; \mathrm{yr}^{-1}$ for at least $10^{8}$ years. Observations of the H$\alpha$ emission \citep[e.g.][]{thronson} are consistent with star formation occuring throughout the disk of NGC 4490. However, H$\alpha$ emission is also present between the galaxies, indicating that tidal interactions are important in determining the location of star formation.

NGC 4490/85 were first imaged in X-rays by \citet{read} using {\it ROSAT}. They found 4 discrete sources in these galaxies as well as diffuse emission, but they were unable to isolate a hot ISM, since the spectrum of the diffuse emission showed a high energy tail that suggested strong contamination by unresolved point sources. The first {\it Chandra} observation, by \citet{roberts}, had an exposure time of 20 ks. They were able to resolve 29 discrete point sources in NGC 4490 and one in NGC 4485, including 6 ultraluminous X-ray sources (ULXs). They also detected diffuse emission that contributed $10\%$ of the total X-ray luminosity. They fitted an absorbed MEKAL model to the X-ray spectrum of the diffuse component, finding a temperature $kT = 0.64_{-0.10}^{+0.05}$ keV, an abundance $Z = 0.05 \pm 0.02 \; \mathrm{Z}_{\odot}$ and an absorbing hydrogen column density $N_{H} = 5.1_{-3.3}^{+6.0} \times 10^{20} \; \mathrm{cm}^{-2}$. Two more {\it Chandra} observations and an observation with {\it XMM-Newton} have been used by \citet{gladstone} to further investigate the ULX population in these galaxies.

In this paper we use all of the archival X-ray data from the {\it Chandra} observatory \citep{rots} to study in detail the properties of the hot interstellar medium (ISM) gas in these galaxies. The sub-arcsecond resolution of the {\it Chandra} mirror assembly \citep{vanSpeybroeck} enables us to remove emission from discrete sources (X-ray binary systems or supernova remnants), and also allows us to separate the remaining diffuse emission from the hot gas in different regions within the galaxies. We extract the X-ray spectra from these regions and use spectral models to probe the properties of the hot gas. We also use these X-ray spectra to study the chemical enrichment of the hot ISM by determining the abundances of various elements in different regions within the galaxies \citep[see][for studies following a similar approach]{baldi_a,wang}.

This paper is divided into the following sections. After the introduction (\S1), in \S2 we describe the details of the {\it Chandra} observations and the data reduction. We analyse the diffuse X-ray morphology in \S3, and then we conduct a spectral analysis of the different regions of the hot ISM in NGC 4490 in \S4. The results of this spectral analysis are used to derive the physical properties of the hot gas in \S5. In \S6 we discuss the morphology of the hot ISM and compare this with the morphology of the cooler phases of the ISM seen in other wavelengths, we discuss the nature of the hot gaseous ISM and we consider the energy input and chemical enrichment of this medium from supernovae. Finally, we summarize our results and present our conclusions in \S7.

\section{Observations and Data Reduction}

NGC 4490/85 have been observed three times with the ACIS-S detector on {\it Chandra}, for a total exposure time of $\sim 100$ ks. The details of these observations are summarized in Table~\ref{obs}. We reprocessed the Level 1 event files from these observations using the latest versions of the {\it Chandra} Interactive Analysis of Observations software (CIAO 4.1)\footnote{http://cxc.harvard.edu/ciao/} and calibration database (CALDB 4.1)\footnote{http://cxc.harvard.edu/caldb/}, and filtered them using the Good Time Interval files to obtain new Level 2 event files. More recent calibration files have been released since we started this analysis (CALDB 4.2), however we found that the improvements in the calibration were small compared to the uncertainties in the data, resulting in the same spectral results, so we did not redo the analysis with CALDB 4.2. To identify any background flares, we created a lightcurve of the background emission in each observation, excluding any point sources detected by the CIAO task {\it wavdetect}, and used the sigma-clipping algorithm {\it lc\_sigma\_clip} in CIAO to identify any time intervals when the lightcurve rose to more than $3 \sigma$ above the mean value. We only found background flares in ObsID 4726, with a total duration of 1.2 ks. These intervals were excluded from the event file.

\begin{figure}[h!]
\centering
\includegraphics[width=160mm]{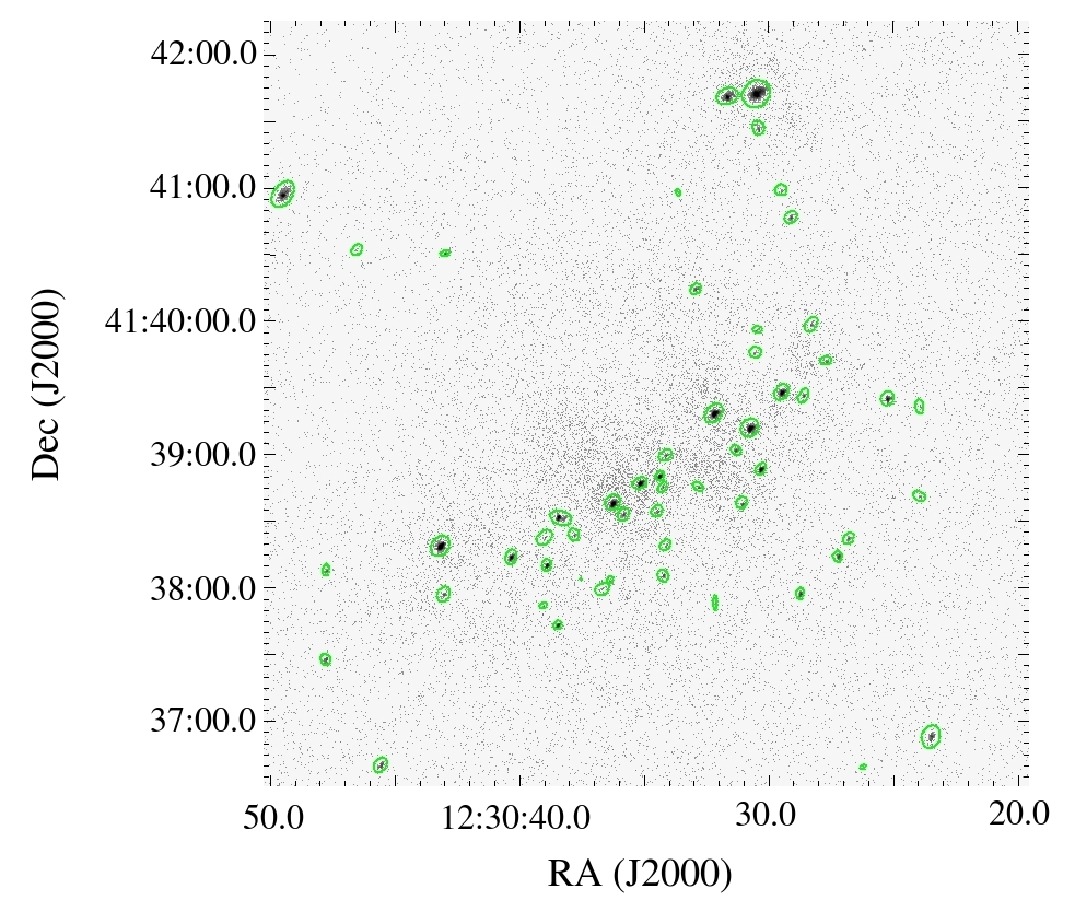}
\caption{$0.3-6.0 \mathrm{keV}$ merged {\it Chandra} X-ray image of NGC 4490/85. Sources detected by {\it wavdetect} are highlighted with their $3\sigma$ ellipses in green.}
\label{galaxies1}
\end{figure}

We used the CIAO software to reproject the World Coordinate System (WCS) coordinates of the event files to match the ObsID 4726 observation and corrected the aspect solutions for small shifts in the coordinates between observations (determined from the position of a bright source found in all three observations). We then merged these reprojected event files using the CIAO task {\it dmmerge} and extracted an image 700 x 700 pixels (5.74 arcmin across) centered on the galaxies in the $0.3-6.0$ keV energy band. This image is shown in fig.~\ref{galaxies1}.

\section{Diffuse X-Ray Morphology}

\subsection{Point Source Removal}

To look at the morphology of the diffuse X-ray emission from the hot gaseous ISM we need to remove the discrete point sources. Following \citet{baldi_a}, we split the X-ray image of the galaxies into three energy bands: $0.3-0.65$ keV, $0.65-1.5$ keV and $1.5-6.0$ keV, and then ran the CIAO task {\it wavdetect} on these three images and the image of the entire energy band, $0.3-6.0$ keV. Using wavelet scales of 2 and 4 pixels, we find a total of 55 sources at a threshold of 3$\sigma$. To confirm that these are genuine point sources, and not extended sources, we compared the radial extent of each source with the radius of the {\it Chandra} ACIS point spread function (PSF) at the same off-axis angle as the source, using an enclosed count fraction of 85\%, which we obtained from the PSF library on the HRMA calibration page\footnote{http://cxc.harvard.edu/cal/Hrma/psf/index.html}. As the off-axis angles of each source are different in the three observations it was necessary to measure the sources in the individual observations and not the merged image. Four sources appeared to have a larger radius than the PSF, however they were all extended at a significance level of below 2$\sigma$, so we cannot confidently determine whether these sources are true point sources or extended objects. We therefore decided to exclude these four sources along with the 51 `true' point sources in our analysis of the hot ISM. All 55 detected sources are listed in appendix~\ref{point_sources}, where the four uncertain sources are also highlighted.

\begin{figure}[h!]
\centering
\mbox{
	\includegraphics[width=80mm]{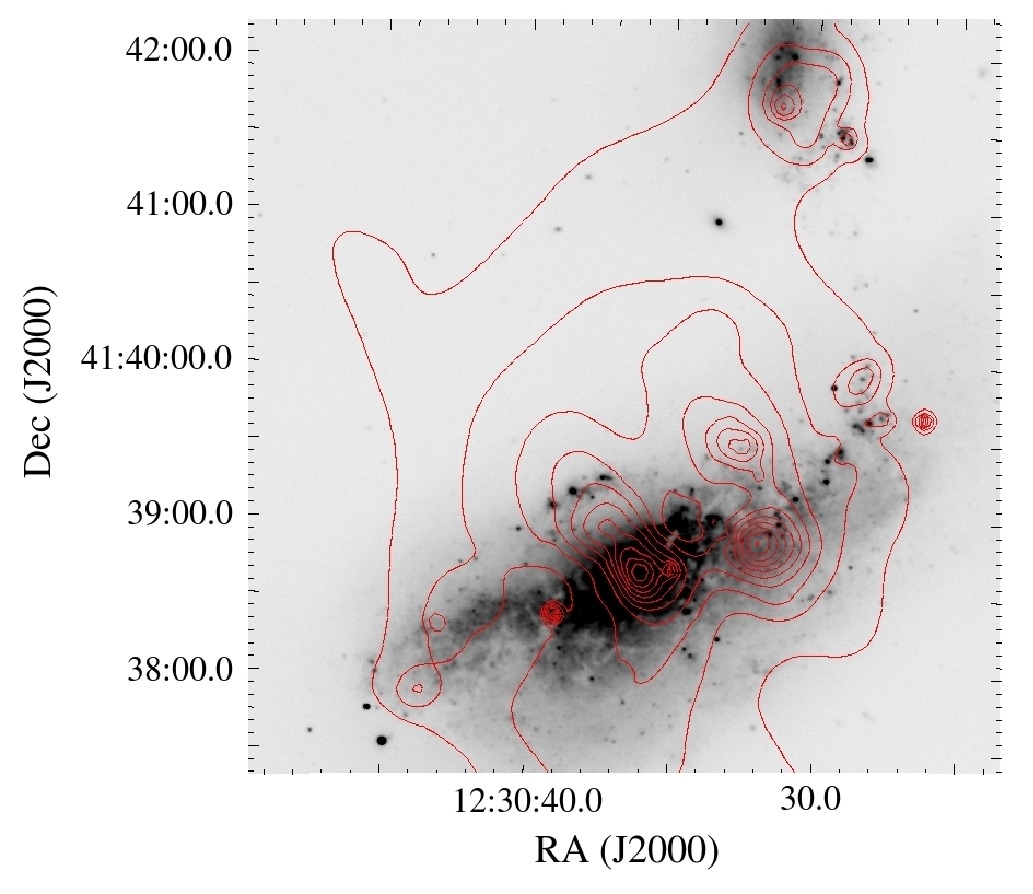}
	\includegraphics[width=80mm]{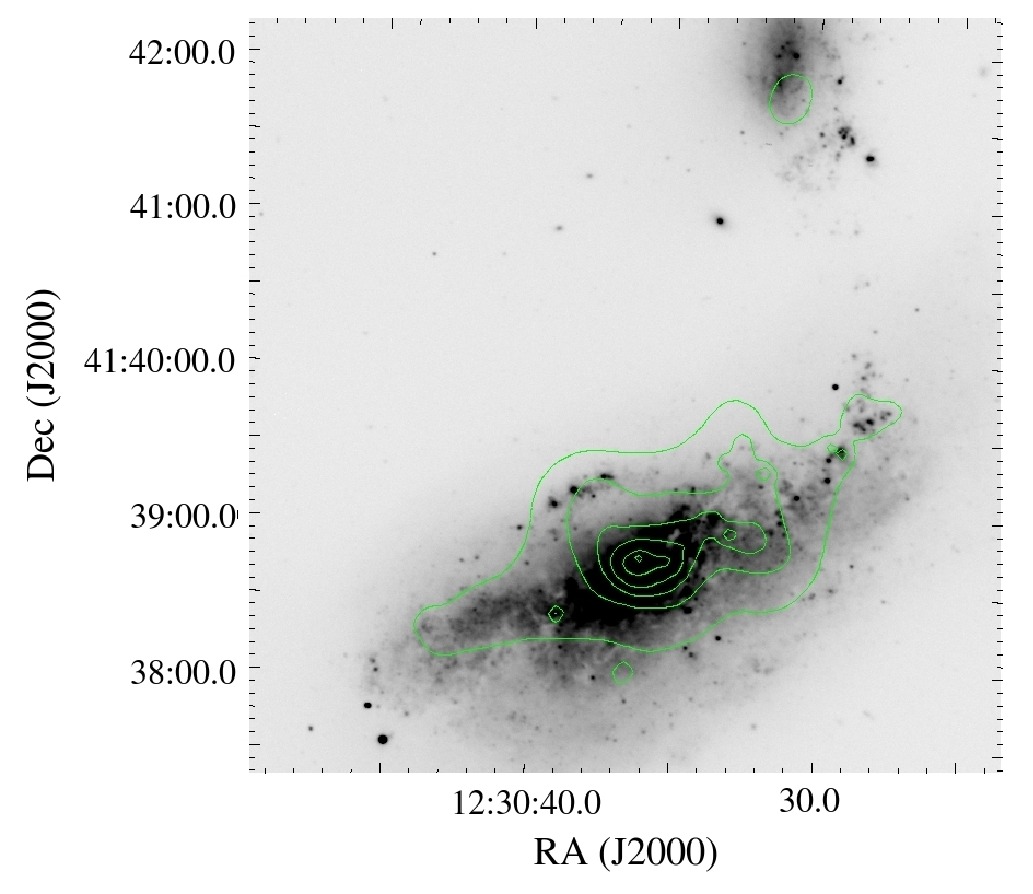}}
\mbox{
	\includegraphics[width=80mm]{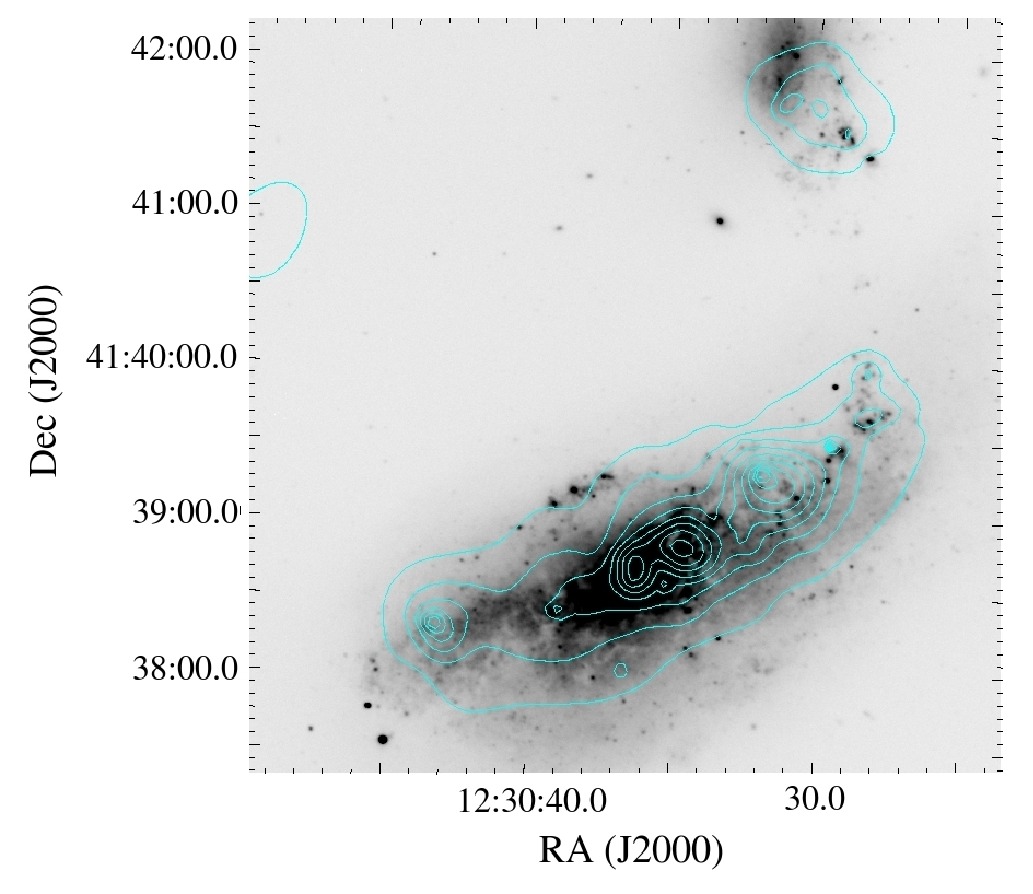}}
\caption{Contours of the average count rates of diffuse X-ray emission in the soft energy band ($0.3-0.65$ keV) (top left), medium energy band ($0.65-1.5$ keV) (top right) and hard energy band ($1.5-6.0$ keV) (bottom), superposed on an optical R-band image of NGC 4490/85, taken by \citet{kennicutt08} using the Steward Observatory's 2.3m Bok telescope on Kitt Peak.}
\label{optical_xray}
\end{figure}

To create images of the diffuse emission (but not for the spectral analysis), these detected point sources were subtracted from the merged images in each of the separate energy bands, and the CIAO task {\it dmfilth} was used to fill in each source region using values randomly sampled from an elliptical annular region around the source, using the source region as the inner boundary and an ellipse with twice the radius as the outer boundary of the elliptical annulus. The remaining emission includes the diffuse emission from the hot gaseous ISM, although it is very likely that there is also contamination from unresolved point sources that could not be detected with the current {\it Chandra} observations. The contours of the diffuse emission in these three energy bands are shown in fig.~\ref{optical_xray}.

\subsection{Mapped-Color X-Ray Image}

\begin{turnpage}
\begin{figure}[h!]
\mbox{
	\includegraphics[width=105mm]{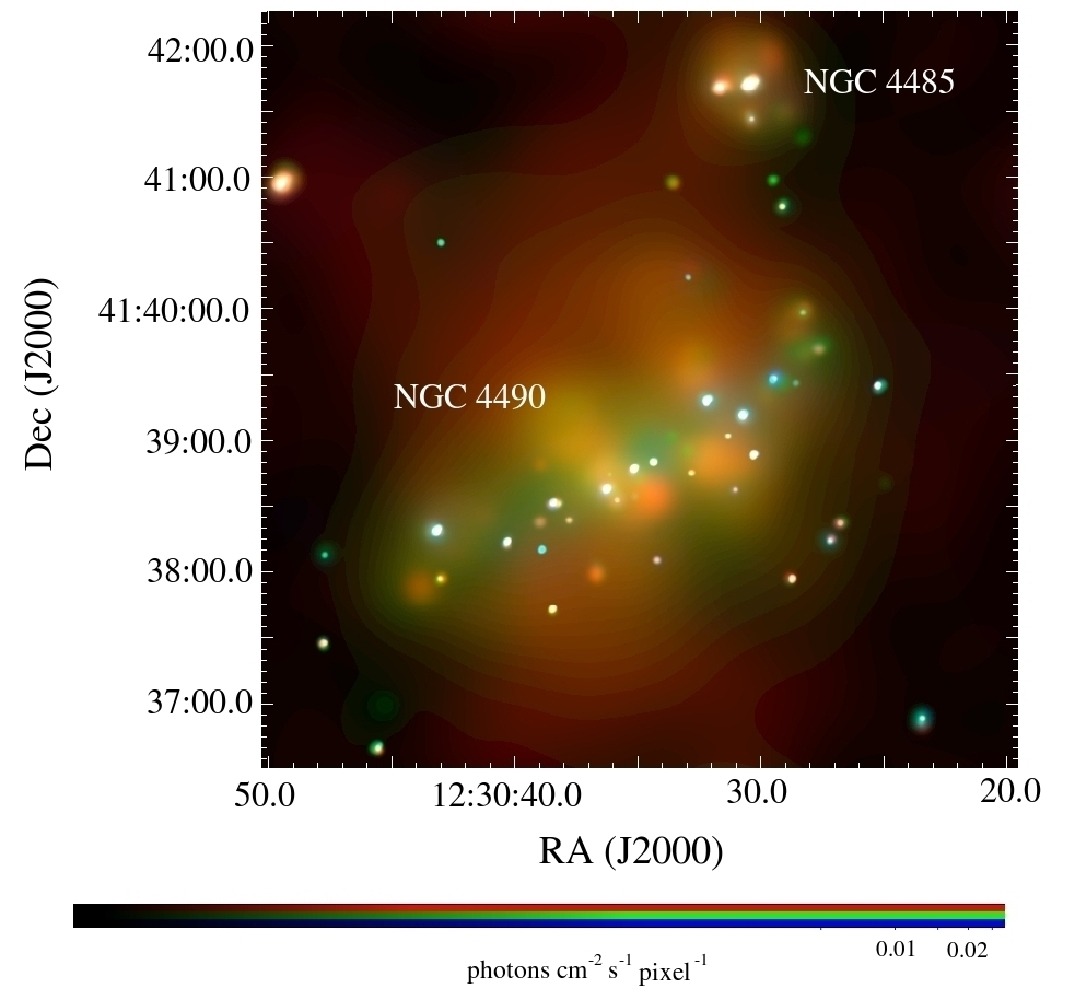}
	\includegraphics[width=105mm]{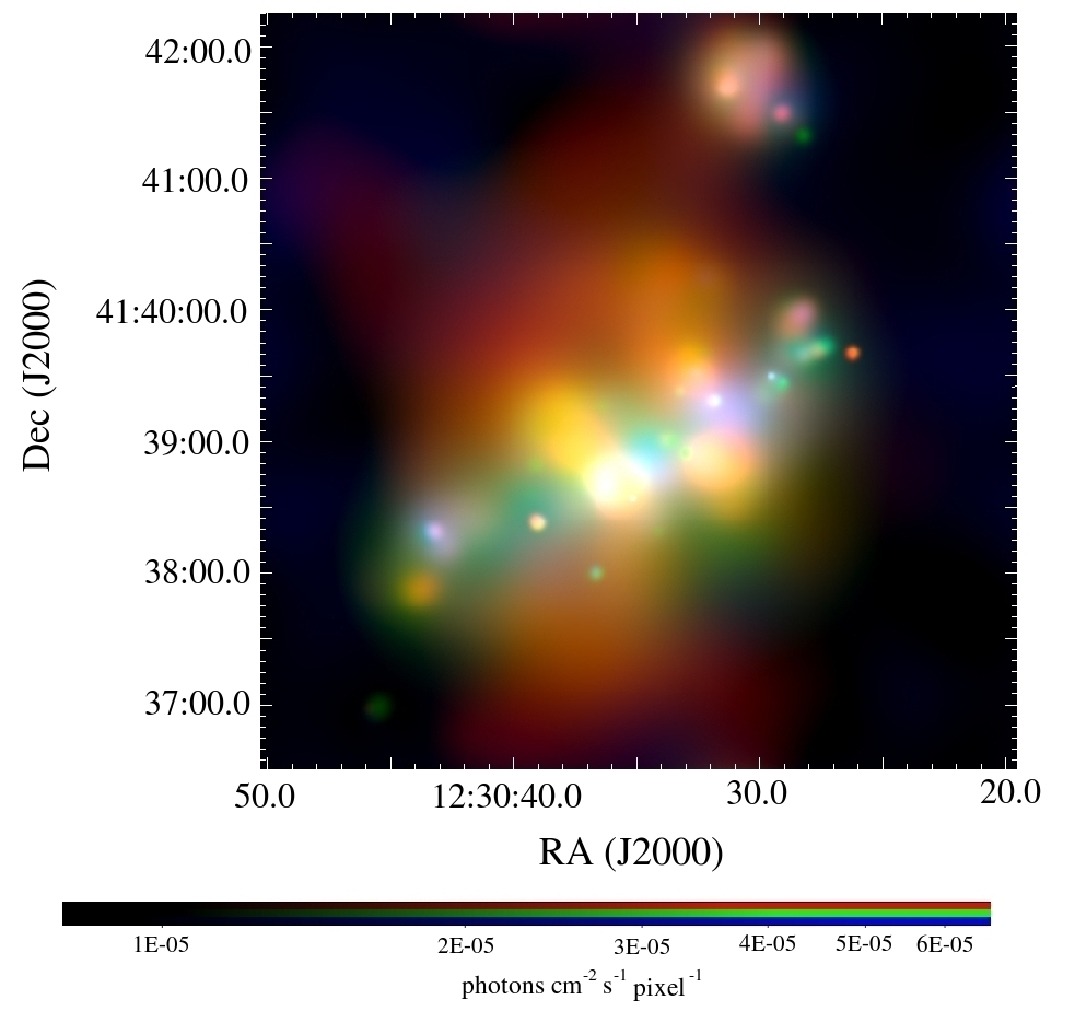}}
\caption{Adaptively smoothed, exposure corrected X-ray images of NGC 4490  and NGC 4485. Red corresponds to emission in the $0.3-0.65 \mathrm{keV}$ energy band, green to the $0.65-1.5 \mathrm{keV}$ band and blue to the $1.5-6.0 \mathrm{keV}$ band. The image on the left includes point sources and diffuse emission, whereas the image on the right contains only the diffuse emission. Both images use logarithmic color scales, although different scales are used in each image as the point sources are significantly brighter than the diffuse emission. The color scales are shown below each image.}
\label{smoothed}
\end{figure}
\end{turnpage}

The point source-subtracted and filled images in each energy band were adaptively smoothed using the {\it csmooth} task in CIAO, which uses a Gaussian kernel that varies in size to obtain a minimum signal/noise ratio beneath the kernel at each pixel. We used a minimum S/N ratio of 2.4, with the maximum S/N ratio set at 5 to prevent oversmoothing on larger scales; and kernel sizes between 1 and 40 pixels. The medium energy band image was used to determine the smoothing scales, as it contained the largest number of counts, and then the same smoothing scales were applied to the other bands. We created exposure maps in each energy band, which were smoothed using the same smoothing scales, and then each image was divided by its corresponding smoothed exposure map. Finally, we combined these smoothed, exposure-corrected images from the three energy bands to create a mapped-color image of the diffuse X-ray emission, which is shown in fig.~\ref{smoothed}. We also created a smoothed image of all X-ray emission, including the point sources, using the same method, which is also shown in fig.~\ref{smoothed} for comparison. 

These images show a soft (red) extended halo and four regions of extended bright emission in NGC 4490. The comparison of soft diffuse X-ray and R-band optical emission in fig.~\ref{optical_xray} shows that two of these regions of extended bright emission emanate from the nucleus of NGC 4490, while the other two emanate from the base of the spiral arm that extends towards the companion galaxy, NGC 4485. These regions are reminiscent of the outflows found in starburst nuclei (e.g. NGC 253 \citep{fabbiano84, strickland02}; M82 \citep{strickland04}). The disk of NGC 4490 appears blue in this image, suggesting considerable absorption by the cold ISM.

\subsection{Radial Profiles Of The Halo And Outflows}

\begin{figure}[h!]
\centering
\mbox{\includegraphics[width=160mm]{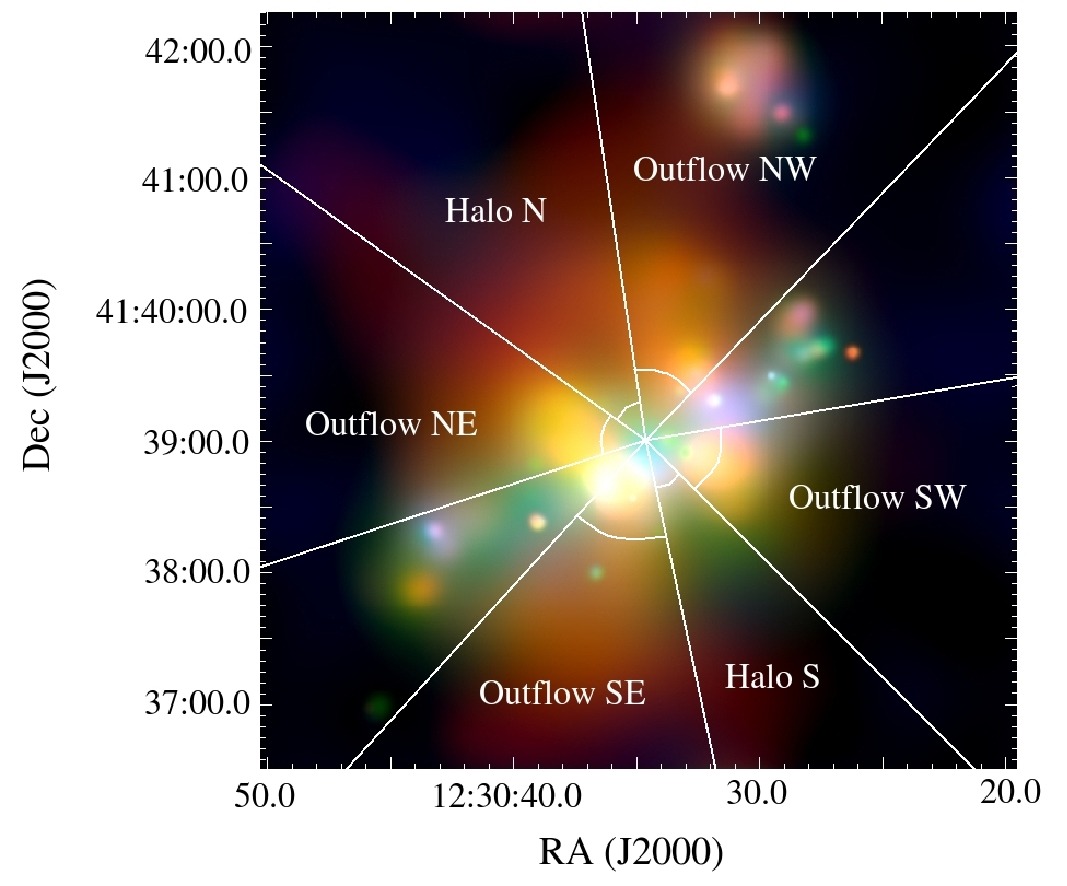}}
\caption{Mapped-color, adaptively-smoothed X-ray image of NGC 4490/85, shown with the six azimuthal sectors used to extract radial X-ray surface brightness profiles of the halo of NGC 4490. The minimum radius of each profile, which excludes the plane region, is highlighted in each sector.}
\label{sectors}
\end{figure}

We divided the halo of NGC 4490 into six azimuthal sectors - four containing the outflows and two containing the regions of the halo between the outflows. These sectors are shown in fig.~\ref{sectors}. We found that the thermal component of the X-ray emission dominates in the energy band $0.3-2.0$ keV, whereas above 2.0 keV the X-ray emission is dominated by the power-law component from unresolved point sources (see \S6.2). Therefore, to investigate the radial profiles of the hot gas in the halo and outflows we used X-ray emission in the $0.3-2.0$ keV band. To enable us to study the spectral hardness ratios (which we describe in \S3.3.2) as well as the total X-ray surface brightness profiles in this band, the X-ray image was divided into two energy ranges, a soft band at $0.3-1.0$ keV and a hard band at $1.0-2.0$ keV. We divided each sector into circular annuli 10 pixels wide centered between the bases of the two pairs of outflows in NGC 4490, and then used the CIAO task {\it dmextract} to extract the number of counts in each radial bin of each sector in the two energy bands, starting from the radius at which the plane region ends and the halo begins in each sector. The North East corner of the image was not covered by the ACIS-S3 chip in all three {\it Chandra} observations, so the radial profiles are strongly affected by exposure in this region. To correct for this effect, we created exposure maps in the two energy bands and created radial profiles of them in each sector. In the three sectors on the South side of the galaxy the profiles of the exposure maps remain approximately constant, however on the North side they decrease significantly above a radius of 100 arcsec. For each radial bin above 100 arcsec in these Northern sectors we calculated the ratio between the exposure map in that bin and the mean value within 100 arcsec, and then used these ratios to weight the corresponding radial bins in the X-ray surface brightness profiles, thereby correcting these profiles for exposure. 

\begin{figure}[h!]
\centering
\mbox{\includegraphics[width=120mm]{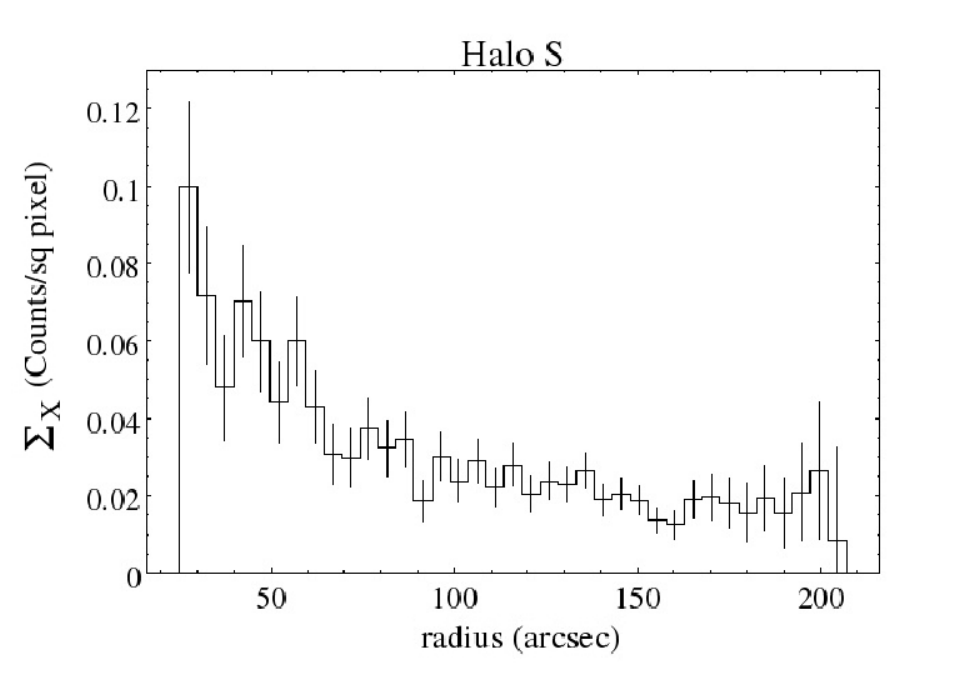}}
\caption{Radial X-ray surface brightness profile of the Halo S sector in the energy band $0.3-2.0$ keV before background subtraction.}
\label{sb_profile_with_bkg}
\end{figure}

These brightness profiles reach a constant background level at a radius of 140 arcsec (see fig.~\ref{sb_profile_with_bkg}), which corresponds to a distance of 5.3 kpc at the distance to NGC 4490 of 7.8 Mpc. Therefore for the background region we used a segment of a circular annulus between radii of 140 arcsec to 220 arcsec, in the three Southern sectors. We did not include the Northern sectors in the background region because they were affected by exposure, and we did not include the regions between the Northern and Southern sectors because the brightness profiles will extend further out radially along the major axis of the galaxy. We were limited by the area of the merged ACIS-S3 image covered by all three observations, however the uncertainties in the net counts in each radial bin were dominated by uncertainties in the total measured counts and not the background, so using a larger background region would have little effect on our results.

Most of the radial bins in each sector contained a small number of counts, so for bins in which there were fewer than 50 counts we obtained the uncertainties using the tables in \citet{gehrels}, who uses Poisson and binomial statistics to calculate the confidence limits of small numbers for which the standard approximation of $\sqrt{N}$ is not valid.

\subsubsection{X-Ray Surface Brightness Profiles}

\begin{figure}[h!]
\centering
\mbox{
	\includegraphics[width=80mm]{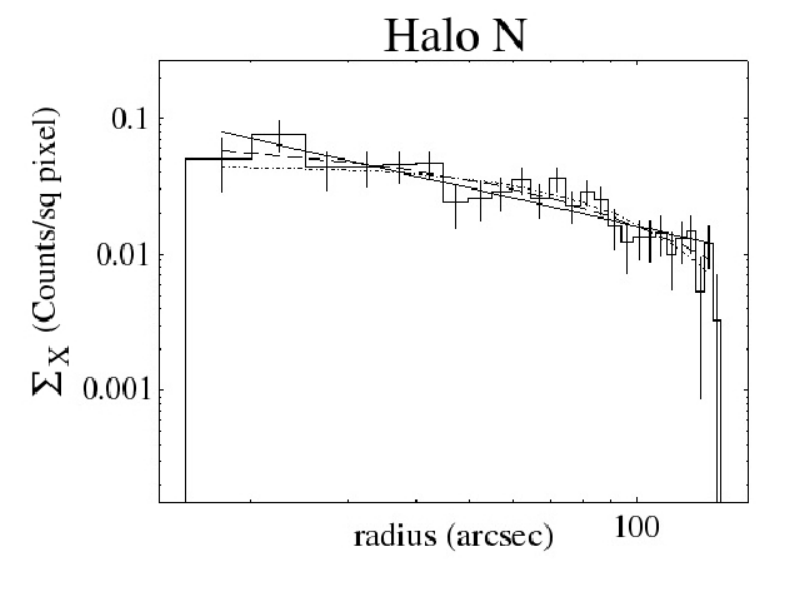}
	\includegraphics[width=80mm]{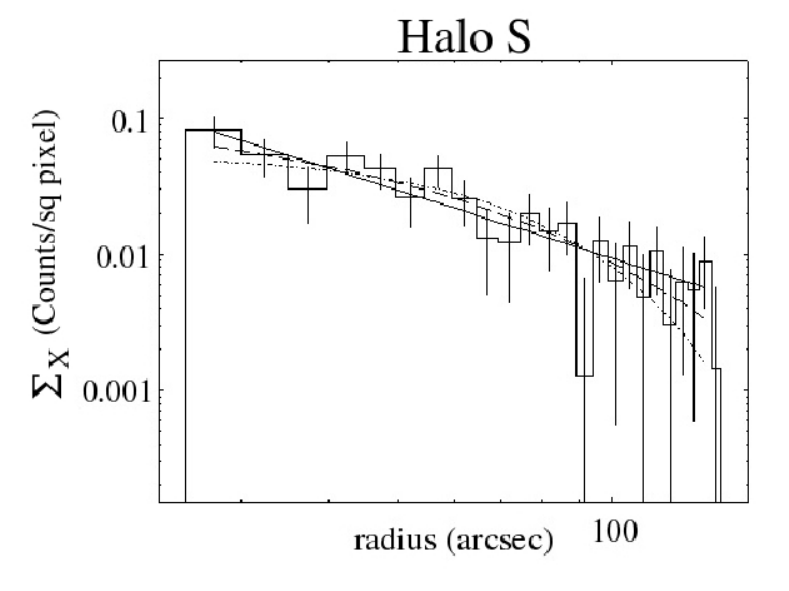}}
\mbox{
	\includegraphics[width=80mm]{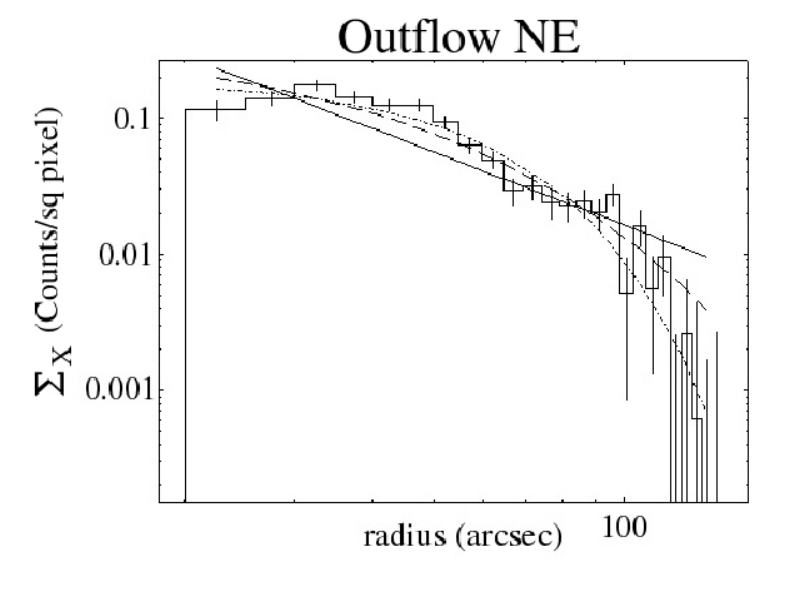}
	\includegraphics[width=80mm]{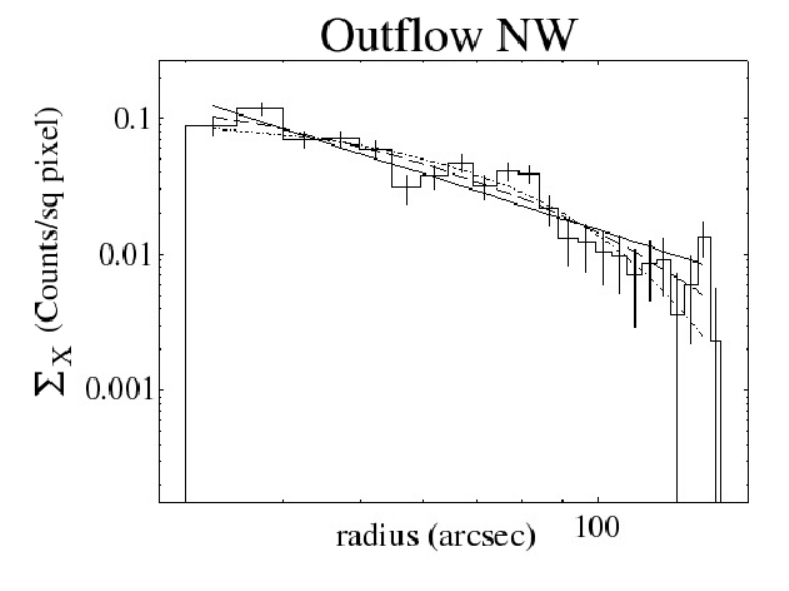}}
\mbox{
	\includegraphics[width=80mm]{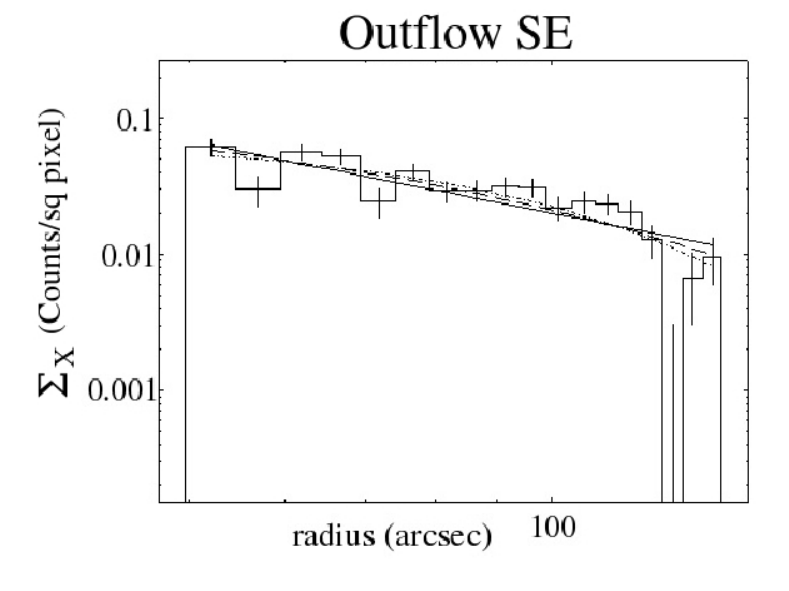}
	\includegraphics[width=80mm]{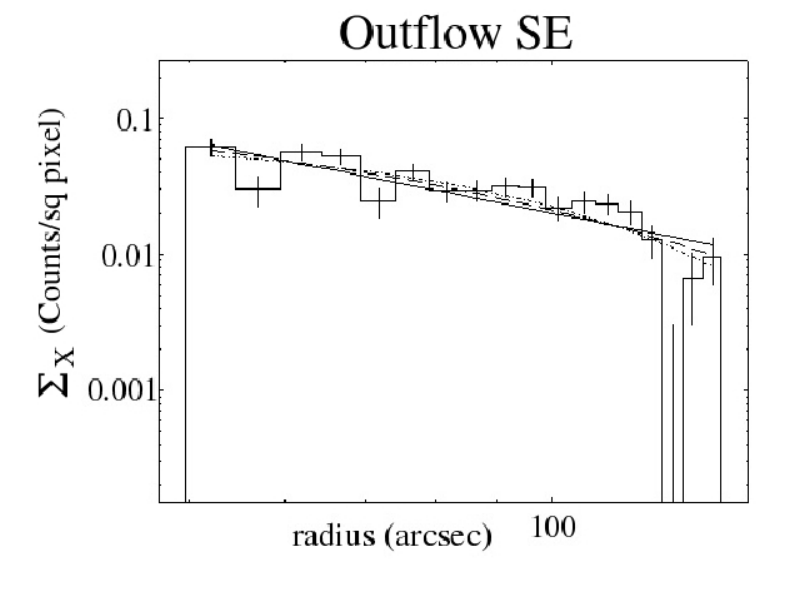}}
\caption{Radial X-ray surface brightness profiles of the two halo sectors (top row) and four outflow sectors (middle and bottom rows), plotted with their best-fit power-law model (solid lines), exponential model (dashed lines) and Gaussian model (dot-dashed lines).}
\label{brightness_profiles}
\end{figure}

The exposure-corrected, background-subtracted profiles in the energy bands $0.3-1.0$ keV and $1.0-2.0$ keV were combined to create the radial X-ray surface brightness profiles of each azimuthal sector in the energy band $0.3-2.0$ keV. We then used CIAO's modeling and fitting software package, {\it Sherpa}\footnote{http://cxc.harvard.edu/sherpa/}, to fit power-law, exponential and Gaussian models to these profiles, from the radius at which the halo region begins in each sector out to a radius of 140 arcsec, where the background region begins. These profiles are shown with their best fit power-law, exponential and Gaussian models in fig.~\ref{brightness_profiles}, and the best fit parameters are summarized in Table~\ref{profile_pars}.

The X-ray surface brightness profiles in the halo are well fit by all three models, with $\chi^{2}_{\nu} < 1$ (where $\chi^{2}_{\nu}$ is the reduced $\chi^{2}$ for $\nu$ degrees of freedom), however in the outflows the power-law model is a poorer fit than the exponential or Gaussian models. The exponential model gives a fairly good fit in the NW and SW outflows, with $\chi^{2}_{\nu} < 1.5$, however in the NE and SE outflows neither model gives a good fit, both giving $\chi^{2}_{\nu} > 2$.

\subsubsection{Spectral Hardness Ratio Profiles}

The exposure-corrected, background-subtracted profiles in the energy bands $0.3-1.0$ keV and $1.0-2.0$ keV were also used to calculate the radial profiles of the spectral hardness ratio, which is defined as: $$Q = \frac{S_{1.0-2.0} - S_{0.3 - 1.0}}{S_{0.3-1.0} + S_{1.0-2.0}}$$ Where $S_{0.3-1.0}$ and $S_{1.0-2.0}$ are the number of counts in the $0.3-1.0 \mathrm{keV}$ and $1.0-2.0 \mathrm{keV}$ energy bands respectively. The uncertainties in Q were relatively large, so we increased the size of the radial bins to 20 pixels per bin up to a radius of 100 arcsec, and we combined all bins from 100 arcsec to 140 arcsec into a single bin. These profiles are shown in fig.~\ref{hardness_ratios}. They suggest that in general the temperatures are declining less steeply than would be expected for an adiabatically expanding gas (see \S6.2).

\section{Spectral Analysis}

\subsection{X-Ray Emission Line Strength Map}

\begin{figure}[h!]
\centering
\mbox{\includegraphics[width=160mm]{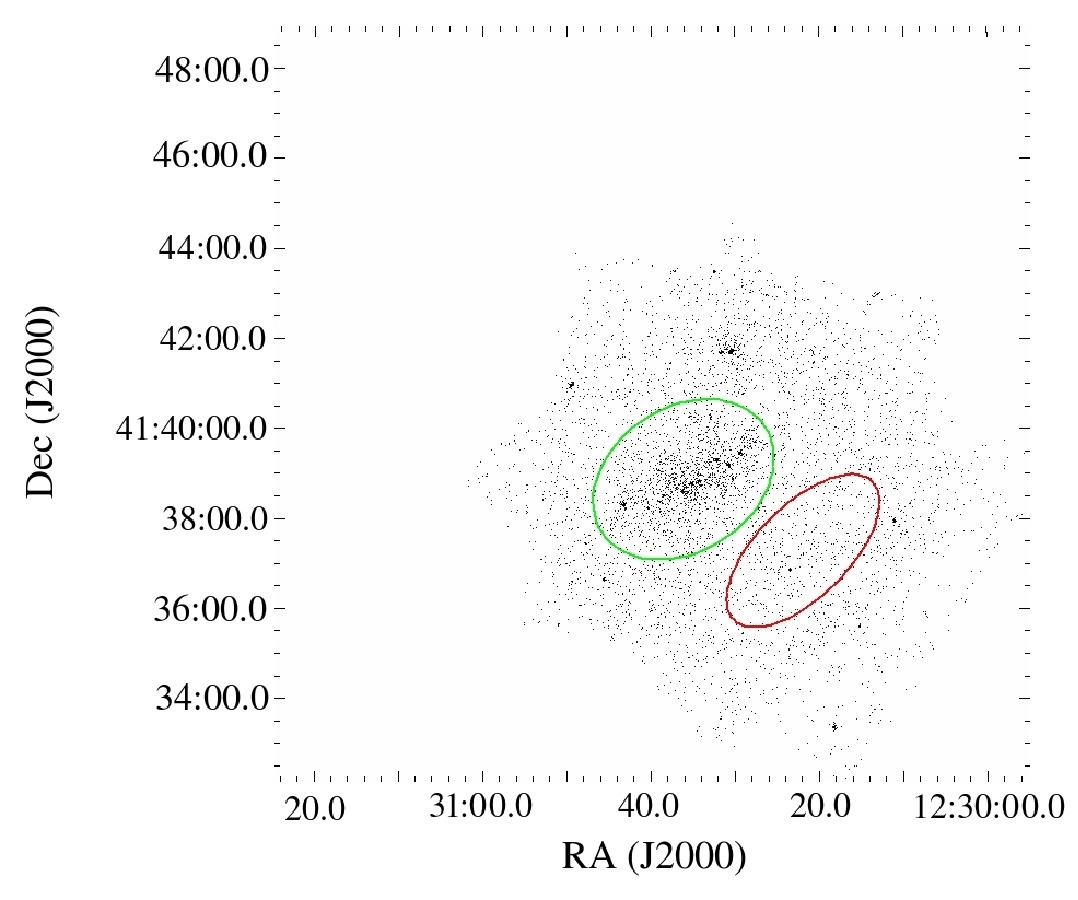}}
\caption{$0.3-6.0$ keV merged {\it Chandra} ACIS-S3 X-ray image of NGC4490/85. The region used to extract the X-ray spectrum of NGC 4490 is highlighted in green, and the region used to extract the background spectrum is highlighted in red.}
\label{4490_spec_regions}
\end{figure}

We extracted the spectrum of the diffuse emission of NGC 4490 from an elliptical region with a semi-major axis of 130 arcsec (corresponding to 4.9 kpc at a distance of 7.8 Mpc) centered on NGC 4490 and excluding the point source regions. We used the CIAO script {\it specextract} to extract the X-ray spectra from each {\it Chandra} observation along with the response matrix files (RMFs) and the ancillary response files (ARFs). The background spectrum was extracted from a region to the South West of NGC 4490 with a total area of $3.05 \; \mathrm{arcmin}^{2}$. These source and background regions are shown in fig.~\ref{4490_spec_regions}. The spectra from the separate observations were added using the FTOOL {\it mathpha}, and the response files from each observation were weighted by the exposure times and added using the FTOOLs {\it addrmf} and {\it addarf}. The spectra were then grouped to contain a minimum of 20 counts per bin, using the `ADAPTIVE' option in the CIAO tool {\it dmgroup}, which leaves bright regions ungrouped and groups low signal to noise regions of the spectrum until they contain the minimum number of counts.

The effective area of the ACIS detector on {\it Chandra} is known to degrade over time, particularly at softer energies, which could create significant differences between data taken from observations separated by several years. We use observations from 2000 and 2004, so to investigate whether this effect is significant we also considered fitting the spectra from the 2000 and 2004 observations separately. We found that this had no significant impact on our results as the response files accurately account for the degradation in the effective area of the ACIS detector, so we continue to use the combined spectra in the following analysis.

\begin{figure}[h!]
\centering
\includegraphics[width=160mm]{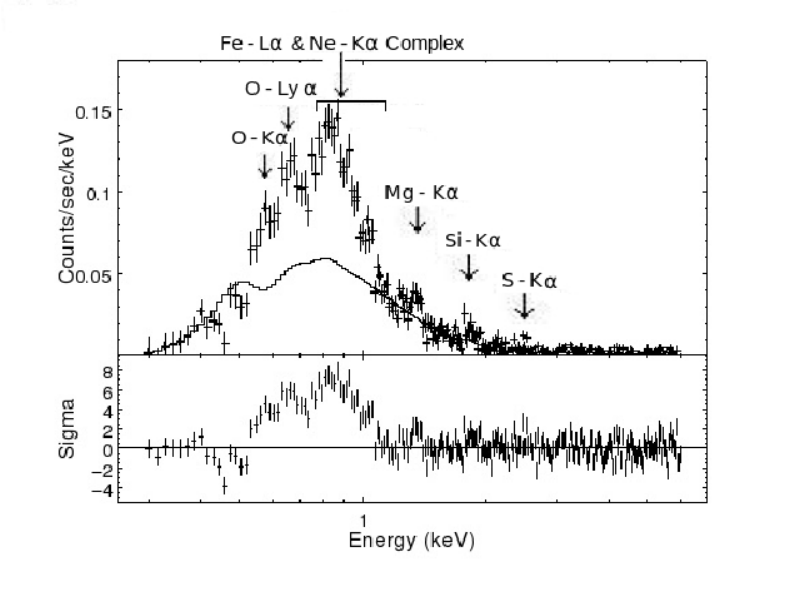}
\caption{X-ray spectrum of NGC 4490 plotted with the best fit continuum model of absorbed Bremsstrahlung plus power-law. The strongest emission lines that can be seen in the residuals have been labeled - the energies of these lines are given by \citet{mewe}.}
\label{spectrum_all}
\end{figure}

To identify any emission lines in the spectrum of the diffuse emission of NGC 4490 we used {\it Sherpa} to fit the continuum emission with an absorbed Bremsstrahlung model plus a power-law component to account for unresolved point sources that may still contaminate the diffuse emission. After an initial fit in the energy range $0.3-6.0$ keV this spectrum showed clear evidence for excesses above the continuum fit, attributable to emission lines, including the (Fe+O+Ne) blend in the energy band $0.6-1.16$ keV, Mg at $1.27-1.38$ keV and Si at $1.75-1.95$ keV. To fit this model to only the continuum emission this fit was repeated excluding these energy bands that include strong line emission. These lines are highlighted in the spectrum in fig.~\ref{spectrum_all}.

\begin{figure}[h!]
\centering
\includegraphics[width=160mm]{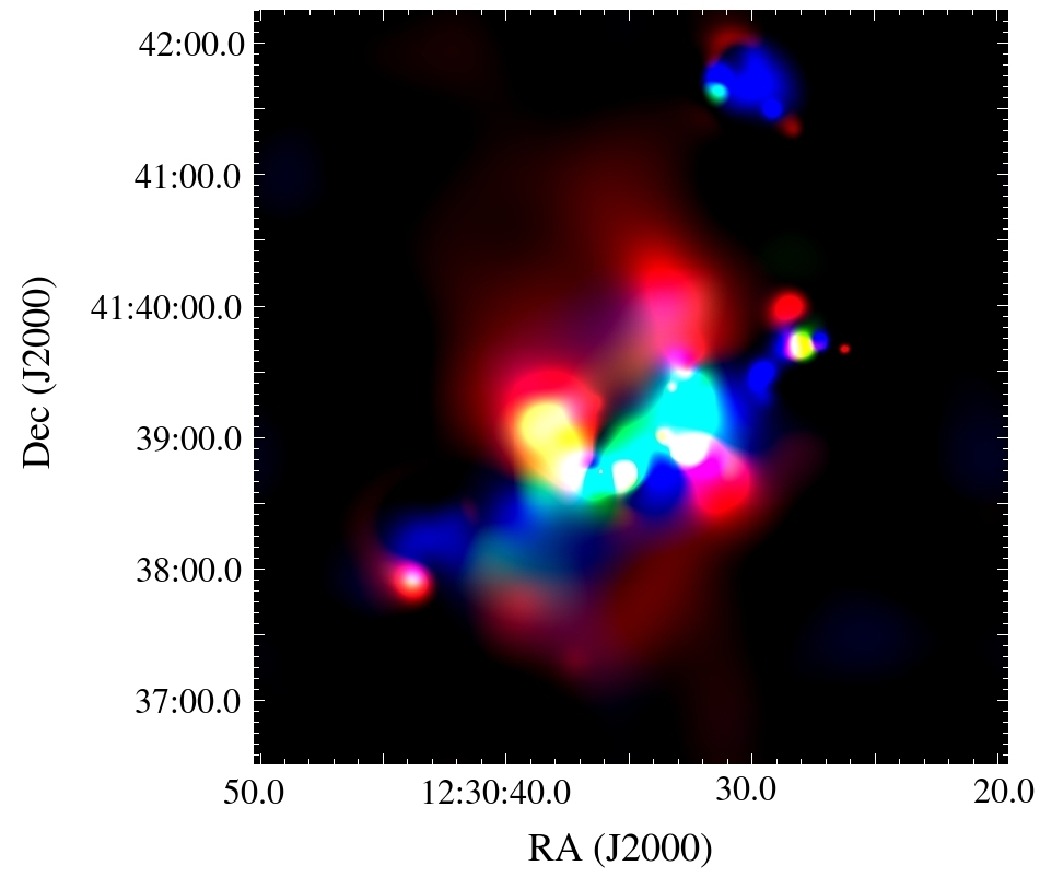}
\caption{Mapped-color emission line strength map of NGC 4490/85. Red corresponds to emission from (Fe+O+Ne), green to emission from Mg and blue to emission from Si (see text).}
\label{line_map}
\end{figure}

Following the method used by \citet{baldi_a} and \citet{wang}, we extracted X-ray images in these energy bands containing strong line emission and also continuum images in the energy bands $1.4-1.65$ keV and $2.05-3.05$ keV. We used the best fit continuum model to weight the continuum images to the emission line energy bands and subtracted them to create images of the line emission from (Fe+O+Ne), Mg and Si. These images were then combined to create a mapped-color emission line strength map of these galaxies, which is shown in fig.~\ref{line_map}. Note that the relative lack of (Fe+O+Ne) emission in the galaxy plane (which appears blue in fig. \ref{line_map}) may be in fact due to the higher $N_{H}$ in this region. This is taken fully into account in our spectral fitting, where we separate different regions.

\subsection{Spectral Fitting}

\begin{turnpage}
\begin{figure}[h!]
\centering
\mbox{\includegraphics[width=105mm]{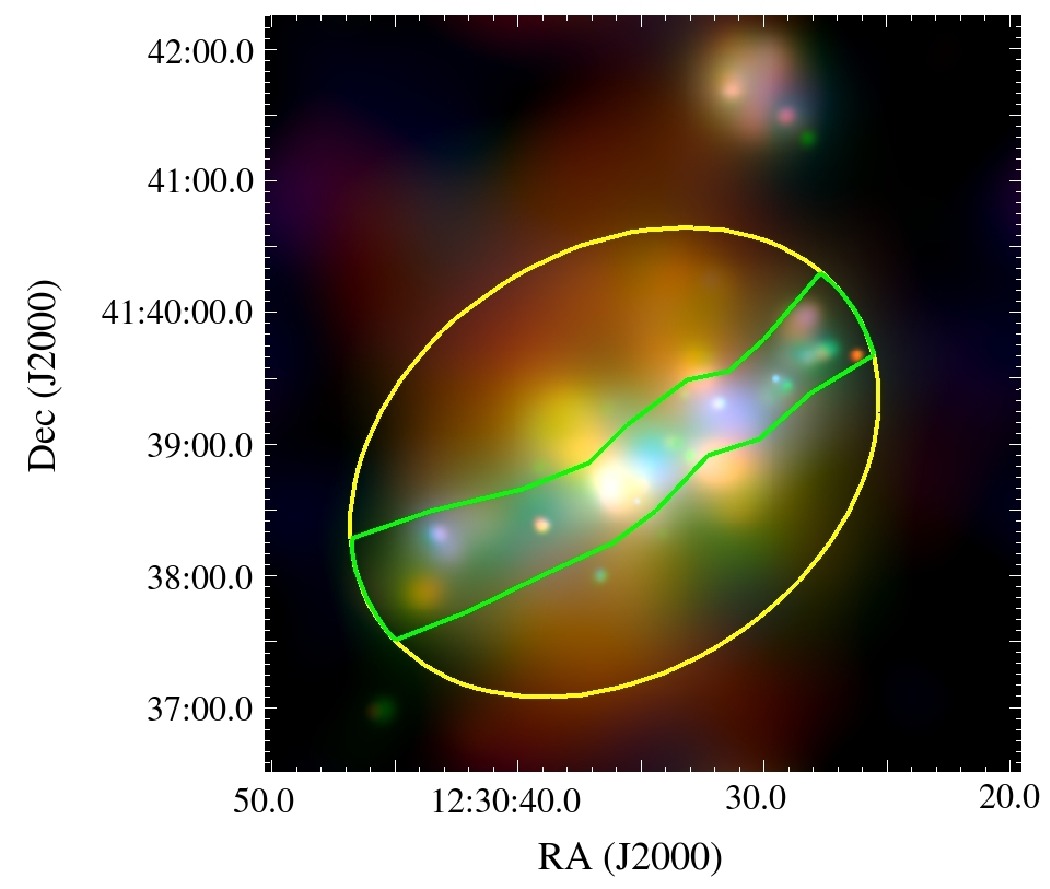}}
\mbox{\includegraphics[width=105mm]{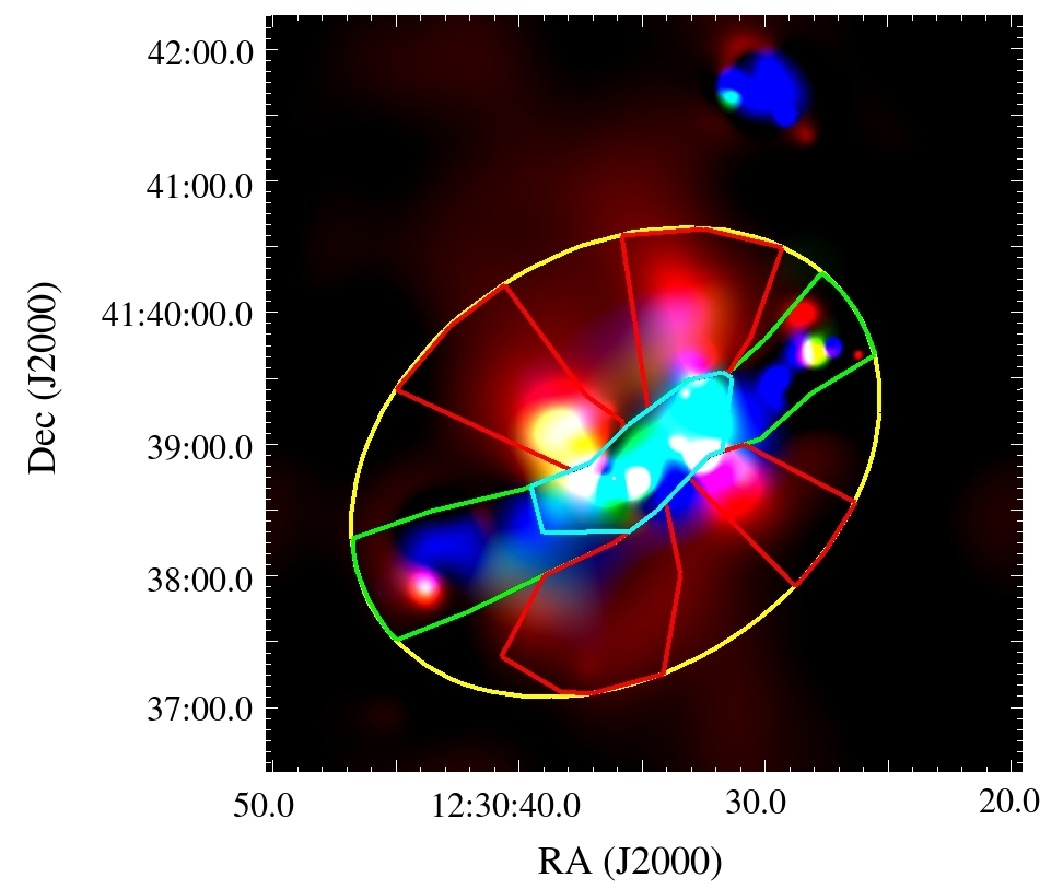}}
\caption{\textbf{Left:} Adaptively smoothed image of the diffuse X-ray emission, showing the plane (green) and halo (yellow) regions used to extract the spectra. \textbf{Right:} Line strength map showing the outflow (red) and central plane (blue) subregions. Note that the four separate outflows shown here were treated as a single subregion.}
\label{regions}
\end{figure}
\end{turnpage}

Using the diffuse X-ray image and the emission line strength map, we divided the diffuse X-ray emission from NGC 4490 into two regions, the halo and the plane of the galaxy. We also divided these regions into two further subregions each that appear to show different spectral properties in these images - the halo was split into a combined subregion containing all outflows, and the rest of the halo; and the plane was split into the central and outer plane subregions. These regions and subregions are highlighted in fig.~\ref{regions}. There were too few counts from the companion galaxy, NGC 4485, to be able to fit a model to its X-ray spectrum. In each region and subregion we extracted the X-ray spectra and their response files using the same method described in \S4.1. Since the line strength map in fig.~\ref{line_map} suggests that the individual outflows show different spectral lines, in particular in the NE outflow, we also extracted X-ray spectra from each individual outflow separately and compared them. The individual outflows contained too few counts to distinguish these different line structures in their separate spectra, so we will use their comined spectrum in the following analysis. 

To model the spectrum of the hot ISM we used the Astrophysical Plasma Emission Code (APEC) \citep{smith}, which is based on the thermal plasma emission model introduced by \citet{raymond_smith} but makes use of modern computing techniques and more accurate atomic data. The spectrum may also be contaminated by unresolved point sources such as X-ray binaries, which, in the hard/low state, typically have power-law spectra with a photon index $\Gamma$ in the range $1.5 < \Gamma < 2.0$ \citep{done}. We therefore added a power-law component with a fixed photon index of $\Gamma=1.8$ to account for these unresolved point sources. \citet{baldi_a} found similar contamination from unresolved point sources in several regions of the hot ISM in NGC 4038/9 (`the Antennae'), which they were able to characterize with a power-law. In many of these regions they fixed the photon index at a value of $\Gamma = 1.88$, although they also considered allowing it to vary freely. To confirm that the value of $\Gamma=1.8$ that we used is consistent with our spectra we extracted the X-ray spectrum of all emission from NGC 4490 (including point sources) in the energy range $2.0 - 5.0$ keV and fitted a power-law using {\it Sherpa}. This fit gave a photon index $\Gamma = 1.92 \pm 0.05$, which is close to the value that we used. We found that using this value instead of $\Gamma = 1.8$ did not make a significant difference to the spectral results, so it was not necessary to redo the analysis using this value.

The thermal and power-law components were both multiplied by a photoelectric absorption model ({\it xswabs} in {\it Sherpa}) to account for both galactic and intrinsic absorption. We then used {\it Sherpa} to fit this absorbed APEC plus power-law model to the spectra from each region and subregion, using the {\it neldermead} optimization method in {\it Sherpa} (a simplex algorithm created by \citet{neldermead}). The APEC component dominates the spectrum at softer energies, while the power-law dominates at harder energies, so we initially fitted just the power-law component in the energy range $2.0-6.0$ keV, where the thermal component is negligible - at energies above $6.0$ keV the background dominates. The power-law parameters were then fixed at their best fit values in this energy range and the remaining components were fit in the energy range $0.3-3.0$ keV - at energies below $0.3$ keV the {\it Chandra} data is poorly calibrated. The abundances of Fe, Ne, Mg and Si in the APEC model were allowed to vary, as they show strong emission lines. We also considered varying the abundance of O, which showed strong line emission as well, however we found that this abundance was poorly constrained. Therefore the abundances of O and all other elements were fixed at their solar values (as measured by \citet{anders}).

At energies above 2.0 keV the spectra have a poor S/N ratio. We considered increasing the binning of the spectra, with up to a minimum of 50 counts per bin, however we found that this made little difference to the fits, with the parameters changing by no more than 10\%. We therefore continue to use spectra binned with a minimum of 20 counts per bin, which prevents the bin sizes becoming too large around the emission lines, which generally have good S/N ratios.

\subsubsection{Results}

\begin{figure}[h!]
\centering
\mbox{
	\includegraphics[width=80mm]{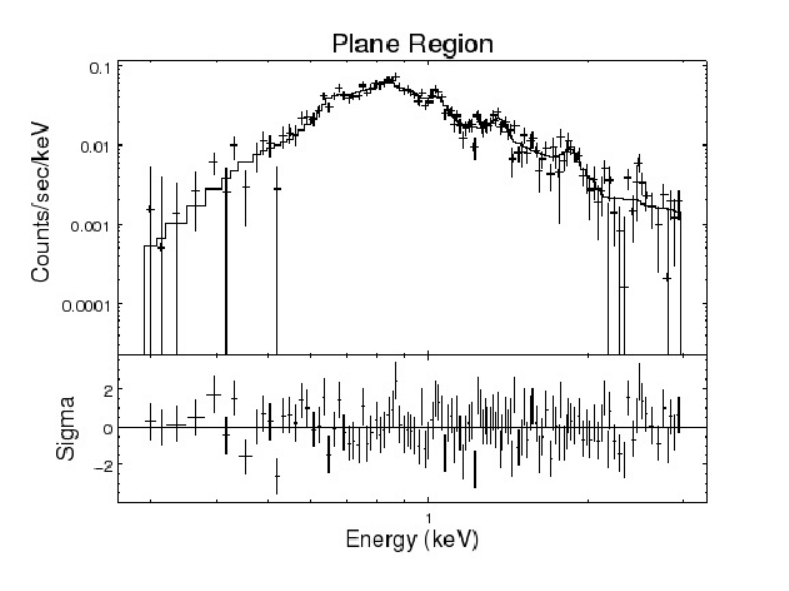}
	\includegraphics[width=80mm]{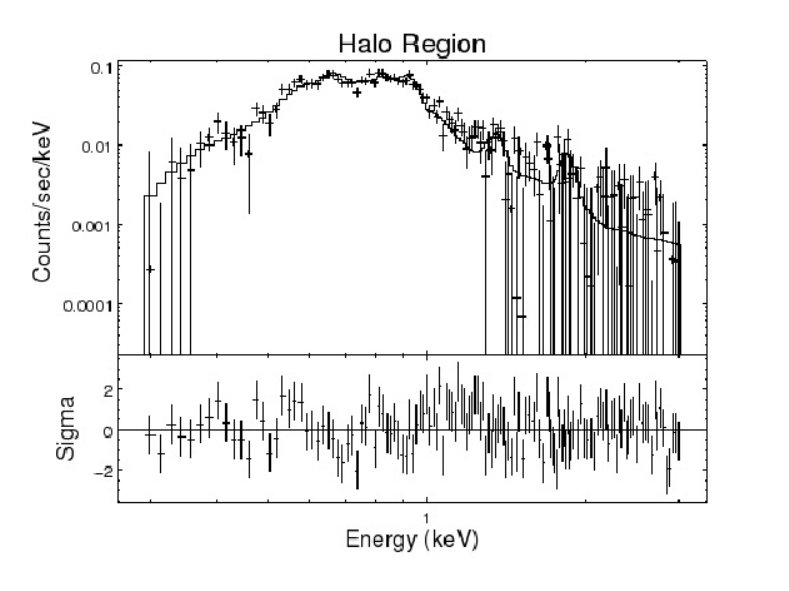}}
\mbox{
	\includegraphics[width=80mm]{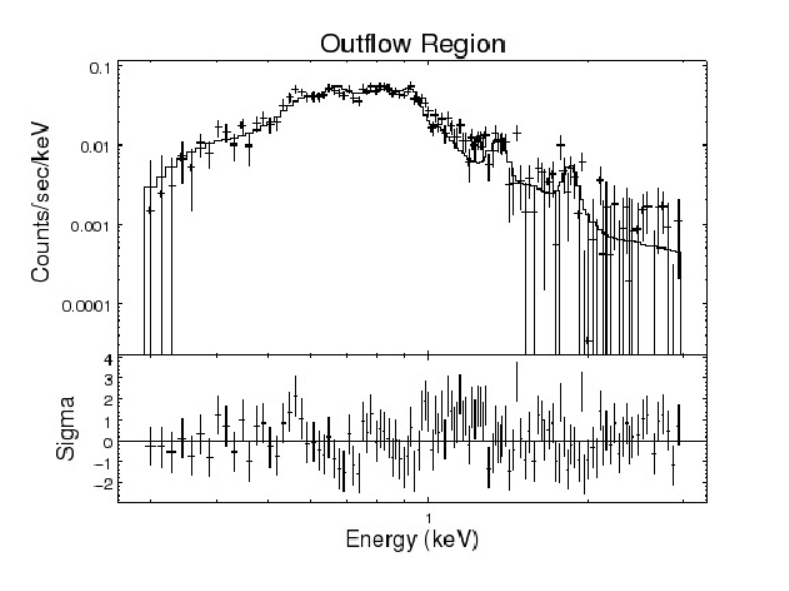}
	\includegraphics[width=80mm]{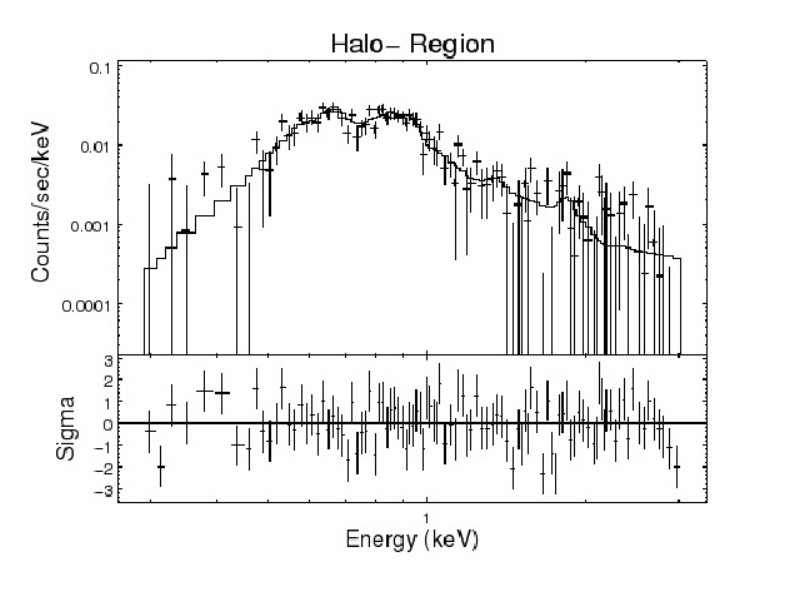}}
\mbox{
	\includegraphics[width=80mm]{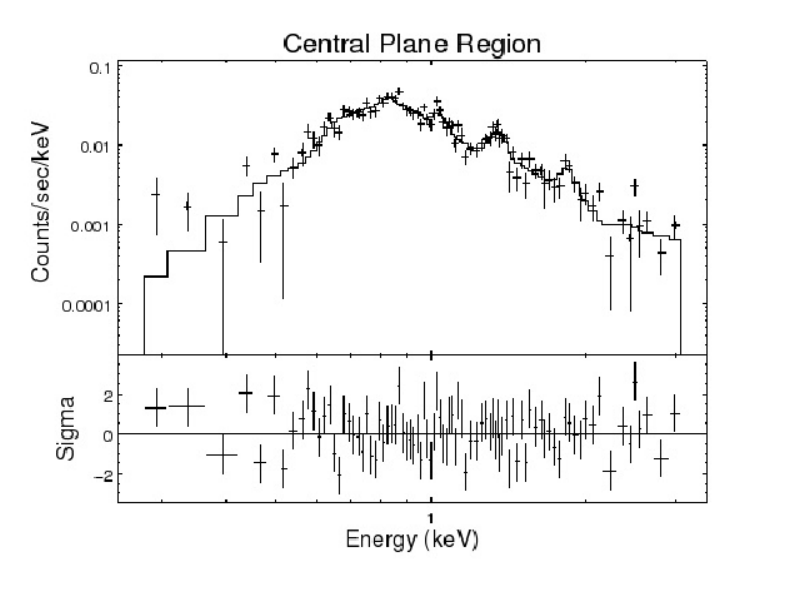}
	\includegraphics[width=80mm]{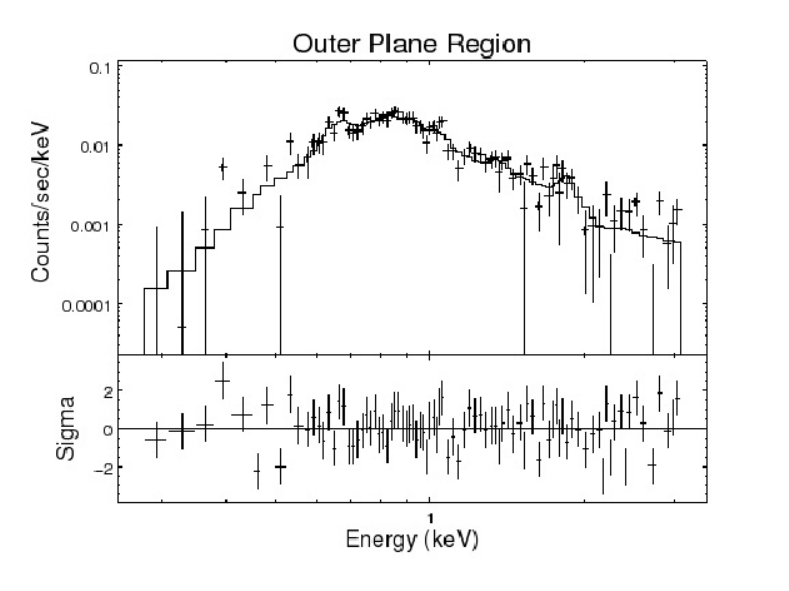}}
\caption{X-ray spectra extracted from the plane and halo regions and their subregions, plotted with their best-fit absorbed APEC plus power-law models. The fit residuals are shown below each spectrum.}
\label{spectra}
\end{figure}

\begin{figure}[h!]
\centering
\includegraphics[width=160mm]{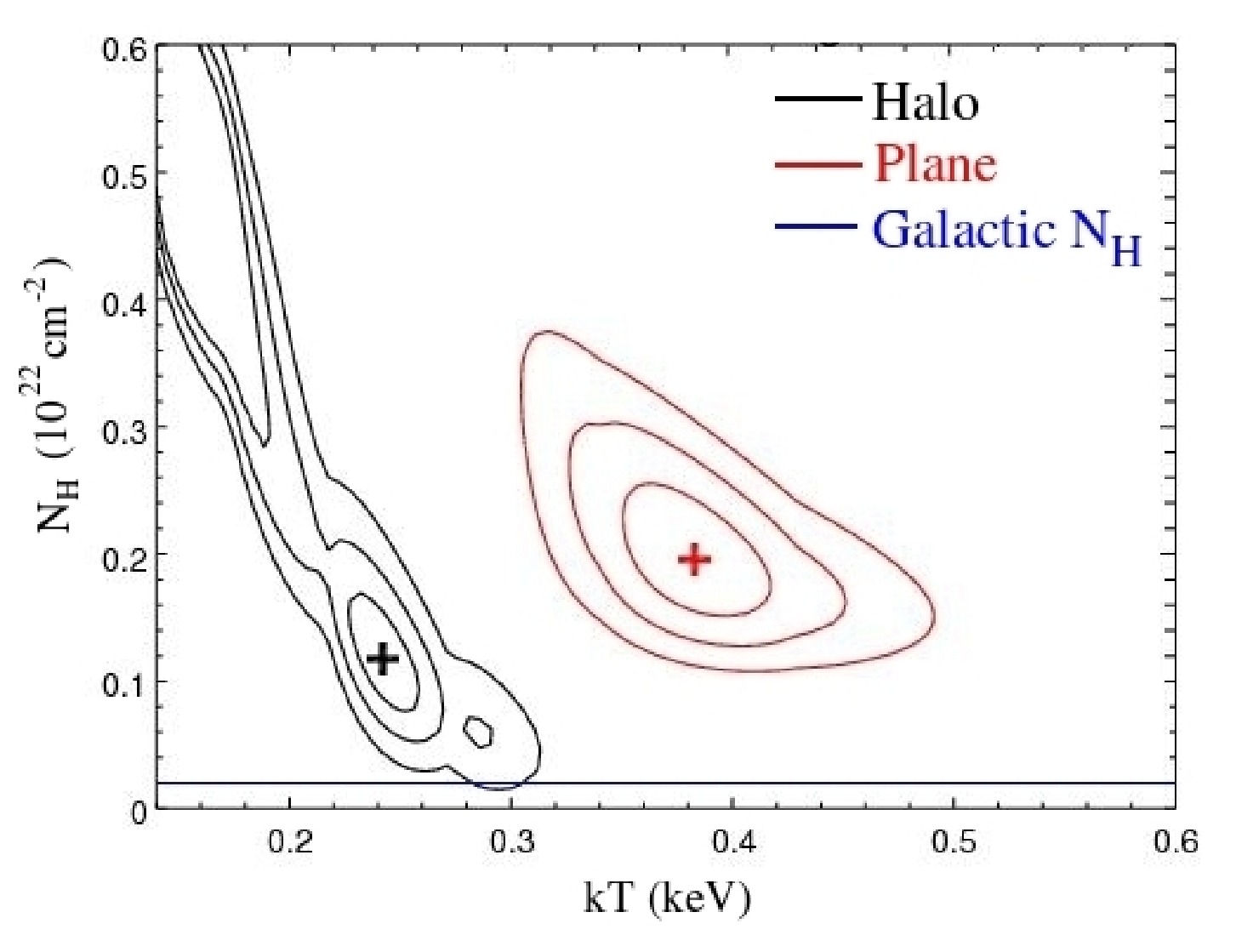}
\caption{1, 2 and 3$\sigma$ confidence level contours in the $kT$ vs $N_{H}$ plane for the halo (black) and plane (red) regions. The galactic line of sight column density towards NGC 4490/85, $N_{H} = 0.018 \times 10^{22} \; \mathrm{cm}^{-2}$, is shown by the blue line.}
\label{nh_kt_contours}
\end{figure}

The spectra are plotted in fig.~\ref{spectra}, and their best fit parameters are summarized in Table~\ref{parameters}. The $1\sigma$ uncertainties in the parameters were calculated using the {\it projection()} command in {\it Sherpa}, which varies each parameter and calculates the change in the $\chi^{2}$ value. To confirm the accuracy of these uncertainties, and also to look at whether any of the parameters were correlated, we looked at the confidence level contours in the planes of both $kT$ and $N_{H}$ plotted against each other and each of the abundances, and also the planes of the Fe abundance against each of the other abundances. In the plane region and subregions these contours were well behaved, showing a single minimum in the $\chi^{2}$ at the best fit values, however in the halo regions and subregions the contours often showed two or three minima. The 1, 2 and 3$\sigma$ confidence level contours in the plane and halo regions are shown for the $kT$ vs $N_{H}$ plane in fig.~\ref{nh_kt_contours}; they show that the gas in the halo region does have a lower temperature than in the plane region, but the hydrogen column density in the halo is uncertain due to the presence of two local minima either side of the best fit value. Therefore, while the best fit parameters do find the global $\chi^{2}$ minimum, the presence of other local minima may make the uncertainties larger than those found by the {\it projection()} command in {\it Sherpa} in the halo regions and subregions.

These fits suggest that the temperature of the gas is higher in the plane of NGC 4490, which could explain the harder color in the plane that is seen in fig.~\ref{smoothed}. We cannot compare $N_{H}$ in the plane and halo from the spectral fitting alone because it is poorly constrained in the halo, however maps of the 21 cm line emission suggest that $N_{H}$ is higher in the plane (see fig.~\ref{h1} in \S6.1). If this is true then this could also contribute to the harder color seen in the plane.

The spectral fits suggest that the metal abundances are higher in the halo, particularly in the outflows, however the large uncertainties in these parameters prevent us from making a conclusive comparison between the abundances in the halo and the plane. The emission line strength map also suggests that there are higher abundances of Fe and Ne in the outflows, however if the absorption is higher in the plane, as suggested by the 21 cm line map (fig.~\ref{h1}), then the Fe and Ne emission may simply be absorbed in the plane. Therefore we cannot conclusively compare the plane and halo based on the emission line strength map either.

\subsubsection{Abundance Ratios}

\begin{turnpage}
\begin{figure}[h!]
\mbox{
	\includegraphics[width=65mm]{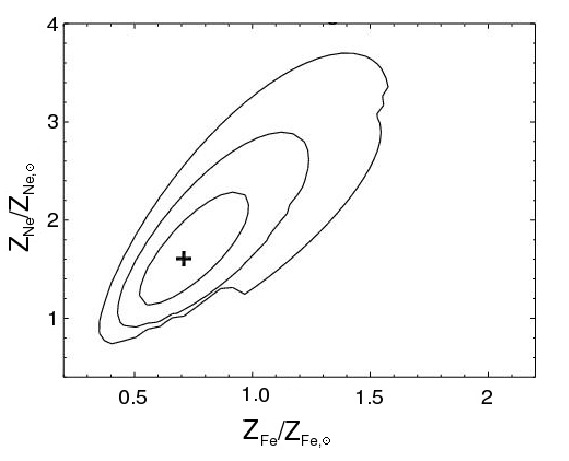}
	\includegraphics[width=65mm]{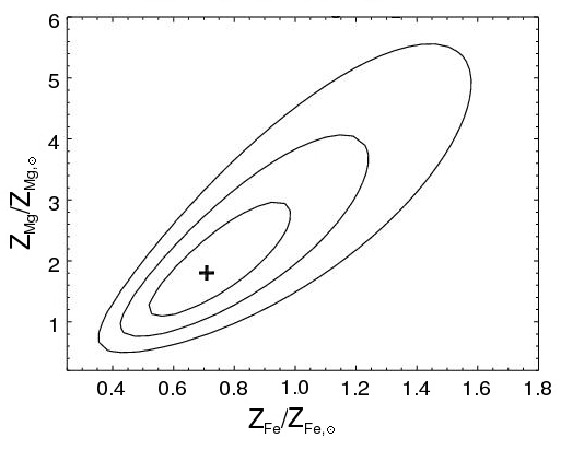}
	\includegraphics[width=65mm]{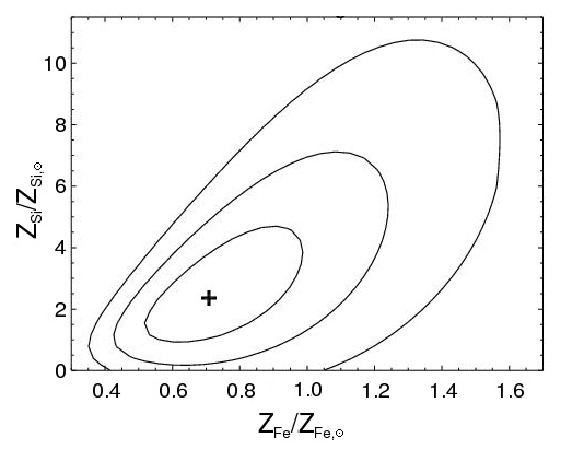}}
\caption{Confidence level contours (1, 2 and 3$\sigma$) from the plane region of NGC 4490 for Fe abundance against Ne abundance (left), Mg abundance (center) and Si abundance (right).}
\label{abun_contours}
\end{figure}
\end{turnpage}

Using the abundances found from the spectral fits in each region and subregion, we calculated the ratios of the Ne, Mg and Si abundances with respect to the Fe abundance. These ratios are summarized in Table~\ref{abundance_ratios}. The uncertainties in the abundances obtained from the spectral fits are asymmetric (see Table~\ref{parameters}), so we cannot use the standard rules for error propagation, which assume that the errors are Gaussian. Therefore, to calculate the uncertainties in the abundance ratios we introduced a new parameter into the {\it Sherpa} model which we linked with the model parameters so that this new parameter was equal to each of the abundance ratios in turn. We were then able to calculate the uncertainties in {\it Sherpa} using the {\it projection()} command. This method correctly accounts for the asymmetric errors, since they are calculated directly from the changes in $\chi^{2}$, and it also accounts for correlations between model parameters, which the error propagation rules cannot as they assume that the parameters are independent. The errors calculated by this method are significantly smaller than if they had been calculated using the standard error propagation rules, which suggests that the abundances are correlated. To investigate this effect further, we ran a Markov Chain Monte Carlo (MCMC) simulation to simulate the distribution of the abundance ratios in our spectral fits. In the plane and central plane regions the results of this simulation agreed with the smaller uncertainties found by {\it Sherpa}, and the confidence level contours in these regions confirm that the abundances are strongly correlated, as they are narrow ellipses in which the individual abundances vary over a large range, while their ratios only vary a little. These confidence level contours from the plane region of NGC 4490 are shown in fig.~\ref{abun_contours}. However, in the other regions the MCMC simulation was strongly affected by the unusual shape of the confidence level contours in these regions, for example getting stuck in local minima, which resulted in confidence intervals that were either strongly asymmetric or did not enclose the best fit values.

\subsection{Non-Equilibrium Ionization}

\begin{figure}[h!]
\centering
\mbox{
	\includegraphics[width=80mm]{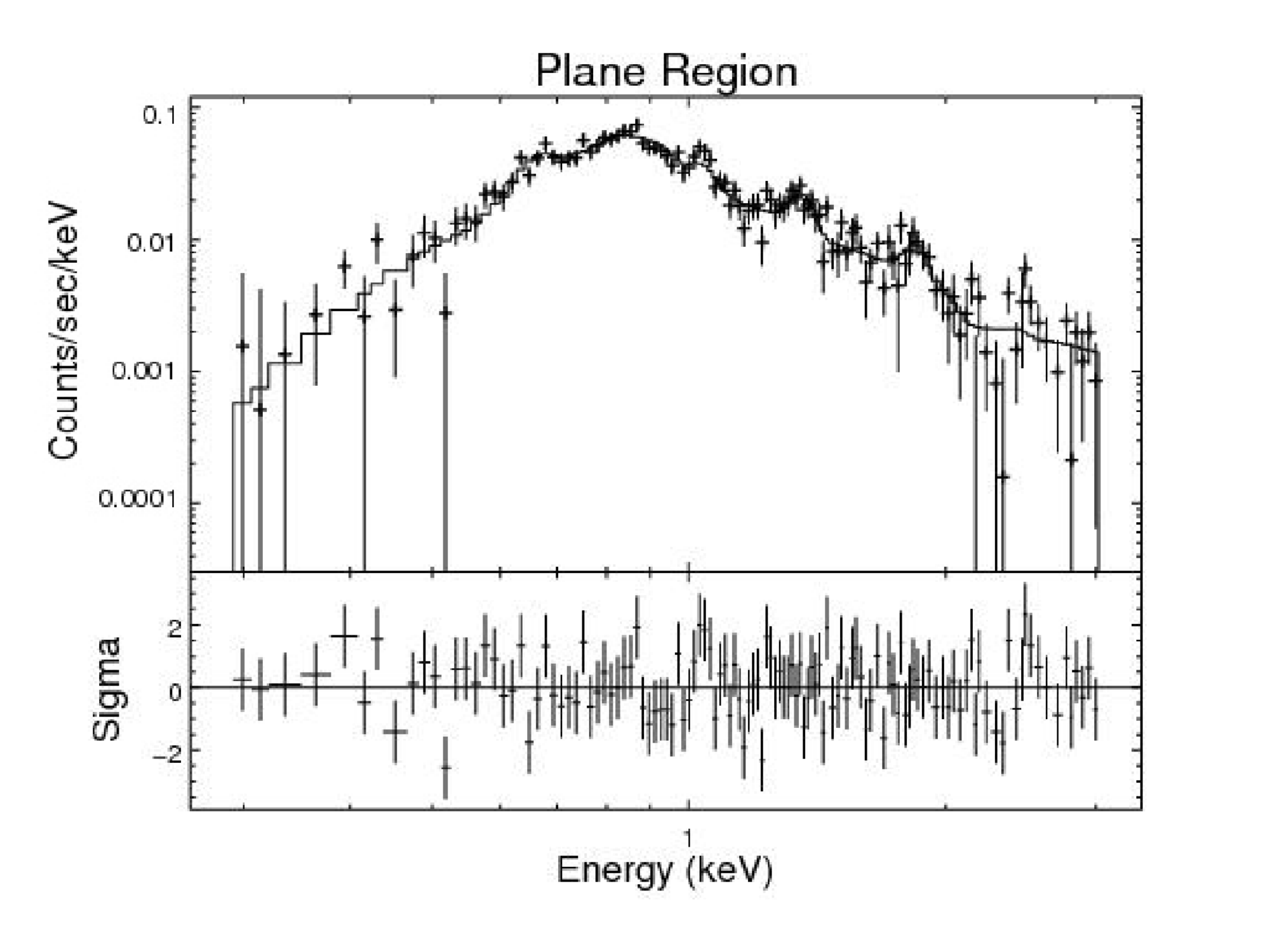}
	\includegraphics[width=80mm]{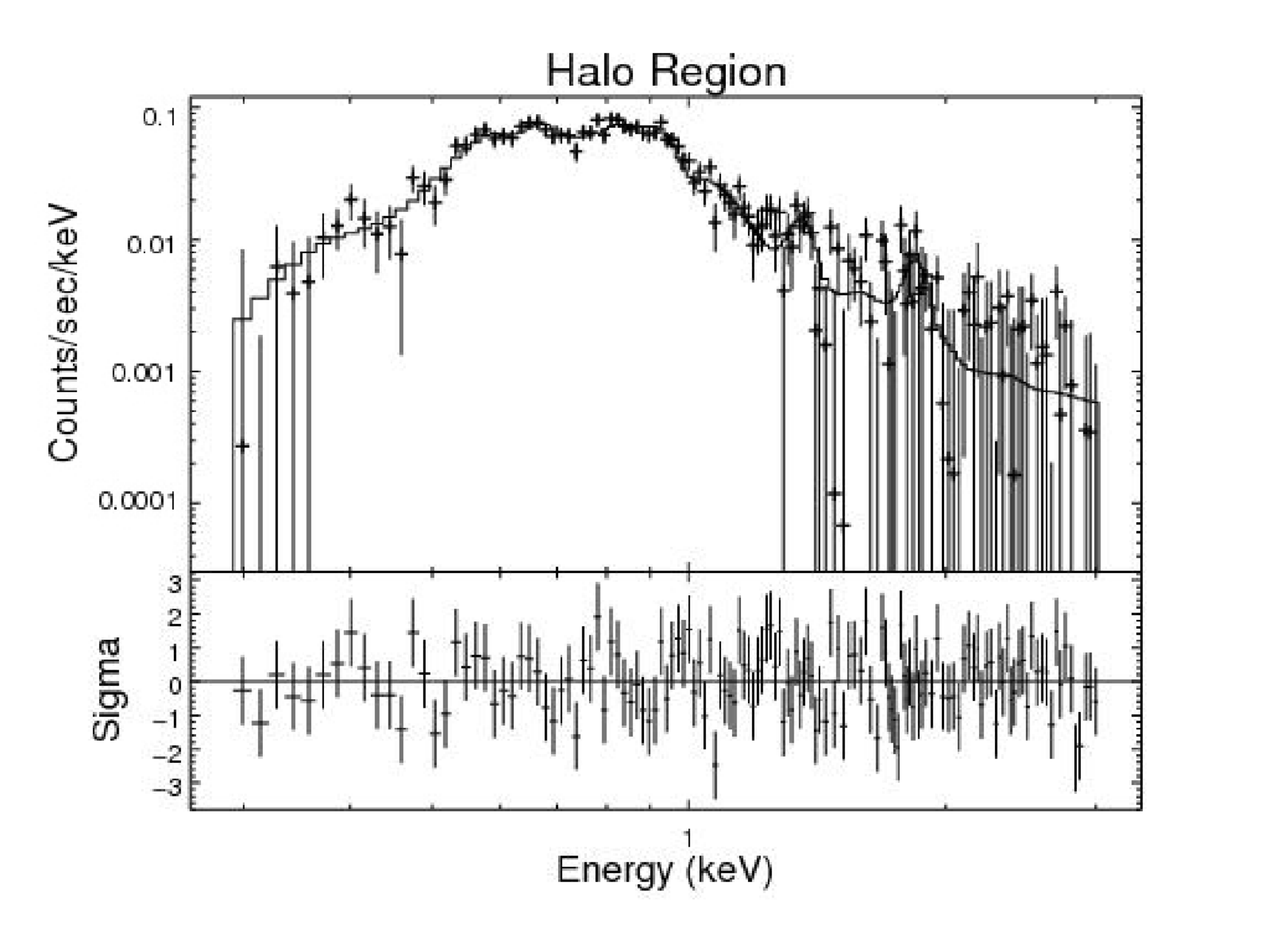}}
\mbox{
	\includegraphics[width=80mm]{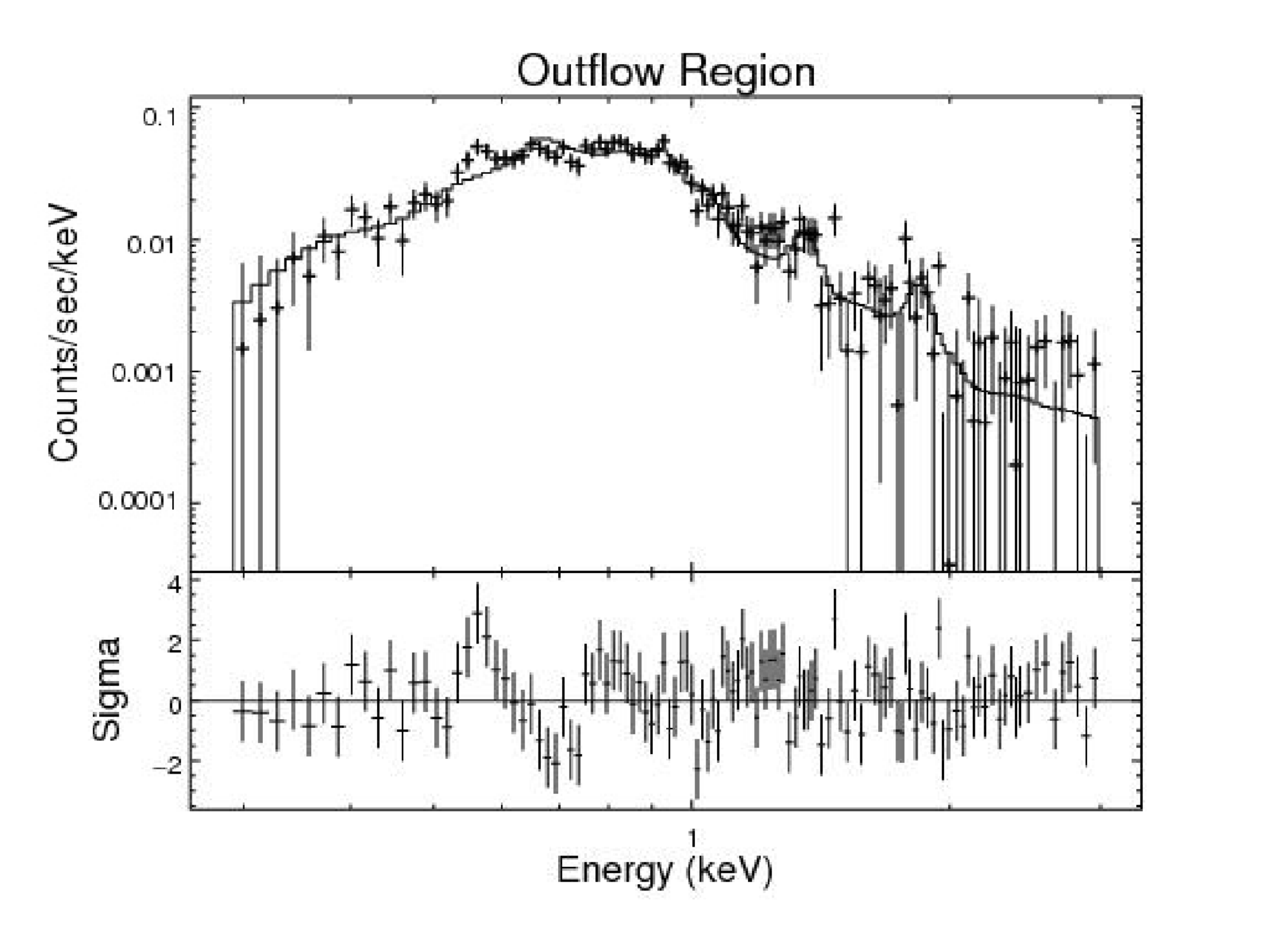}
	\includegraphics[width=80mm]{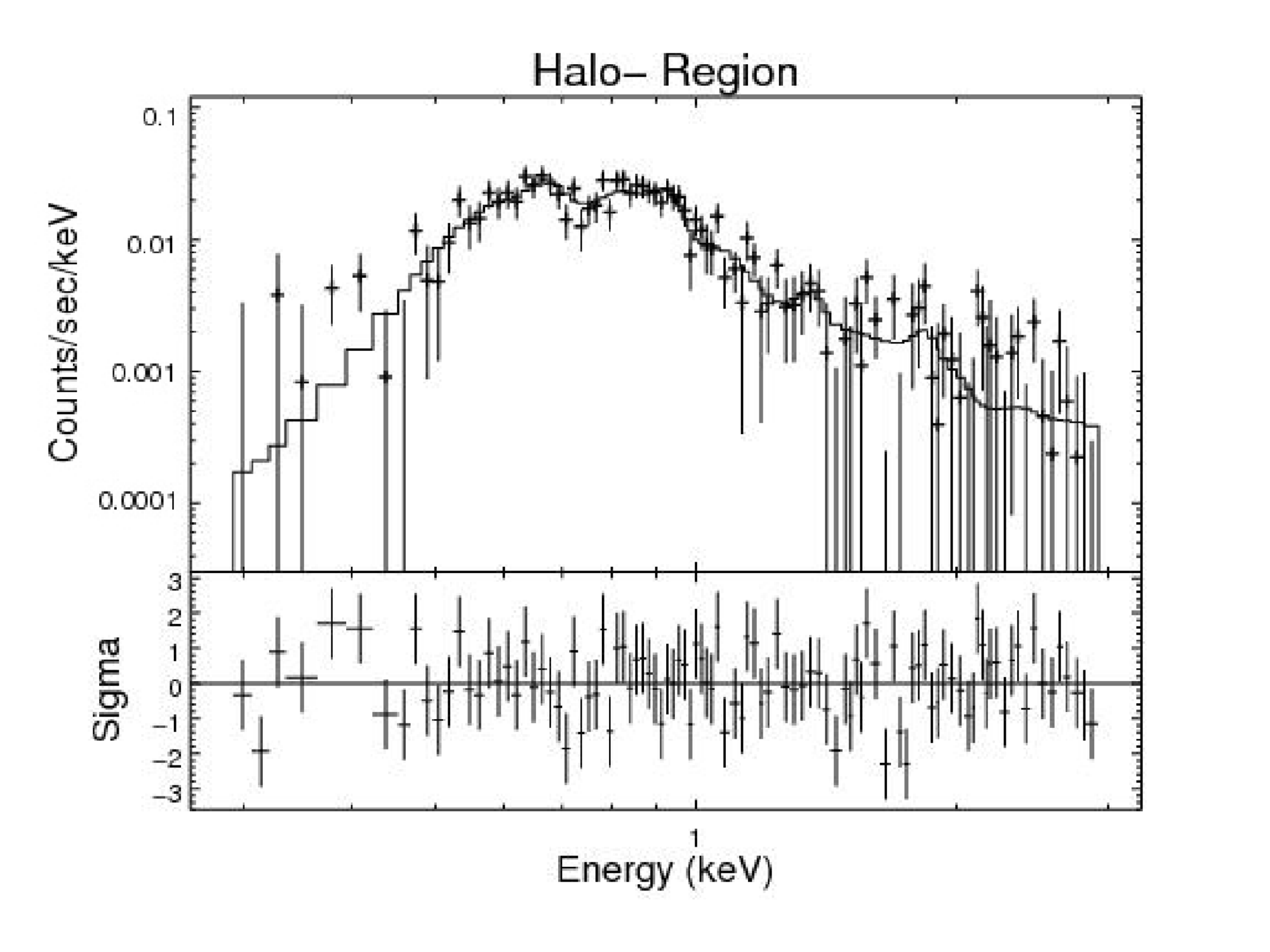}}
\mbox{
	\includegraphics[width=80mm]{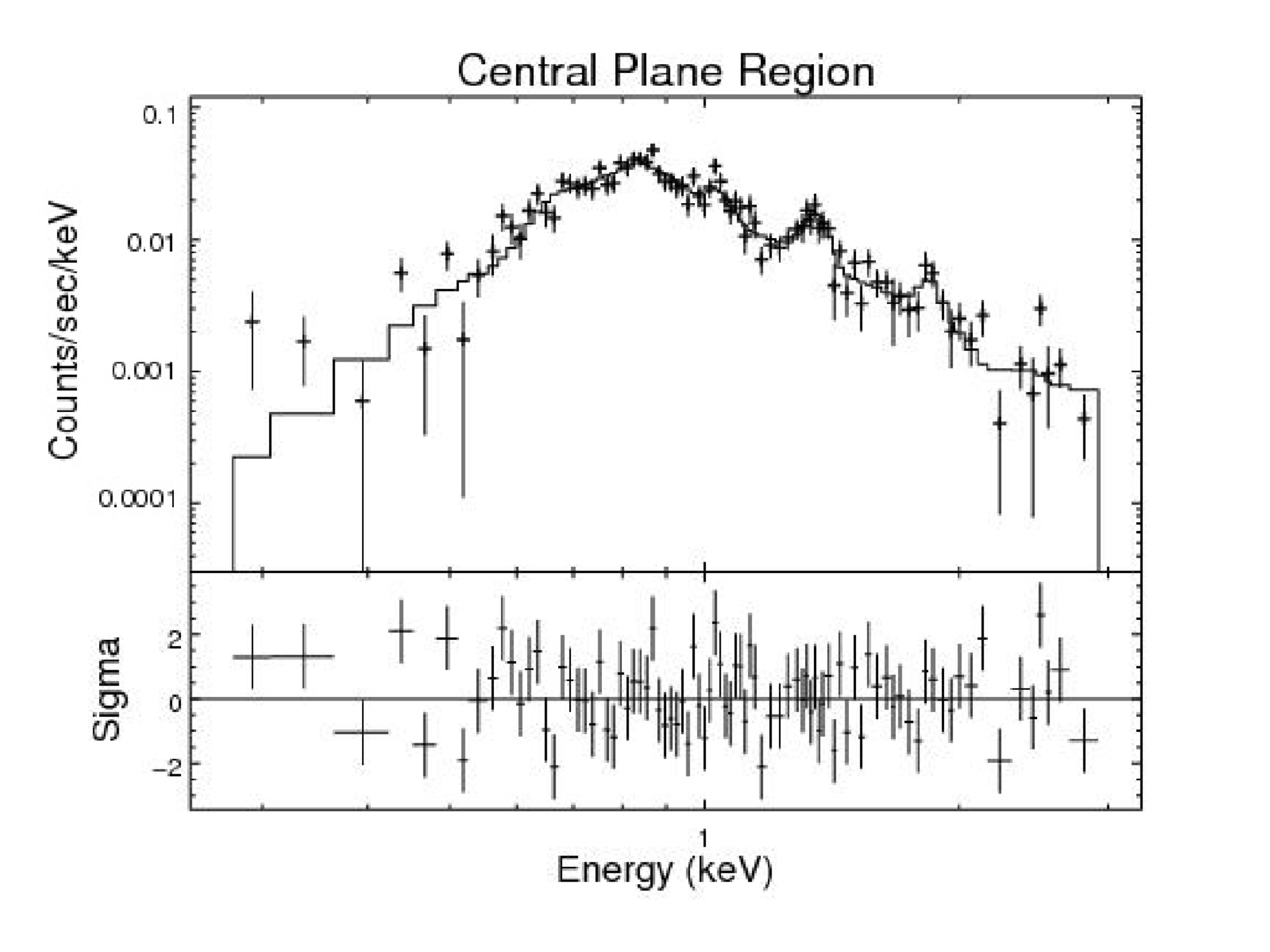}
	\includegraphics[width=80mm]{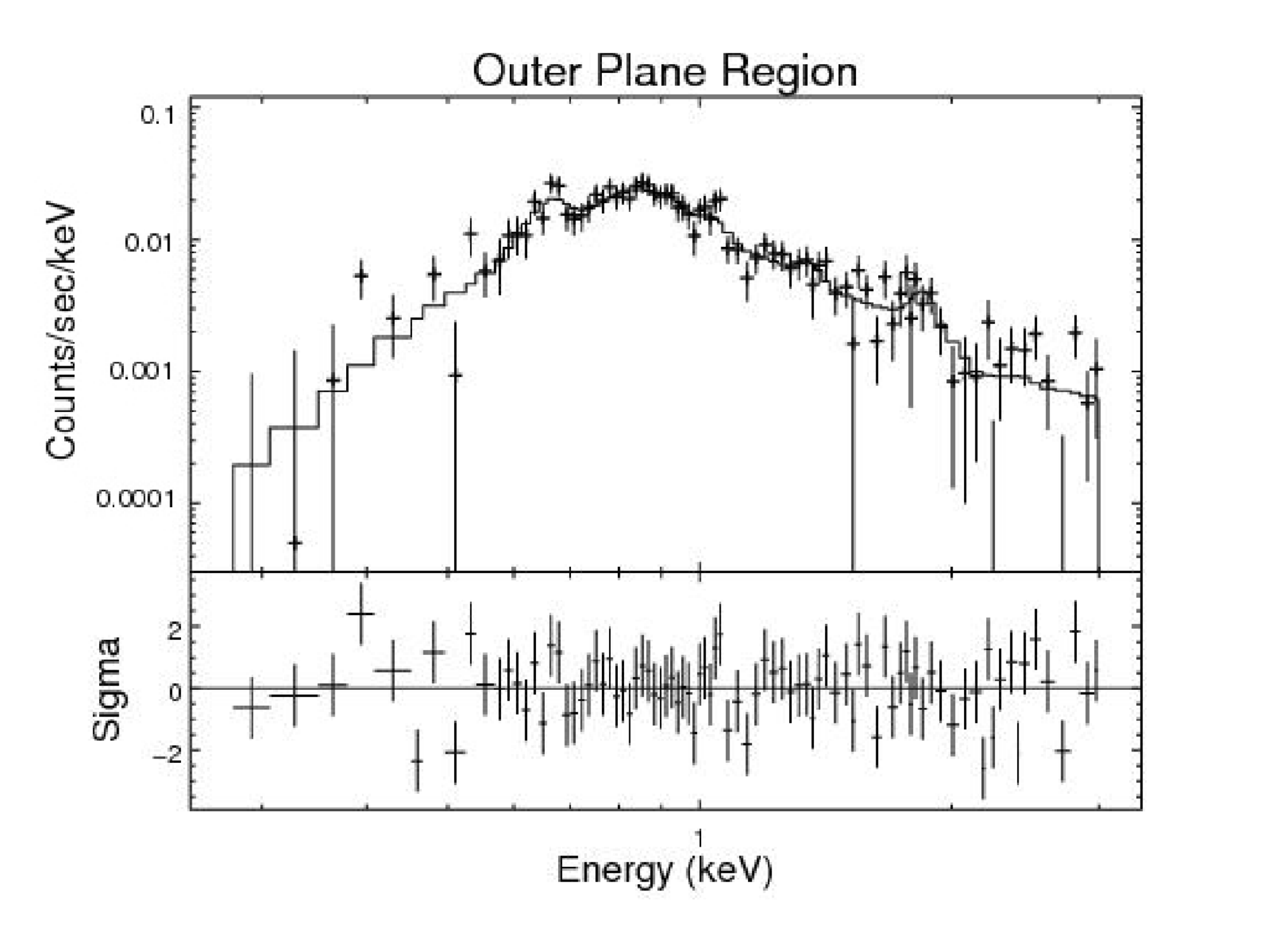}}
\caption{X-ray spectra extracted from the plane and halo regions and their subregions, plotted with their best-fit absorbed NEI plus power-law models. The fit residuals are shown below each spectrum.}
\label{nei_spectra}
\end{figure}

We have characterized the thermal X-ray emission from the hot gaseous ISM using the APEC model, which assumes that the plasma is in a collisional ionization equilibrium. However, the relatively high abundance ratios that we find in the hot ISM suggest that this plasma may originate from young supernova remnants (SNRs). The X-ray emission from this plasma would then be dominated by shocked supernova ejecta for which this assumption of ionization equilibrium may not be true. Therefore we also fitted the spectra of the diffuse X-ray emission from NGC 4490 with an absorbed Non-Equilibrium Ionization (NEI) model ({\it xsvnei} in {\it Sherpa}) plus a power-law component. This model was fitted using the same method described in \S4.2 for the absorbed APEC plus power-law model. The best fit parameters of this model are summarized in Table~\ref{nei_pars} and the spectra are plotted with this model in fig.~\ref{nei_spectra}. The abundance ratios from this model were calculated as described in \S4.2.2 and are summarized in Table~\ref{nei_abund}.

The spectral fits using the NEI model give very similar values for the reduced $\chi^{2}$ as the fits using the APEC model, however the ionization timescale parameter $\tau$ in the NEI fits is typically $\tau \sim 10^{11} \; \mathrm{s} \; \mathrm{cm}^{-3}$. If the plasma was close to a collisional ionization equilibrium then we would expect $\tau \gtrsim 10^{12} \; \mathrm{s} \; \mathrm{cm}^{-3}$ \citep{masai}, so this suggests that the plasma is in a nonequilibrium ionization state. We will therefore use the results of the NEI fits for the rest of the analysis.

The spectral fits using the APEC model suggested that the gas was hotter in the plane of NGC 4490 than in the halo (see \S4.2.1), however the parameters of the NEI model were poorly constrained, and the temperature in the plane region is only a lower limit, so we cannot make a conclusive comparison between the temperatures of the gas in the plane and the halo. The metal abundances in the NEI fits are higher in the halo than in the plane, similar to the results using the APEC model, however the uncertainties from the NEI fits are even larger, so we still cannot conclusively compare the metal abundances in the plane and halo. We also note that the abundance ratios derived from the NEI model are generally consistent with those derived from the APEC model.

\section{Physical Properties Of The Hot ISM Gas}

Using the results from the spectral analysis we can derive properties of the hot ISM gas in each region and subregion. We use the results of the spectral fits with the NEI model in this section as the NEI fits suggested that the plasma is not in a collisional ionization equilibrium. The normalization of the NEI model is proportional to the emission measure, $EM = n^{2}V = \int n_{e} n_{H} \,dV$, where $n$ is the number density and $V$ is the volume. To calculate the electron number density $n_{e}$ we need to make assumptions about the volume of the emitting region. For the volume of the plane we assume that we are viewing the disk of the galaxy edge on, so that the emitting region is a circular disk with thickness equal to the width of the region that we see, and a diameter equal to the length of this region. This gives a volume $V_{plane} = 6.62 \times 10^{66} \; \mathrm{cm}^{3}$. We then assume that the central plane subregion is also a disk, giving a volume $V_{central} = 0.56 \times 10^{66} \; \mathrm{cm}^{3}$, and thus the outer plane has a volume $V_{outer} = 6.06 \times 10^{66} \; \mathrm{cm}^{3}$. For the halo we assume that the emitting region consists of two conical sections either side of the plane, with the base as the galaxy disk, an opening angle of $120^{\circ}$ and extending to the edge of the halo region. This gives a volume $V_{halo} = 215.7 \times 10^{66} \; \mathrm{cm}^{3}$. For the outflow subregion, we assume that the emitting region consists of two pairs of conical sections, one pair extending either side of the nucleus of NGC 4490 and the other pair extending either side of the base of the spiral arm that extends towards the companion galaxy, NGC 4485. This gives a volume of $V_{outflow} = 2.27 \times 10^{66} \; \mathrm{cm}^{3}$, and thus the rest of the halo, excluding the outflows, has a volume of $V_{halo-} = 213.4 \times 10^{66} \; \mathrm{cm}^{3}$.

Using these volumes and the emission measure $EM = n^{2}V$, we can calculate the electron number densities $n_{e}$ and hence the total mass of hot gas, and then using the temperatures $kT$ from the spectral analysis we can calculate the pressure $p = 2n_{e}kT$ and the thermal energy $E_{th} = 3n_{e}VkT$. 

We used the {\it calc\_energy\_flux} command in {\it Sherpa} to calculate the observed X-ray flux in each region in the energy range $0.3-10.0$ keV. By setting the model in {\it Sherpa} to just the absorbed thermal or power-law component, we can calculate the fluxes from each component separately. The uncertainties in these fluxes were calculated using a Monte Carlo simulation to randomly sample the model parameters and produce the resulting distribution of energy fluxes. This simulation requires that the uncertainies in the model parameters are well defined, however the parameters of the NEI model were often poorly constrained, so we were unable to calculate the uncertainties in the fluxes of the thermal component. We then calculated the X-ray luminosities from these fluxes. Finally, we defined the cooling time in each region and subregion as the ratio of the thermal energy $E_{th}$ to the X-ray luminosity of the thermal component, $L_{0.3-10.0 keV}^{therm}$: $\tau_{c} = E_{th}/L_{0.3-10.0 keV}^{therm}$. Since we were unable to calculate the uncertainties in the fluxes of the thermal component, we were also unable to calculate the uncertainties in the luminosities of the thermal component and the cooling times. The emission measures, fluxes and luminosities are summarized in Table~\ref{em_par}, and the physical properties of the hot gas are summarized in Table~\ref{phys_par}. Since the volume of the gas will also depend on the filling factor $\eta$ we present the results in Table~\ref{phys_par} as functions of $\eta$.

The uncertainties in the best fit model parameters from which these hot gas properties are derived are asymmetric, so to calculate the uncertainties in the hot gas properties we used the same method that we used for the abundance ratios, introducing a new parameter into the {\it Sherpa} model that was linked to the other model parameters to make it equal to each gas property in turn and then using the {\it projection()} command in {\it Sherpa} to calculate the 1$\sigma$ uncertainties.

\section{Discussion}

\subsection{The Multi-Phase ISM of NGC 4490/85}

The smoothed diffuse X-ray image in fig.~\ref{smoothed} shows that there is a halo of hot, X-ray emitting gas surrounding NGC 4490, and the emission from this halo gas appears softer than the X-ray emission from the plane of NGC 4490. This could be due to hotter intrinsic emission or stronger absorption in the plane, although the temperatures and absorbing hydrogen column densities from the spectral analysis were too poorly constrained to compare these parameters in the plane and halo (see Table~\ref{nei_pars}). We can also see structures in the halo emission that suggest outflows of hot gas originating from the disk of NGC 4490. 

The X-ray emission from NGC 4490/85 gives us information about the hot phase of the ISM in these galaxies, but to study how the hot ISM interacts with other phases of the ISM we need to compare the X-ray emission to emission at other wavelengths.

\begin{figure}[h!]
\centering
\includegraphics[width=160mm]{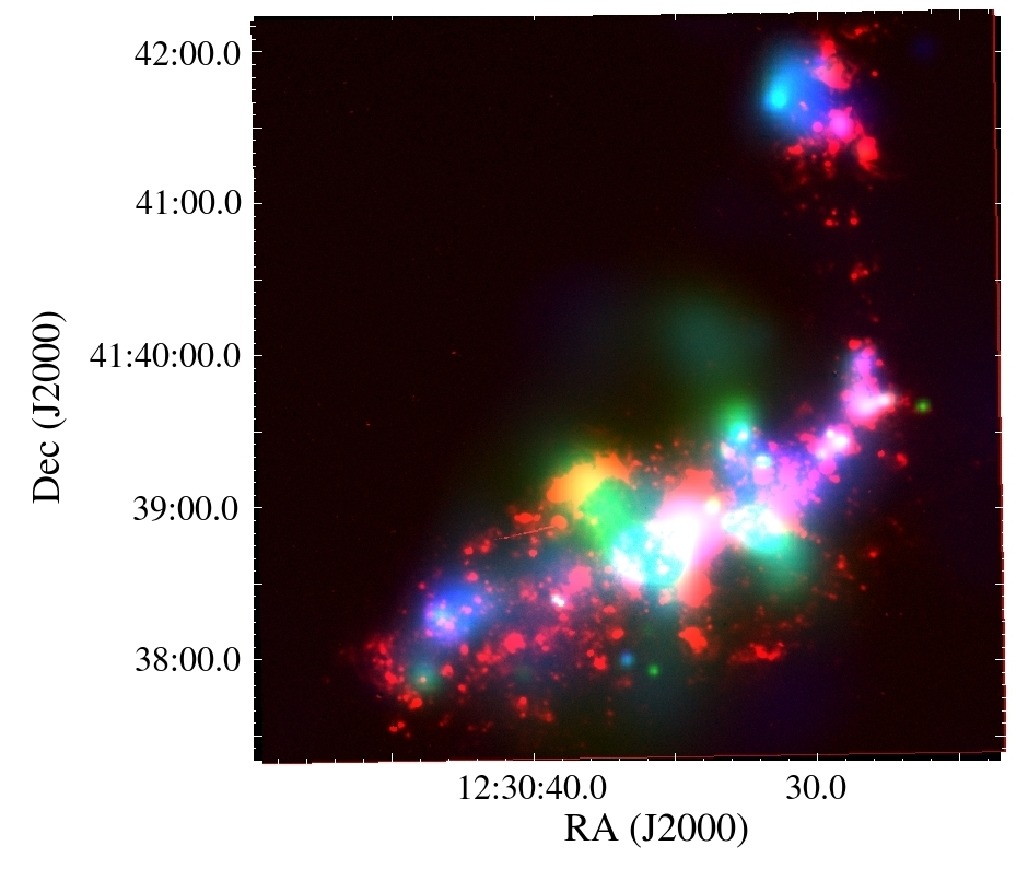}
\caption{Composite color image of NGC 4490/85. H$\alpha$ emission is shown in red, soft X-ray emission from the energy band $0.3-1.5$ keV is shown in green and hard X-ray emission from the energy band $1.5-6.0$ keV is shown in blue. The {\it Chandra} data has a spatial resolution of $\sim0.7$ arcsec and the astrometry in the H$\alpha$ image was accurate to within $\sim0.5$ arcsec, however there are insufficient common sources to cross-match the astrometry for this field.}
\label{halpha}
\end{figure}

NGC 4490/85 have been imaged in H$\alpha$ by \citet{kennicutt08} using the Steward Observatory's 2.3m Bok telescope on Kitt Peak. We used this image and the {\it Chandra} X-ray data to create a composite color image of H$\alpha$ and X-ray emission, which is shown in fig.~\ref{halpha}. This image suggests that there is an anticorrelation between H$\alpha$ and X-ray emission - in particular, the outflows that are seen in the soft X-ray band appear to be bounded by filaments of H$\alpha$ emission, suggesting that the hot gas in these outflows is either pushing the cooler ionized hydrogen gas aside as it expands into the halo, or it is being confined by the cool gas. These scenarios are consistent with the shape of the X-ray surface brightness profiles and the spectral hardness ratio profiles of the outflows, which suggested that the hot gas was being heated by shocks with cooler gas that was already present in the halo. Similar hot/ionized gas interactions were seen in the Antennae galaxies \citep{fabbiano01, fabbiano03}.

\begin{figure}[h!]
\centering
\includegraphics[width=160mm]{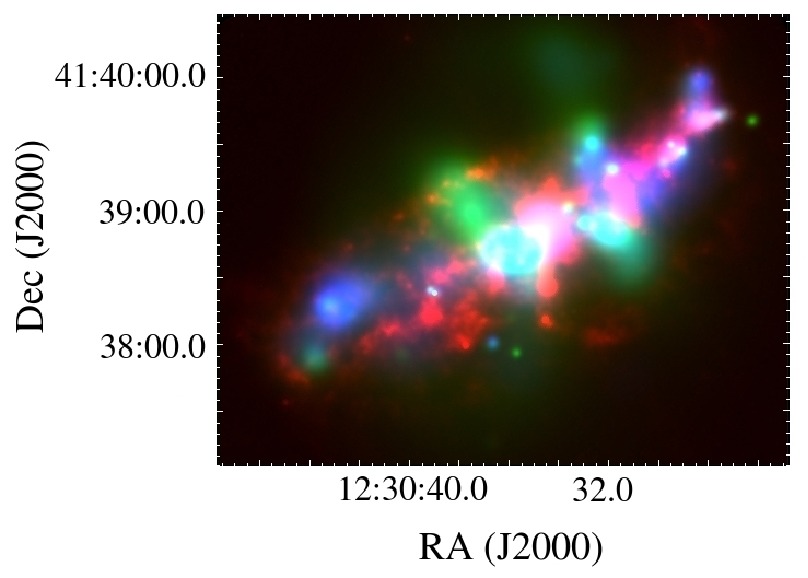}
\caption{Composite color image of NGC 4490. 8$\mu$m infrared emission is shown in red, soft X-ray emission from the energy band $0.3-1.5$ keV is shown in green and hard X-ray emission from the energy band $1.5-6.0$ keV is shown in blue.}
\label{ir}
\end{figure}

We obtained mid-infrared images of NGC 4490 from the Spitzer archive taken with the IRAC instrument on the Spitzer Space Telescope. We combined the image at a wavelength of 8$\mu$m with the {\it Chandra} X-ray data to create a composite color image of mid-infrared and X-ray emission, which enables us to compare the distributions of warm dust and the hot ISM in NGC 4490. This image is shown in fig.~\ref{ir}, and it suggests that there is an anticorrelation between the mid-infrared emission from warm dust and the X-ray emission from the outflows, which suggests that the dust is predominantly found in regions where there is little hot gas. This is similar to what we found in the H$\alpha$ emission, suggesting that the warm dust may be mixed with the cool, ionized hydrogen gas.

\begin{figure}[h!]
\centering
\includegraphics[width=160mm]{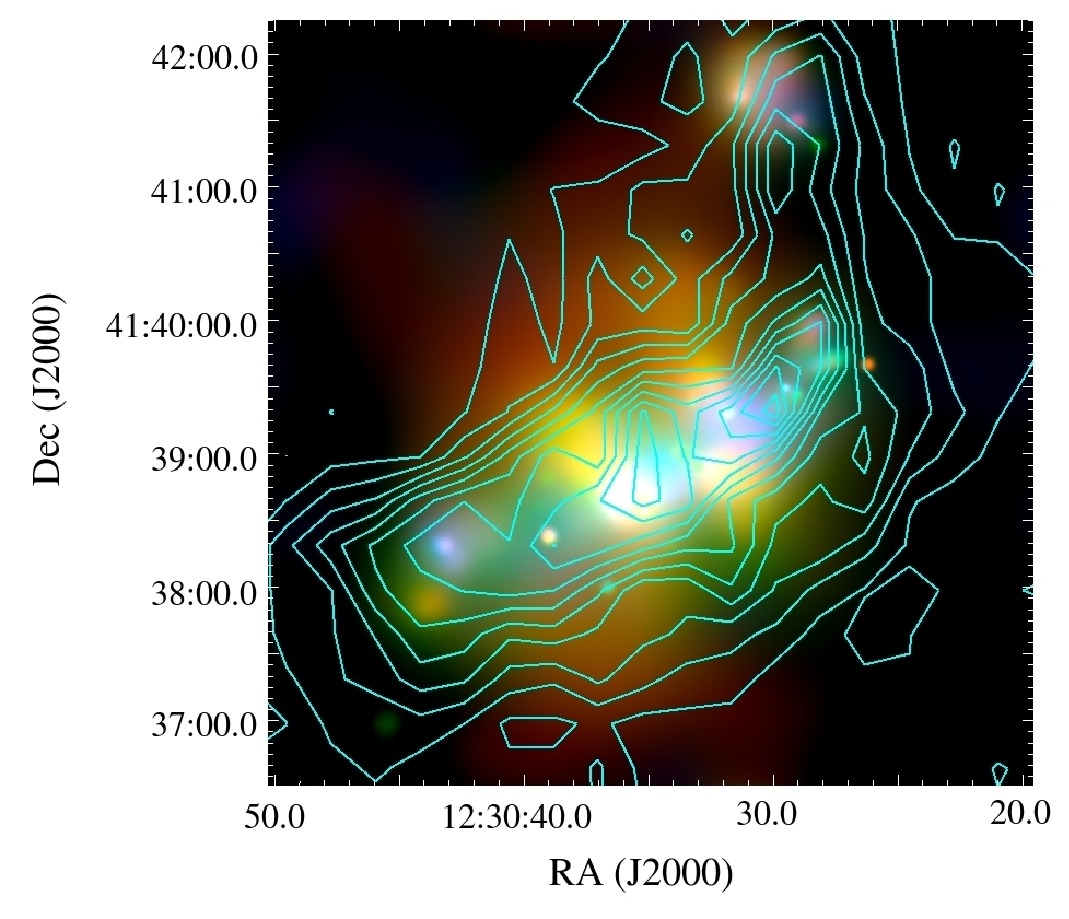}
\caption{Contours of the 21cm line emission from HI with a resolution of 17.9 arcsec. The astrometry of this data is unknown. These contours are superposed over a smoothed, mapped-color image of the diffuse X-ray emission from NGC 4490/85. Soft X-ray emission from the energy band $0.3-0.65$ keV is shown in red, medium X-ray emission from the energy band $0.65-1.5$ keV is shown in green and hard X-ray emission from the energy band $1.5-6.0$ keV is shown in blue.}
\label{h1}
\end{figure}

The 21 cm line emission from NGC 4490/85 has been observed by \citet{van_der_hulst} using the Westerbork Synthesis Radio Telescope with a resolution of 17.9 arcsec, which we obtained from the NASA/IPAC Extragalactic Database (NED)\footnote{http://nedwww.ipac.caltech.edu/}. We used this image to create contours of the line emission from the neutral hydrogen in NGC 4490/85 and superposed these on the smoothed, mapped-color image of the diffuse X-ray emission. This is shown in fig.~\ref{h1}. 

The strong HI source at the center of NGC 4490 does not appear to correspond to any feature in the X-ray emission, although it is only $\sim 18$ arcsec from the base of the two Eastern outflows, which is comparable to the resolution limit of the 21cm data, so it is possible that these features are related. There is also a second strong HI source towards the Western edge of the disk of NGC 4490, which is close to a bright clump of X-ray emission, although they do not coincide exactly. However, the minimum in 21cm line emission between these sources does coincide with the base of the two Western outflows seen in the soft X-ray emission, suggesting that there may be an anticorrelation between the hot gas and the neutral HI gas.

The contours in fig.~\ref{h1} show a bridge of HI gas extending from NGC 4490 to its companion galaxy NGC 4485. Previous studies have found that this HI bridge coincides with a chain of HII regions between the galaxies \citep{viallefond} and also with CO emission \citep{clemens00}, however fig.~\ref{h1} shows that a similar bridge of hot gas seen in the X-ray emission does not coincide with these other phases of the ISM, but is instead further to the North. 

The HI emission from NGC 4490/85 has also been studied using the VLA by \citet{clemens98}, who found that these galaxies are embedded in an unusually large HI envelope that extends to 56 kpc perpendicular to the plane of NGC 4490. They suggest that this envelope may have been formed by star formation driven outflows, and the presence of outflows in the X-ray emission supports this hypothesis.

\citet{clemens00} found that emission from HI, molecular CO and ionised hydrogen gas in the companion galaxy, NGC 4485, has been displaced from the optical emission due to ram-pressure stripping of the ISM as it moves through the extended HI envelope of NGC 4490. Fig.~\ref{h1} shows that the 21cm line emission and diffuse X-ray emission peaks do not coincide, and fig.~\ref{halpha} also shows an anticorrelation between H$\alpha$ and diffuse X-ray emission in NGC 4485. However, comparing the diffuse X-ray emission with the R-band optical emission (see fig.~\ref{optical_xray}) shows that the diffuse X-ray emission is displaced from the optical continuum emission. This suggests that the hot phase of the ISM has also been stripped from NGC 4485, although it has not been displaced as far as the other components of the ISM have been. The presence of an X-ray tail in NGC 4485 was suggested by \citet{roberts}, and is confirmed by this result. Similar examples of ram-pressure stripping of the hot, X-ray emitting ISM have been found in cluster galaxies, for example M86 in the Virgo cluster \citep{forman}, and ESO 137-001 and ESO 137-002 in the rich cluster Abell 3627 \citep{sun}. However, few such examples have been found in isolated pairs of galaxies.

Measurements of the 21cm line emission taken by \citet{viallefond} using the Westerbork Synthesis Radio Telescope show that the peak hydrogen column density $N_{H}$ in the center of NGC 4490 is $N_{H} = 7.6 \times 10^{21} \; \mathrm{cm}^{-2}$. Our spectral analysis found a hydrogen column density in the central plane subregion of NGC 4490 of $N_{H} = 2.2 \times 10^{21} \; \mathrm{cm}^{-2}$, however this is the average column density in the central 4.2kpc of the disk of NGC 4490, and not the peak column density. Also, the column density derived from the X-ray spectra is only that of the hydrogen which is absorbing the X-ray emission, whereas the 21cm line emission is from all of the neutral hydrogen along the line of sight, and so includes hydrogen that is behind the X-ray emitting region. 

\subsection{Nature of the Hot Gaseous ISM}

Following the method used by \citet{strickland04} on a sample of ten star-forming, edge-on disk galaxies, we investigate the nature of the hot gas in the halo and the outflows by studying the shapes of the radial X-ray surface brightness profiles in these regions. For example, if this gas is expanding freely then we would expect the surface brightness $\Sigma_{X}$ to follow a power-law, with $\Sigma_{X} \propto r^{-3}$ \citep{chevalier}. Alternatively, the gas in these outflows may be heated via shocks with cooler gas that is already present in the halo, in which case we would expect the brightness profiles to follow the density profile of this gas. There are several models that could explain the presence of such gas in the halo, for example galactic fountain models \citep[e.g.][]{shapiro,bregman}, and these models predict that the density profile of this gas will be either exponential or Gaussian with height above the plane of the galaxy. In \S3.3.1 we fitted power-law, exponential and Gaussian models to these X-ray brightness profiles.

The brightness profiles in the two halo sectors between the outflows are very well described by all three models (with values of $\chi^{2}_{\nu}$ below 1), so we are unable to distinguish between them; however in the outflows the power-law model is a poorer fit than the exponential or Gaussian models. We found that the exponential model is a fairly good fit in the NW and SW outflows, whereas both the exponential and the Gaussian models still give fairly poor fits in the NE and SE outflows. This suggests that there may be more complex structure in these outflows - for example, the residuals from the exponential and the Gaussian models in the NE outflow show two broad peaks, suggesting that it might be better fit by two components with different scale heights, rather than a single exponential or Gaussian component.

The two outflows on the North side of NGC 4490 have similar scales, with exponential scale heights of just over 1kpc, while the halo sector between these outflows is more extended, with an exponential scale height $H\sim2.5$kpc. On the South side, the halo sector is less extended than this, with $H\sim1.4$kpc, and the outflows show more dissimilar scales, with $H\sim0.8$kpc and $H\sim1.8$kpc in the SE and SW outflows respectively.

To further investigate the nature of this gas, it would also be useful to look at how the temperature varies radially. There are too few counts to extract X-ray spectra from each radial bin, however we can measure the spectral hardness ratio $Q$ in each bin (see \S3.3.2), which depends strongly on the temperature of the gas, although it is also affected by absorption. 

\begin{figure}[h!]
\centering
\mbox{
	\includegraphics[width=80mm]{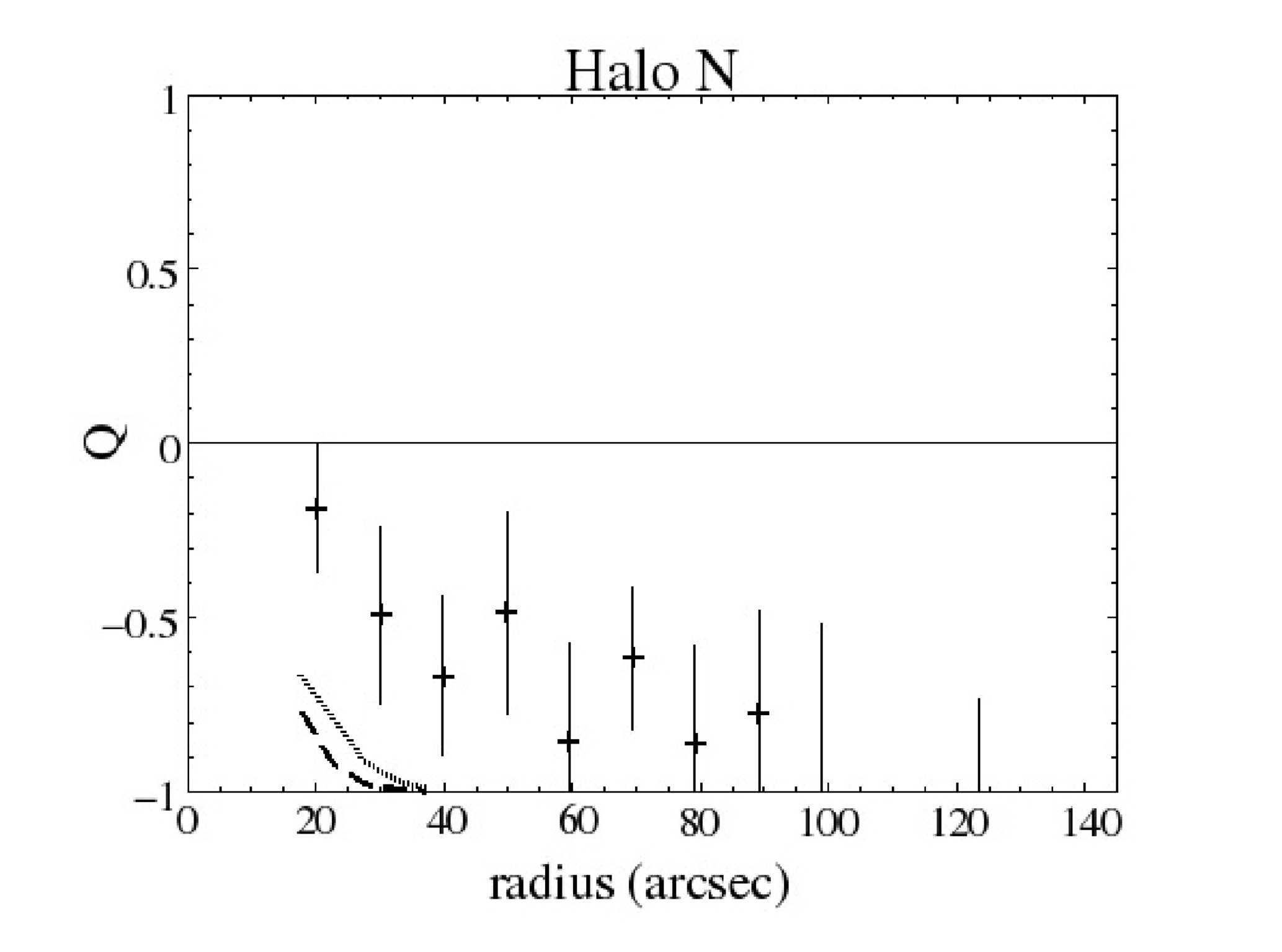}
	\includegraphics[width=80mm]{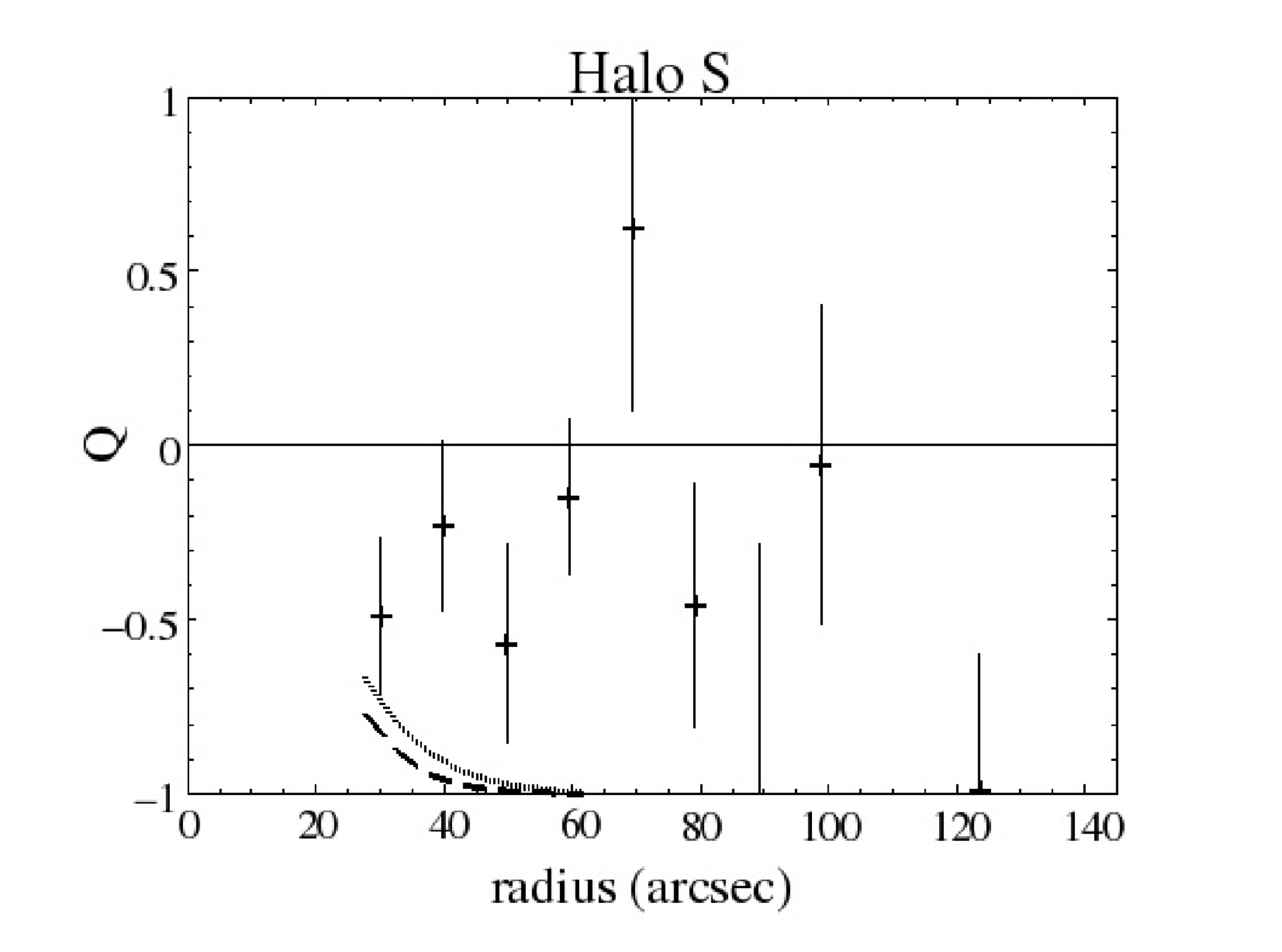}}
\mbox{
	\includegraphics[width=80mm]{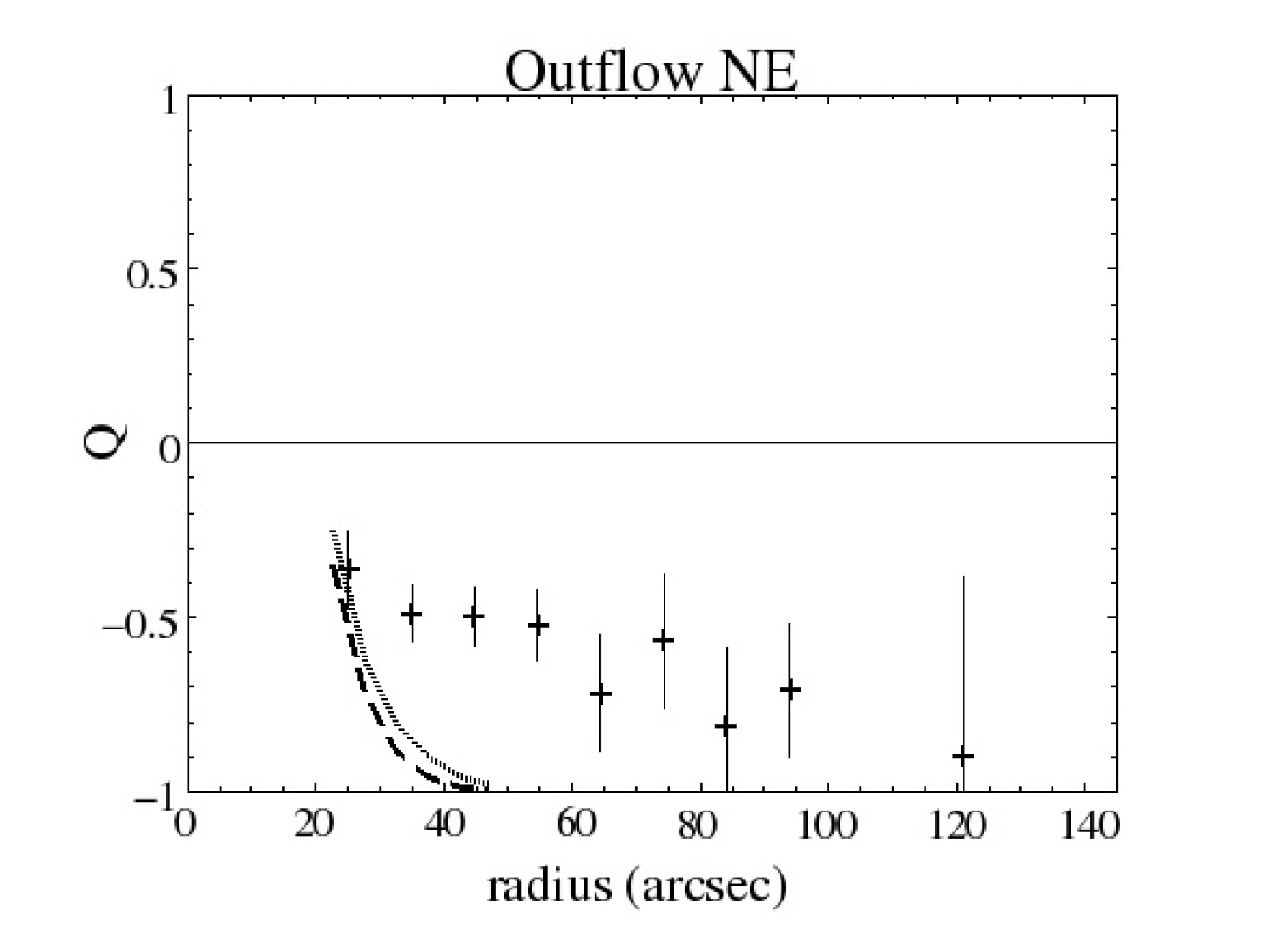}
	\includegraphics[width=80mm]{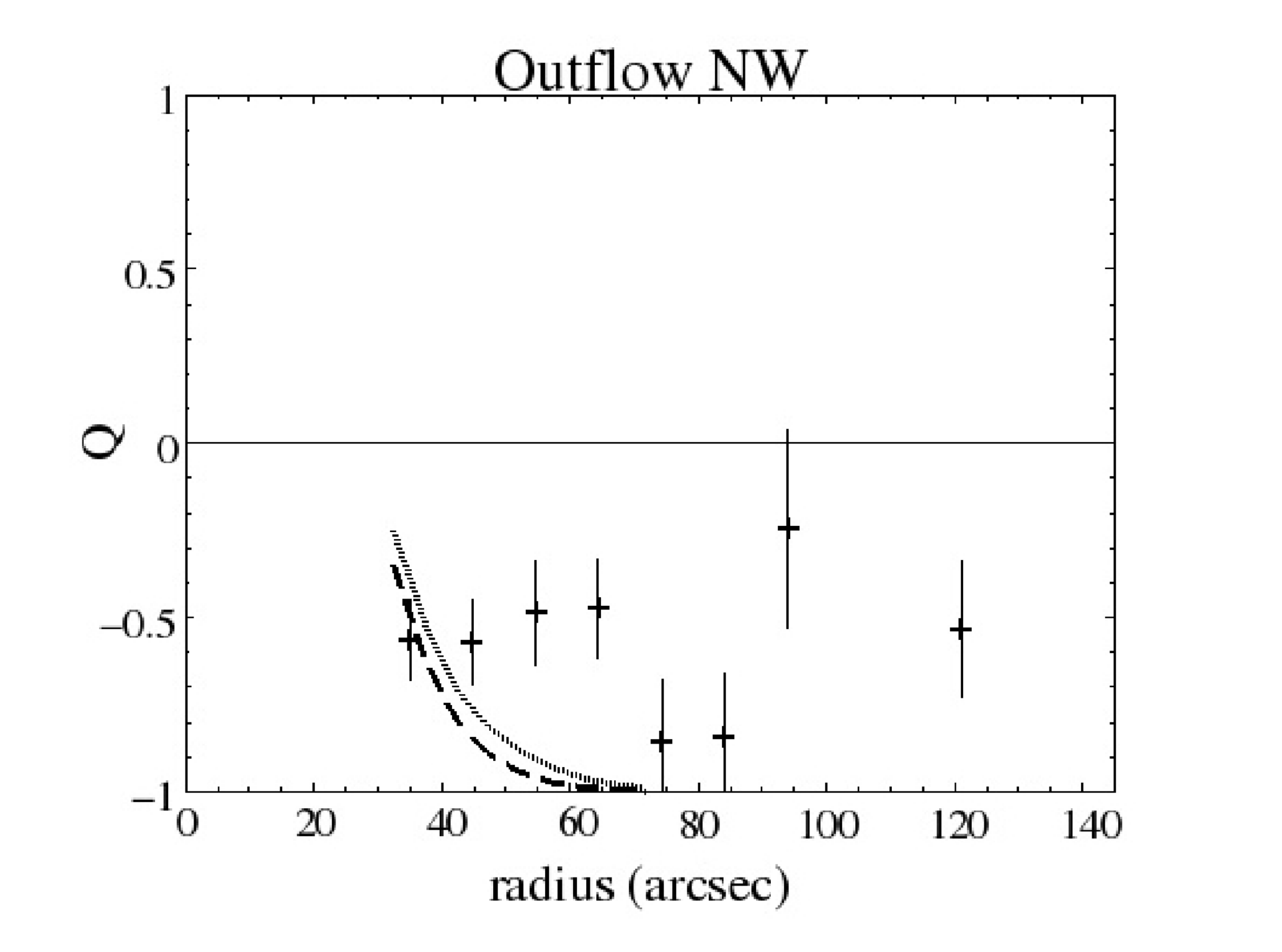}}
\mbox{
	\includegraphics[width=80mm]{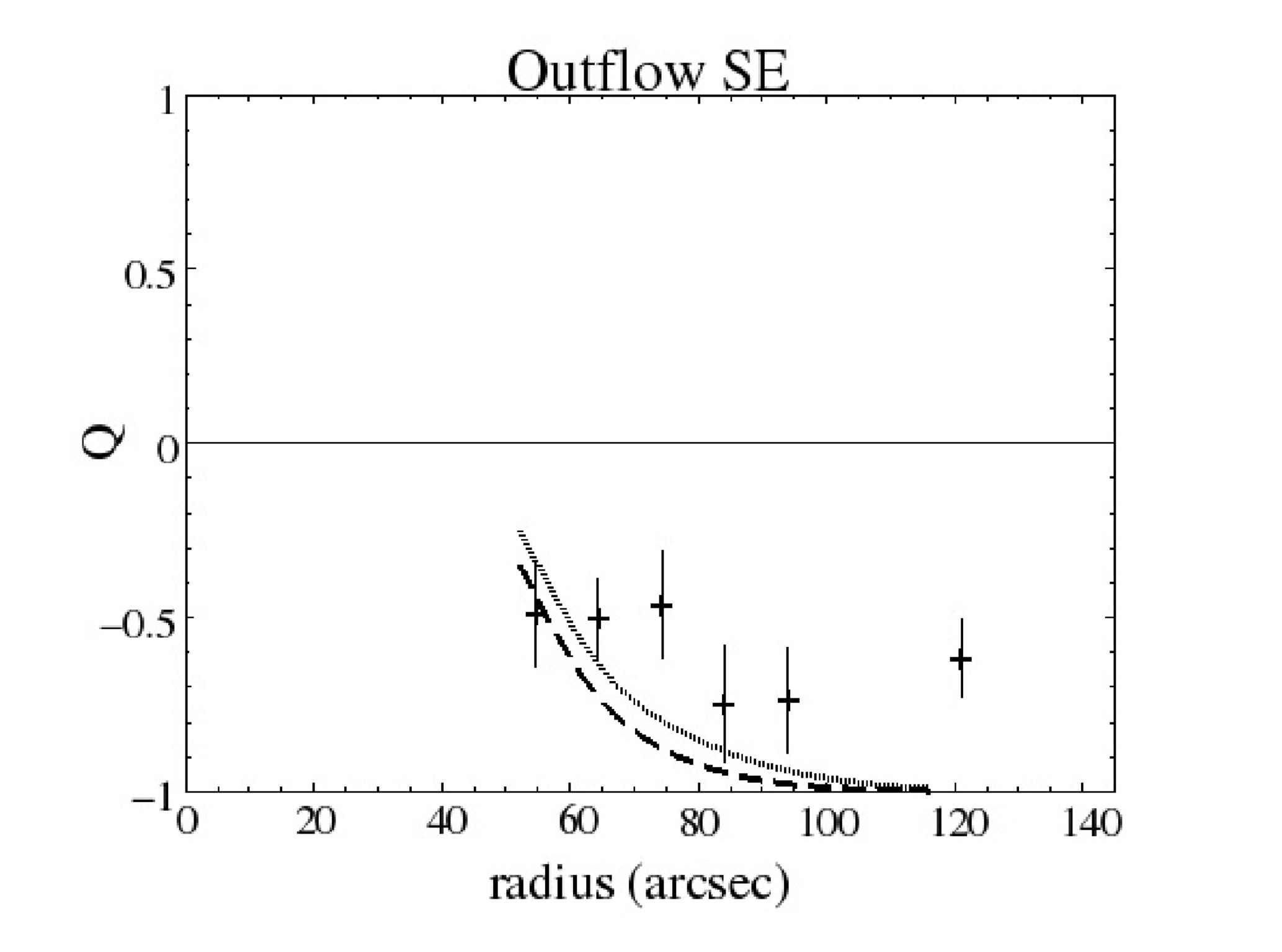}
	\includegraphics[width=80mm]{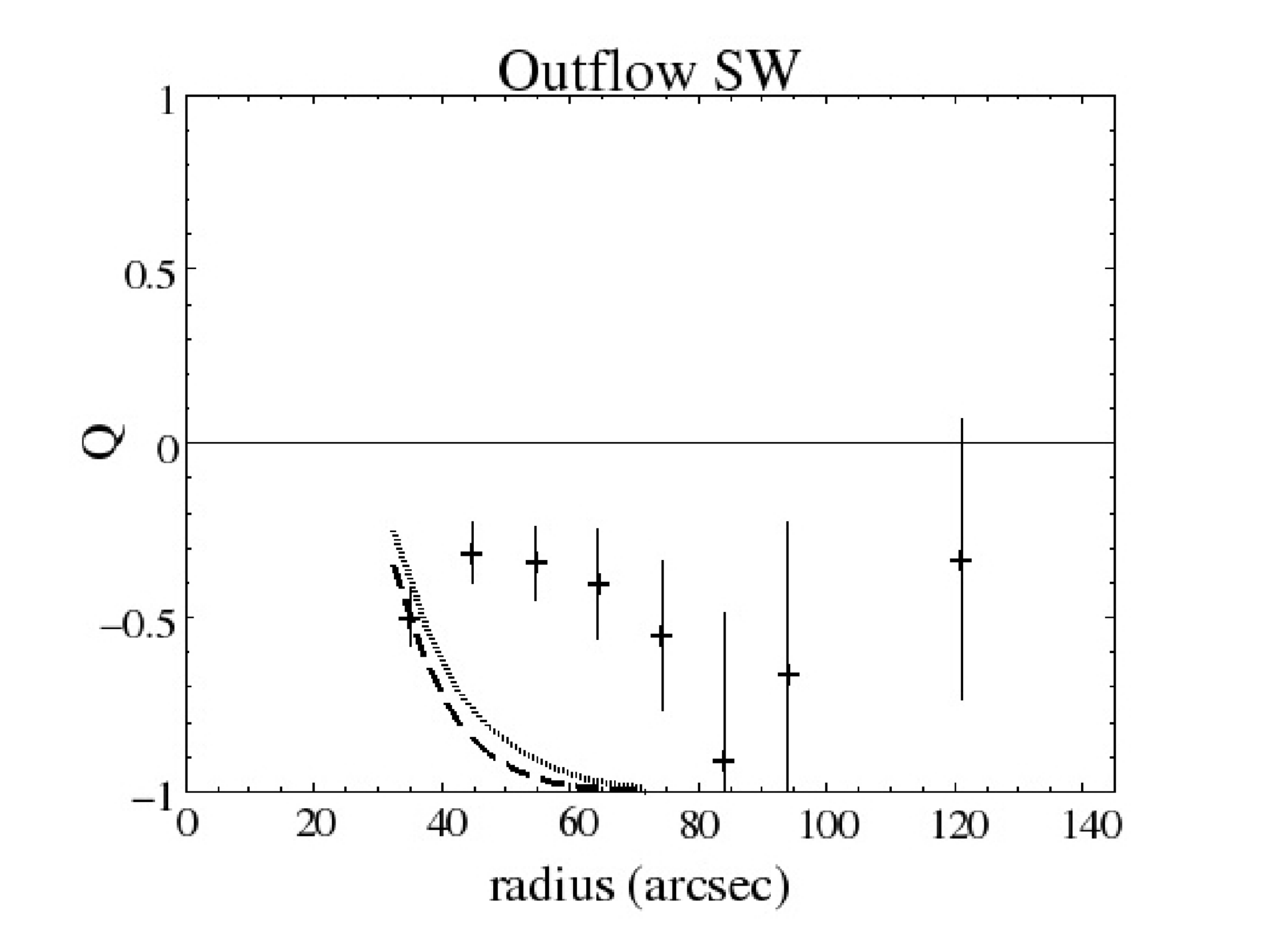}}
\caption{Radial profiles of the spectral hardness ratio in the two halo sectors (top row) and four outflow sectors (middle and bottom rows). The dotted and dashed lines show the profiles that we would expect for an adiabatically expanding gas with a constant $N_{H}$ and linearly decreasing $N_{H}$ respectively (see \S6.2).}
\label{hardness_ratios}
\end{figure}

In most of the azimuthal sectors the spectral hardness ratio decreases with radius (fig.~\ref{hardness_ratios}), suggesting that the temperature is also decreasing with radius. To determine how much these results differ from the scenario of an adiabatically expanding gas, we calculated the radial profiles of $Q$ that we would expect in each sector if the gas was expanding adiabatically. To calculate these profiles, we assumed that the temperature of the gas at the base of the halo in each sector was equal to the temperature in the plane region. From the NEI fits, the temperature of the hot gas in the plane is $kT > 0.41$ keV, so we used an initial temperature of $kT = 0.41$ keV to plot the model profiles, however this is only a lower limit, so we will also need to consider what effect a higher initial temperature will have on these models. The temperature of an adiabatically expanding gas then varies with radius as $T \propto r^{-2}$, so for each radial bin we created an absorbed NEI model in {\it Sherpa} with this temperature, and the abundances and hydrogen column density obtained from the spectral analysis of the appropriate Halo subregion in \S4.3. The normalization of the NEI model was varied so that the X-ray surface brightness profile followed a power-law $\Sigma_{X} \propto r^{-3}$, as expected for an adiabatically expanding gas. The {\it Sherpa} command {\it calc\_model\_sum} was then used to calculate the number of counts from this model in the energy bands $0.3-1.0$ keV and $1.0-2.0$ keV, and hence the spectral hardness ratio. In {\it Sherpa}, the minimum temperature that can be used in the NEI model is $kT = 0.0808$ keV, so the profiles were calculated to the radius at which this temperature is reached. Since the spectral hardness ratio is also affected by absorption, which may not be constant in the halo, we also modeled the profiles of an adiabatically expanding gas for which the absorbing hydrogen column density $N_{H}$ decreases linearly with radius.

These model profiles are plotted in fig.~\ref{hardness_ratios} along with the measured profiles of $Q$, and they demonstrate that in general the measured profiles decrease less steeply than would be expected for an adiabatically expanding gas. Since the initial temperature used in these model profiles is a lower limit, taken from the NEI fits, we also need to consider what effect a higher initial temperature has on these models. We found that this shifted the model profiles upwards in these plots, as we would expect, however the observed profiles could not be matched by a single adiabatic model over their entire radial extent. To match the profiles at large radii typically required an initial temperature of $kT \sim 1-3$ keV, however in these cases the models overestimated $Q$ at smaller radii. We also found that the model profiles in which $N_{H}$ is decreasing radially are steeper than those in which $N_{H}$ is constant. For the measured profiles to be consistent with the scenario of an adiabatically expanding gas $N_{H}$ would need to increase from the plane to the halo, however such an increase is not seen in the map of the 21 cm line emission (fig.~\ref{h1}). This suggests that the hot gas in the halo is not expanding adiabatically, but is instead being reheated as it expands into the halo, most likely via shocks with cooler gas that is already present in the halo.

In the emission line strength map in fig.~\ref{line_map} we can see that the outflows contain strong line emission from the (Fe+O+Ne) blend. Fig.~\ref{line_map} also shows stronger line emission from Mg in the center of the plane of NGC 4490. In \S4.2 and \S4.3 we analysed the spectra of these individual subregions as well as the halo and plane of NGC 4490 in more detail. The results of this spectral analysis using the NEI model (Table~\ref{nei_pars}) suggest that the plasma is in a nonequilibrium ionization state, as the values of the ionization parameter in these fits are generally lower than we would expect for a plasma that is in a collisional ionization equilibrium. The temperatures in the NEI fits were poorly constrained, so we are unable to compare the temperatures of the hot gas in the plane and halo of NGC 4490. We were able to obtain the abundances of Fe, Ne, Mg and Si and found that most of the abundances are close to or slightly above their solar values, although there are large uncertainties in many of these abundances. Previous studies \citep[e.g.][]{weaver,strickland02} have demonstrated that deriving abundances from the X-ray spectra can be difficult due to degeneracy between abundance and temperature, and also due to a dependence of abundance on which spectral model is used, for example whether we use single or multiple temperature components. However, \citet{baldi_a} show how using small regions for the spectral analysis, based on the morphology of the line emission, gives more accurate results than averaging large regions of emission.

To investigate the possibility of contamination by unresolved point sources in our analysis we calculated the energy fluxes in the energy band $0.3-10.0$ keV from the power-law component, which we used to characterize the spectrum of these unresolved point sources (see \S4.2), and the NEI component. These fluxes are summarized in Table~\ref{em_par}. We find that the observed fluxes from these two components are similar, which confirms that the diffuse emission is still significantly contaminated by unresolved point sources that could not be removed. In our analysis of the radial profiles in the halo we are unable to distinguish between emission from the two components. By calculating the energy fluxes of these two components in the energy bands $0.3-2.0$ keV and $2.0-6.0$ keV using the entire diffuse X-ray spectrum of NGC 4490 we find that the flux from the NEI component is greater than the flux from the power-law component by a factor of $\sim 2.6$ in the soft band, whereas in the hard band the power-law component has a greater flux than the NEI component by a factor of $\sim 30$. Therefore the power-law component dominates the spectrum at energies above 2 keV, so we can minimize the effect of the contamination by the power-law component by considering emission from the energy band $0.3-2.0$ keV, although there is still significant contamination in this band. 

In the spectral analysis we account for this contamination by including a power-law component in the model, however it is possible that some unresolved point sources have a thermal spectrum. Such sources would introduce contamination into the thermal component in the spectral analysis and would introduce further contamination into the radial profile analysis. One example of thermal point sources are X-ray binaries in a soft/high state \citep{done}, however such disk-dominated X-ray binaries typically have higher temperatures than we have observed in the hot ISM (with $kT \sim 1.0-1.5$ keV). These sources can be approximated by the power-law component, given the quality of our data, and so should not contaminate the softer thermal spectrum of the hot ISM. Another source of thermal X-ray emission is from stellar sources such as stellar coronae and cataclysmic variables (CVs). \citet{revnivtsev08} measure the X-ray emissivity (per stellar mass) excluding low mass X-ray binaries in the elliptical galaxy NGC 3379, which they argue is produced by CVs and coronally active binaries (ABs), and compare this with similar measurements from M32 \citep{revnivtsev07}, the bulge of M31 \citep{li07} and the local solar neighbourhood \citep{sazonov}. They find that the X-ray emissivity per stellar mass from CVs and ABs is approximately constant in these galaxies, so we can use this value to estimate the X-ray luminosity from these stellar sources in NGC 4490. The absolute B-band magnitude of NGC 4490 is $M_{B} = -19.55$ \citep{devaucouleurs}, so assuming a mass to light ratio of 1.03 (determined from \citet{bell} based on an approximate $B-V$ color of 0.55 as used by \citet{elmegreen}) we estimate the stellar mass of NGC 4490 to be $1.1 \times 10^{10} \; \mathrm{M}_{\odot}$. The average X-ray emissivity per stellar mass of NGC 3379, M32 and the bulge of M31 is $L_{0.5-2.0 \; \mathrm{keV}}/M_{\ast} = (7.0 \pm 2.9) \times 10^{27} \; \mathrm{ergs} \; \mathrm{s}^{-1} \; \mathrm{M}_{\odot}^{-1}$ \citep{revnivtsev08} in the energy band $0.5-2.0$ keV, so we estimate the X-ray luminosity of CVs and ABs in NGC 4490 to be $L_{0.5-2.0 \; \mathrm{keV}} \sim 8 \times 10^{37} \; \mathrm{ergs} \; \mathrm{s}^{-1}$. \citet{revnivtsev08} note that the X-ray emissivity would only be constant for old stellar populations, so the value that we have used does not include the contribution from young stellar coronae, which depends on the star formation history of the galaxy. Since NGC 4490 has a relatively high star formation rate the contribution from young stellar coronae may be significant. Using measurements of the H$\alpha$ luminosity, \citet{thronson} estimate the mass of newly formed stars in NGC 4490 to be $2.5 \times 10^6 \; \mathrm{M}_{\odot}$. If a typical O-type star has a mass of $40 \; \mathrm{M}_{\odot}$ and a luminosity of $4 \times 10^4 \; \mathrm{L}_{\odot}$ then the B-band luminosity from O-type stars in NGC 4490 is $L_{B} \sim 1.5 \times 10^{42} \; \mathrm{erg} \; \mathrm{s}^{-1}$. The ratio of X-ray to optical flux from O-type stars is $f_{x}/f_{B} \sim 10^{-6} - 10^{-7}$ \citep{vaiana,fabbiano82}, so the X-ray luminosity from O-type stars in NGC 4490 is $L_{0.2-4.0 \mathrm{keV}} \sim 10^{36} \; \mathrm{erg} \; \mathrm{s}^{-1}$ in the energy band $0.2-4.0$ keV. Assuming a thermal APEC spectrum with $kT \sim 0.6$ keV, this gives a luminosity of $L_{0.5-2.0 \mathrm{keV}} \sim 9 \times 10^{35} \; \mathrm{erg} \; \mathrm{s}^{-1}$ in the energy band $0.5-2.0$ keV. Therefore we estimate that the total X-ray luminosity from stellar sources, including the old stellar population and young stellar coronae, is $L_{0.5-2.0 \mathrm{keV}} \sim 8.1 \times 10^{37} \; \mathrm{erg} \; \mathrm{s}^{-1}$. In comparison, the luminosity of the thermal component of all diffuse emission in NGC 4490 from the spectral analysis is $L_{0.5-2.0 \; \mathrm{keV}} = 2.5 \times 10^{39} \; \mathrm{ergs} \; \mathrm{s}^{-1}$, so we estimate that stellar sources contribute $\sim 3-4\%$ to the total thermal component of diffuse X-ray emission. We also note that these stellar sources will mostly be in the disk of NGC 4490 and so should not affect the halo region.

Using the results of the spectral analysis, we were able to derive the physical properties of the hot gas, which are summarized in Table~\ref{phys_par}. We can only find an upper limit for the pressure in the plane, however the pressure in the central plane subregion is significantly higher than in the halo, and the electron density is also higher in the plane than in the halo. These results are consistent with the scenario of an expanding halo. In the plane, both the electron density and the pressure are higher in central subregion than the outer plane, and in the halo they are higher in the outflows than in the rest of the halo. We also find that the gas in the halo has a longer cooling time than in the plane, with $\tau_{c} \sim 3.3 \eta^{1/2} \times 10^{9}$ years in the halo and $\tau_{c} \sim 1.0 \eta^{1/2} \times 10^{9}$ years in the plane, however we were unable to calculate the uncertainties in these cooling times, so we do not know how significant this comparison is.

\subsection{Energy Input From Supernovae}

In \S5 we derived the thermal energy $E_{th}$ of the hot gas in each region and subregion, which we can compare with the energy input from supernovae (SN). The SN rate in the galaxy can be calculated from the star formation rate above $5 \; \mathrm{M}_{\odot}$, $SFR(M \geq 5 \; \mathrm{M}_{\odot})$ \citep{wilding,clemens98}:
\begin{equation}
f_{SN} = 0.041[SFR(M \geq 5 \; \mathrm{M}_{\odot})/\mathrm{M}_{\odot} \; \mathrm{yr}^{-1}] \; \mathrm{yr}^{-1}
\end{equation}
\citet{clemens99} used measurements of the thermal component of the radio continuum emission from NGC 4490 to estimate a SFR (for stars with mass $0.1 \leq M \leq 100 \; \mathrm{M}_{\odot}$) of $4.7 \; \mathrm{M}_{\odot} \; \mathrm{yr}^{-1}$. We assume an extended Miller-Scalo initial mass function \citep{miller_scalo,kennicutt83}:
\begin{equation}
\begin{split}
\phi(M) & \propto M^{-1.4} \; \; \; \; (0.1 \leq M < 1.0 \; \mathrm{M}_{\odot}) \\
& \propto M^{-2.5} \; \; \; \; (1.0 \leq M \leq 100 \; \mathrm{M}_{\odot})
\end{split}
\end{equation}
Then we can calculate the SFR for masses above $5 \; \mathrm{M}_{\odot}$:
\begin{equation}
\begin{split}
SFR(M & \geq 5 \; \mathrm{M}_{\odot}) = \\
& SFR(0.1 \leq M \leq 100 \; \mathrm{M}_{\odot}) \frac{\int_{5}^{100} M \phi(M) \, dM}{\int_{0.1}^{100} M \phi(M) \, dM}
\end{split}
\end{equation}
This gives $SFR(M \geq 5 \; \mathrm{M}_{\odot}) = 1.07 \; \mathrm{M}_{\odot} \; \mathrm{yr}^{-1}$, and hence the SN rate in NGC 4490 is $f_{SN} = 0.044 \; \mathrm{yr}^{-1}$.

\citet{condon} give an alternative method for estimating the SFR, using the non-thermal radio luminosity $L_{NT}$:
\begin{equation}
L_{NT} \; (\mathrm{W} \; \mathrm{Hz}^{-1}) \sim 1.3 \times 10^{23} \left( \frac{\nu}{1 \; \mathrm{GHz}} \right)^{-\alpha} f_{SN} \; (\mathrm{yr}^{-1})
\end{equation}
\citet{clemens99} find a non-thermal radio luminosity of $L_{NT} = 5.06 \times 10^{21} \; \mathrm{W} \; \mathrm{Hz}^{-1}$, giving a SN rate of $f_{SN} = 0.039 \; \mathrm{yr}^{-1}$, which is consistent with our previous result.

\citet{clemens99} show that the current epoch of active star formation in NGC 4490 must have been progressing for at least $10^{8}$ years. Interestingly, this is of the order of the cooling time we derive for the hot gas in the disk of NGC 4490 (Table~\ref{phys_par}). Following previous authors \citep[e.g.][]{clemens98,baldi_b} we assume that each SN releases $10^{51}$ ergs of mechanical energy, then the total energy from supernovae over this period of active star formation (using $f_{SN} = 0.044 \; \mathrm{yr}^{-1}$) is $E_{SN} = 4.39 \times 10^{57}$ ergs. In comparison, the total thermal energy in the hot ISM is $E_{th} = 2.25 \times 10^{56}$ ergs, thus this energy can be explained by SN if $\sim 5 \%$ of the energy from SN went into heating the ISM.

\subsection{Chemical Enrichment}

\begin{figure}[H]
\begin{minipage}{180mm}
\centering
\mbox{
    \includegraphics[width=80mm]{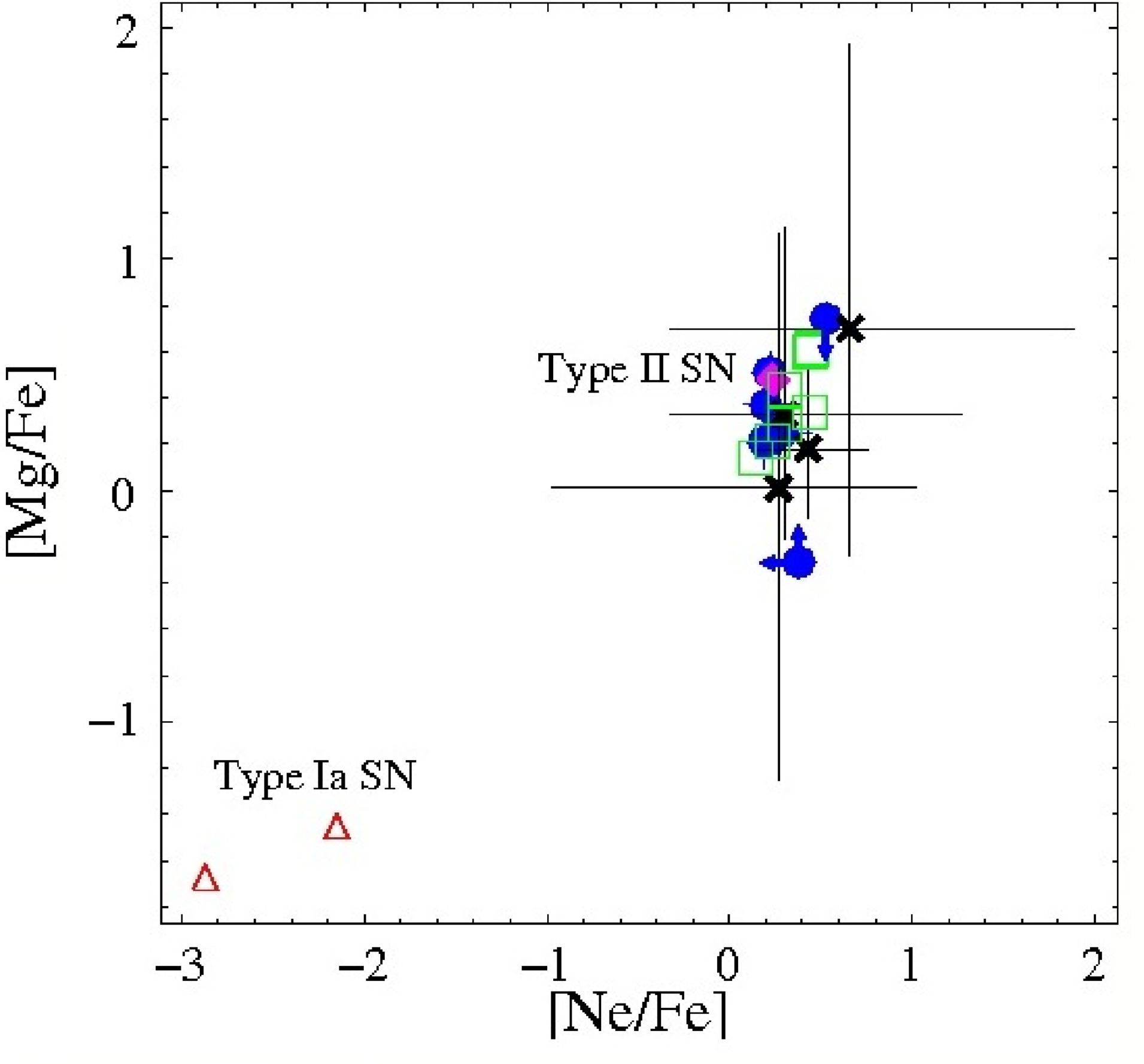}
    \includegraphics[width=85mm]{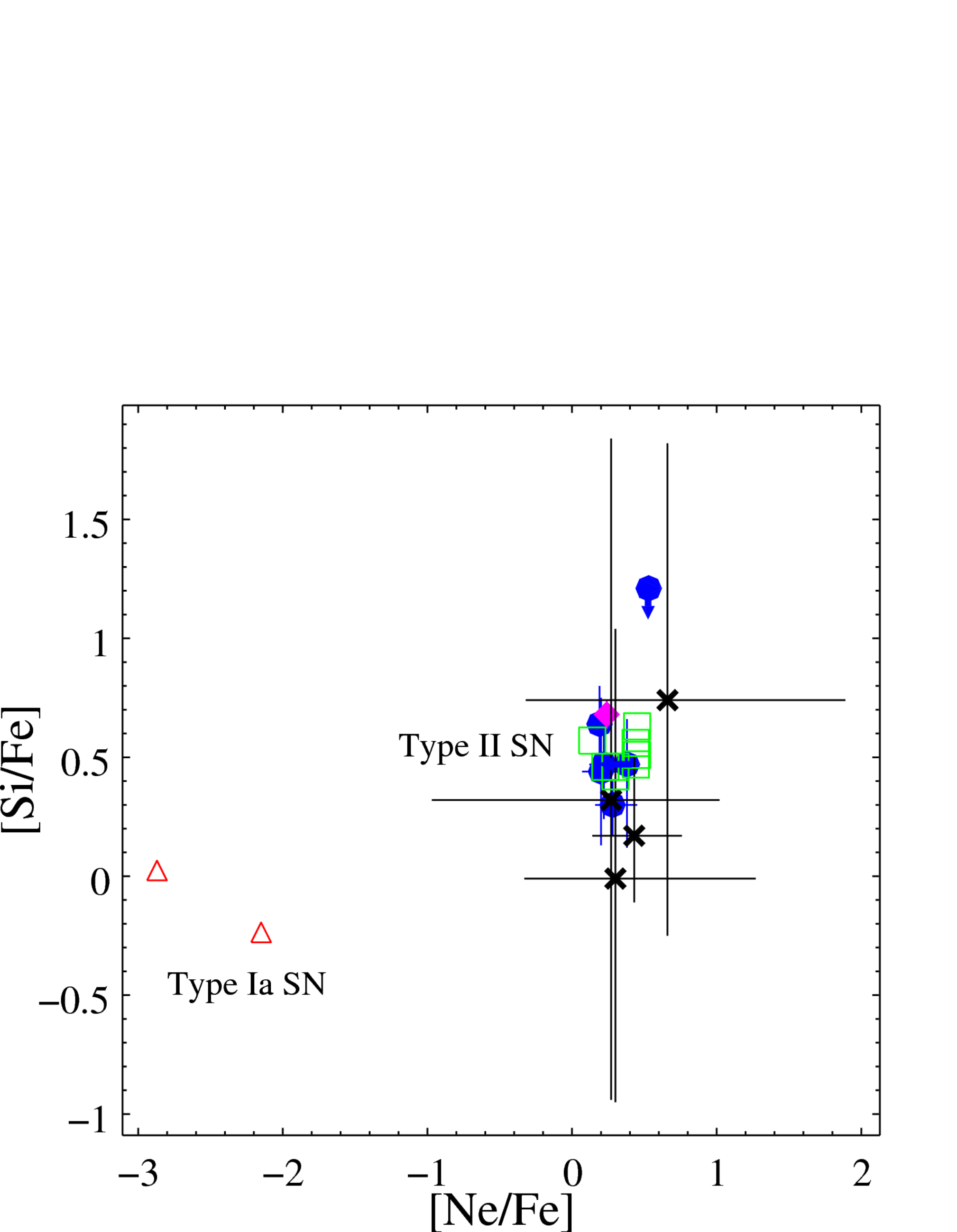}}
\mbox{
    \includegraphics[width=85mm]{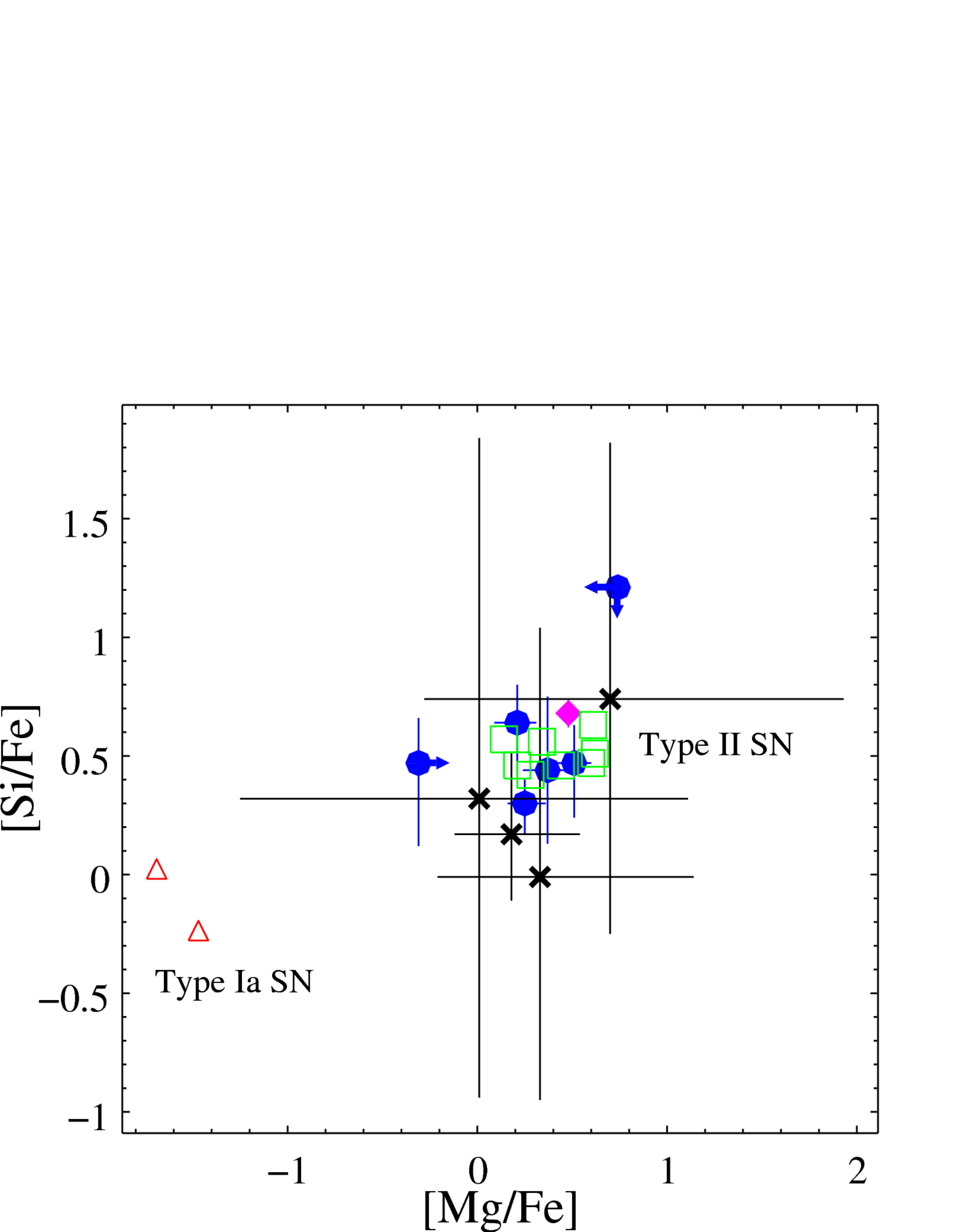}}
\caption{Comparison of the abundance ratios [Ne/Fe] against [Mg/Fe] (top left), [Ne/Fe] against [Si/Fe] (top right) and [Mg/Fe] against [Si/Fe] (bottom), where square brackets indicate logarithmic values, and solar abundances are taken from \citet{anders}. The regions of NGC 4490 are represented by blue circles, M82 by a magenta diamond and the regions of NGC 4038/9 by black crosses. The theoretical models of type Ia SN are represented by empty red triangles, while the type II SN models are represented by empty green squares (see text for references). For the galaxies from other papers, we calculated the uncertainties in the abundance ratios using the standard error propagation rules, as we do not have enough data to use the method described in \S4.2.2 that we used for the abundance ratios in NGC 4490, and so they appear significantly larger than the uncertainties in our results.}
\label{abundances}
\end{minipage}
\end{figure}

We can further investigate the role of SN in the formation of the hot ISM by comparing the abundance ratios calculated in \S4.3 with the abundance ratios that are predicted by theoretical models of Type Ia and II SN. The ratios [Ne/Fe], [Mg/Fe] and [Si/Fe] are plotted against each other in fig.~\ref{abundances} for the regions and subregions in NGC 4490 and the theoretical type Ia and II SN models that are summarized in \citet{nagataki}. For comparison, we also plot the abundance ratios of the starburst galaxy M82 \citep{ranalli} and four regions of the hot ISM in the interacting galaxies NGC 4038/9 \citep{baldi_a} - the 'overlap' region between the galaxies (region 7 in \citet{baldi_a}), the nuclear regions of NGC 4039 (regions 8a and 8b) and the nuclear region of NGC 4038 (region 15).

Fig.~\ref{abundances} shows that the abundance ratios in all regions of NGC 4490 are consistent with those predicted by the theoretical models of type II SN, and are dissimilar from those predicted by the type Ia SN models, suggesting that the hot ISM has been enriched by type II supernovae. The abundance ratios in NGC 4490 are also similar to those found in the interacting galaxies NGC 4038/9 and the starburst galaxy M82.

\section{Summary and Conclusions}

In this paper we have conducted a detailed study of the diffuse X-ray emission from the hot gaseous ISM in NGC 4490/85, using three {\it Chandra} ACIS-S observations with a total exposure time of $\sim 100$ ks. To summarize our main results:

\begin{enumerate}
\item We created a smoothed, mapped color image of the diffuse emission, which showed soft X-ray emission from an extended halo of hot gas around NGC 4490. The X-ray emission from the plane of NGC 4490 appeared harder, which could be due to hotter intrinsic emission and possibly higher absorption in the disk, although these parameters were poorly constrained in our spectral fits. We can also see structures in the extended halo that suggest outflows of hot gas originating from the disk of NGC 4490. The presence of these outflows supports the hypothesis by \citet{clemens98} that the extended HI envelope around NGC 4490/85 was formed by star formation driven outflows. Comparing the X-ray data with 21cm line emission from HI \citep{van_der_hulst}, we also found that the bridge of HI gas between NGC 4490 and its companion galaxy, NGC 4485, which is coincident with similar features seen in H$\alpha$ and CO emission \citep{viallefond,clemens00}, is not coincident with the bridge of hot gas that is seen in the X-ray emission. We created a composite color image of H$\alpha$ \citep{kennicutt08} and X-ray emission, and found that the H$\alpha$ appears to be anticorrelated with the X-ray emission, which suggests that the hot gas is bounded by filaments of cool, ionized hydrogen. We found a similar result between mid-infrared (from the Spitzer archive) and X-ray emission, suggesting that the warm dust is mixed with the cool, ionized hydrogen gas and not the hot gas.
\item The X-ray surface brightness profiles of the outflows are best described by either an exponential or a Gaussian model, suggesting that this gas is being heated via shocks with cooler gas that was already present in the halo. The radial profiles of the spectral hardness ratio in the halo suggest that the temperature of the gas in these outflows is decreasing less steeply than would be expected for an adiabatically expanding gas, confirming that the gas is being reheated as it expands into the halo. This scenario is also supported by the comparison of H$\alpha$ and X-ray emission, which suggested that these outflows are bounded by filaments of cooler, ionized hydrogen gas.
\item The temperatures of the hot gas and the absorbing hydrogen column densities were poorly constrained in the spectral analysis of the diffuse X-ray emission in NGC 4490, so we are unable to compare these parameters in the plane and the halo. Fe, Ne, Mg and Si are detected both in the plane and (less constrained) in the halo. The low values of the ionization timescale parameter from the NEI fits suggest that the hot gas is in a nonequilibrium ionization state.
\item The physical properties of the hot ISM gas were calculated from the results of the spectral analysis using the NEI model. We found that the gas in the halo of NGC 4490 has a higher electron density than in the plane, and a higher pressure than in the central plane subregion.
\item The abundance ratios were found to be similar in all regions and subregions of NGC 4490, and are consistent with the abundance ratios predicted by theoretical models of type II supernovae. Furthermore, the thermal energy in the hot ISM could have originated from supernovae if $\sim 5\%$ of the mechanical energy from these supernovae went into heating the ISM. Therefore, it is likely that the hot ISM has been enriched and heated by type II supernovae.
\end{enumerate}

\acknowledgments
We thank the anonymous referee for their many useful comments that improved this paper. The data analysis was supported by the CXC CIAO software and CALDB. We have used the NASA NED and ADS facilities, and have extracted archival data from the {\it Chandra} and {\it Spitzer} archives. We thank Aneta Siemiginowska and Tom Aldcroft for helpful discussions on {\it Sherpa}. We also thank Robert Kennicutt and Janice Lee for providing the H$\alpha$ images. This research was partially supported by NASA contract NAS8-39073 (CXC). This paper is based on work performed by AR while visiting CfA as a part of the visiting student program sponsored by Southampton University. These results will be part of AR's MSc thesis.

{}

\eject

\begin{table}
\centering
\begin{minipage}{110mm}
\caption{Summary of the {\it Chandra} observations of NGC 4490/85.}
\begin{tabular}{ccccc}
\hline
ObsID & Date & Instrument & Exposure Time (ks)\tablenotemark{a} & Data Mode \\
\hline
\dataset[ADS/Sa.CXO#obs/01579]{1579} & $2000-11-03$ & ACIS-S & $19.8$ & FAINT \\
\dataset[ADS/Sa.CXO#obs/04725]{4725} & $2004-07-29$ & ACIS-S & $39.0$ & VFAINT \\
\dataset[ADS/Sa.CXO#obs/04726]{4726} & $2004-11-20$ & ACIS-S & $38.9$ & VFAINT
\label{obs}
\end{tabular}
\tablenotetext{1}{After filtering for GTIs and background flares.}
\end{minipage}
\end{table}

\begin{table}[h!]
\centering
\begin{minipage}{160mm}
\caption{Best fit parameters of the radial X-ray surface brightness profiles. Errors are quoted at 1$\sigma$ for 1 interesting parameter.}
\begin{tabular}{lcccccccccc}
\hline
& Net Counts & \multicolumn{3}{c}{Power Law} & \multicolumn{3}{c}{Exponential} & \multicolumn{3}{c}{Gaussian} \\
Sector & ($0.3-2.0$ keV) & $\Gamma$\tablenotemark{a} & $\chi^{2}$ & $\nu$ & $H$ (kpc)\tablenotemark{b} & $\chi^{2}$ & $\nu$ & FWHM (kpc)\tablenotemark{c} & $\chi^{2}$ & $\nu$ \\
\hline
Halo N & $830 \pm 72$ & $0.92_{-0.11}^{+0.10}$ & $17.7$ & $23$ & $2.52 \pm 0.34$ & $13.3$ & $23$ & $6.30_{-0.44}^{+0.53}$ & $14.9$ & $23$ \\
Halo S & $289 \pm 43$ & $1.65 \pm 0.20$ & $12.7$ & $21$ & $1.40 \pm 0.21$ & $13.3$ & $21$ & $4.33_{-0.47}^{+0.53}$ & $17.4$ & $21$ \\
Outflow NE & $853 \pm 61$ & $1.78 \pm 0.06$ & $148.1$ & $22$ & $1.08 \pm 0.06$ & $63.1$ & $22$ & $3.60_{-0.11}^{+0.12}$ & $48.2$ & $22$ \\
Outflow NW & $1211 \pm 65$ & $1.87 \pm 0.11$ & $38.3$ & $20$ & $1.30 \pm 0.09$ & $29.4$ & $20$ & $4.41_{-0.19}^{+0.20}$ & $34.3$ & $20$ \\
Outflow SE & $854 \pm 57$ & $1.75 \pm 0.16$ & $47.3$ & $16$ & $1.80 \pm 0.17$ & $39.80$ & $16$ & $5.77_{-0.28}^{+0.32}$ & $35.0$ & $16$ \\
Outflow SW & $1281 \pm 56$ & $2.78_{-0.11}^{+0.12}$ & $43.8$ & $20$ & $0.76 \pm 0.05$ & $18.0$ & $20$ & $2.99_{-0.11}^{+0.12}$ & $21.7$ & $20$ \\
\label{profile_pars}
\end{tabular}
\tablenotetext{1}{Power-law index}
\tablenotetext{2}{Scale height}
\tablenotetext{3}{Full width at half maximum}
\end{minipage}
\end{table}

\eject

\begin{table}[h!]
\centering
\begin{minipage}{180mm}
\caption{Best fit parameters of the APEC model in all regions and subregions of NGC 4490. Errors are quoted at $1\sigma$ for 1 interesting parameter.}
\begin{tabular}{lcccccccc}
\hline
 & Net Counts \\
Region & ($0.3-6.0$ keV) & $\chi^{2}$/$\nu$ & $N_{H} (10^{22} cm^{-2})$\tablenotemark{a} & $kT$ (keV) & $Z_{Fe}/Z_{Fe \odot}$\tablenotemark{b} & $Z_{Ne}/Z_{Ne \odot}$\tablenotemark{b} & $Z_{Mg}/Z_{Mg \odot}$\tablenotemark{b} & $Z_{Si}/Z_{Si \odot}$\tablenotemark{b} \\
\hline
Plane & $3888 \pm 76$ & $114.9/115$ & $0.20_{-0.03}^{+0.04}$ & $0.39 \pm 0.02$ & $0.71_{-0.13}^{+0.17}$ & $1.60_{-0.33}^{+0.42}$ & $1.81_{-0.52}^{+0.70}$ & $2.34_{-1.02}^{+1.40}$ \\
Halo & $4778 \pm 108$ & $140.5/133$ & $0.12 \pm 0.03$ & $0.24 \pm 0.01$ & $2.07_{-0.37}^{+0.46}$ & $3.47_{-0.47}^{+0.53}$ & $5.64_{-1.53}^{+1.86}$ & $29.22_{-9.06}^{+11.29}$ \\
Outflow & $3204 \pm 80$ & $123.2/112$ & $0.07_{-0.02}^{+0.03}$ & $0.25 \pm 0.01$ & $2.48_{-0.47}^{+0.59}$ & $4.23_{-0.57}^{+0.65}$ & $9.23_{-2.25}^{+2.77}$ & $37.13_{-11.38}^{+14.12}$ \\
Halo-\tablenotemark{c} & $1491 \pm 67$ & $95.2/95$ & $0.22_{-0.09}^{+0.41}$ & $0.23_{-0.08}^{+0.02}$ & $1.51_{-0.49}^{+3.80}$ & $2.35_{-0.68}^{+0.90}$ & $1.49_{-1.46}^{+2.07}$ & $<35.44$ \\
Central Plane & $2243 \pm 52$ & $104.6/81$ & $0.22_{-0.04}^{+0.05}$ & $0.41 \pm 0.03$ & $0.74_{-0.17}^{+0.22}$ & $1.57_{-0.41}^{+0.56}$ & $2.45_{-0.75}^{+1.10}$ & $2.16_{-0.99}^{+1.46}$ \\
Outer Plane & $1647 \pm 53$ & $93.5/82$ & $0.27_{-0.08}^{+0.09}$ & $0.34_{-0.03}^{+0.04}$ & $0.59_{-0.18}^{+0.26}$ & $1.23_{-0.37}^{+0.55}$ & $0.73_{-0.43}^{+0.71}$ & $3.01_{-1.90}^{+2.83}$ \\
\label{parameters}
\end{tabular}
\tablenotetext{1}{Total hydrogen column density, galactic plus intrinsic, from the photoelectric absorption model.}
\tablenotetext{2}{All solar abundances are taken from \citet{anders}.}
\tablenotetext{3}{The rest of the halo, excluding the outflow subregion.}
\end{minipage}
\end{table}

\begin{table}[h!]
\centering
\begin{minipage}{80mm}
\caption{Abundance ratios from the APEC model.}
\begin{tabular}{lccc}
\hline
Region & [Ne/Fe]\tablenotemark{a} & [Mg/Fe]\tablenotemark{a} & [Si/Fe]\tablenotemark{a} \\
\hline
Plane & $0.35_{-0.07}^{+0.06}$ & $0.41_{-0.09}^{+0.08}$ & $0.52_{-0.21}^{+0.15}$ \\
Halo & $0.22 \pm 0.06$ & $0.44_{-0.11}^{+0.10}$ & $1.15_{-0.14}^{+0.11}$ \\
Outflow & $0.23_{-0.07}^{+0.06}$ & $0.57_{-0.10}^{+0.08}$ & $1.18_{-0.14}^{+0.11}$ \\
Halo- & $0.19_{-0.85}^{+0.13}$ & $-0.01_{-1.65}^{+0.31}$ & $<1.10$ \\
Central Plane & $0.32_{-0.09}^{+0.08}$ & $0.52_{-0.09}^{+0.08}$ & $0.46_{-0.22}^{+0.16}$ \\
Outer Plane & $0.32_{-0.12}^{+0.13}$ & $0.09_{-0.35}^{+0.21}$ & $0.71_{-0.33}^{+0.21}$ \\
\label{abundance_ratios}
\end{tabular}
\tablenotetext{1}{Square brackets indicate logarithmic values}
\end{minipage}
\end{table}

\begin{table}[h!]
\centering
\begin{minipage}{180mm}
\caption{Best fit parameters of the NEI model in all regions and subregions of NGC 4490. Errors are quoted at $1\sigma$ for 1 interesting parameter.}
\begin{tabular}{lcccccccc}
\hline
Region & $\chi^{2}$/$\nu$ & $N_{H} (10^{22} cm^{-2})$\tablenotemark{a} & $kT$ (keV) & $Z_{Fe}/Z_{Fe \odot}$\tablenotemark{b} & $Z_{Ne}/Z_{Ne \odot}$\tablenotemark{b} & $Z_{Mg}/Z_{Mg \odot}$\tablenotemark{b} & $Z_{Si}/Z_{Si \odot}$\tablenotemark{b} & $\tau$\tablenotemark{c} $(10^{11} s \; cm^{-3})$ \\
\hline
Plane	&	127.0/114	&	$>0.13$	&	$>0.41$	&	$0.93_{-0.47}^{+0.13}$	&	$1.32_{-0.62}^{+0.37}$	&	$>1.18$	&	$2.23_{-1.15}^{+3.58}$	&	$1.73_{-0.22}^{+2.65}$	\\
Halo	&	152.1/132	&	$0.09 \pm 0.04$	&	$0.85_{-0.12}^{+0.59}$	&	$1.78_{-0.25}^{+0.38}$	&	$2.80_{-0.45}^{+0.46}$	&	$2.96_{-0.85}^{+0.97}$	&	$8.16_{-3.88}^{+3.42}$	&	$0.17_{-0.06}^{+0.03}$	\\
Outflow	&	132.8/111	&	$<0.30$	&	$<0.38$	&	$<1.03$	&	$2.59_{-0.60}^{+0.81}$	&	$<5.42$	&	$8.78_{-5.53}^{+3.37}$	&	$>4.66$	\\
Halo-\tablenotemark{d}	&	94.1/93	&	$0.33_{-0.28}^{+0.11}$	&	$0.40_{-0.06}^{+1.00}$	&	$0.77_{-0.26}^{+0.72}$	&	$1.49_{-0.28}^{+0.11}$	&	$<1.35$	&	$<5.78$	&	$0.50_{-0.21}^{+0.32}$	\\
Central Plane	&	123.3/79	&	$0.21_{-0.03}^{+0.06}$	&	$>0.43$	&	$0.66_{-0.14}^{+0.22}$	&	$1.07_{-0.33}^{+0.48}$	&	$2.11_{-0.63}^{+0.93}$	&	$1.86_{-0.82}^{+1.38}$	&	$3.32_{-0.68}^{+0.93}$	\\
Outer Plane	&	90.5/80	&	$0.18 \pm 0.04$	&	Unconstr.	&	$0.77_{-0.26}^{+0.17}$	&	$1.26_{-0.64}^{+0.60}$	&	$0.88_{-0.60}^{+0.54}$	&	$2.34_{-1.67}^{+1.89}$	&	$1.41_{-0.26}^{+0.79}$	\\
\label{nei_pars}
\end{tabular}
\tablenotetext{1}{Total hydrogen column density, galactic plus intrinsic, from the photoelectric absorption model.}
\tablenotetext{2}{All solar abundances are taken from \citet{anders}.}
\tablenotetext{3}{The ionization timescale $\tau = n_{e}t$, where $n_{e}$ is the electron number density and $t$ is the age of the gas.}
\tablenotetext{4}{The rest of the halo, excluding the outflow subregion.}
\end{minipage}
\end{table}

\begin{table}[h!]
\centering
\begin{minipage}{80mm}
\caption{Abundance ratios from the NEI model.}
\begin{tabular}{lccc}
\hline
Region & [Ne/Fe]\tablenotemark{a} & [Mg/Fe]\tablenotemark{a} & [Si/Fe]\tablenotemark{a} \\
\hline
Plane	&	$0.20_{-0.13}^{+0.19}$	&	$0.37_{-0.13}^{+0.16}$	&	$0.44 \pm 0.31$	\\
Halo	&	$0.19_{-0.08}^{+0.05}$	&	$0.21_{-0.12}^{+0.10}$	&	$0.64_{-0.22}^{+0.16}$	\\
Outflow	&	$0.53_{-0.09}^{+0.07}$	&	$<0.74$	&	$<1.21$	\\
Halo-	&	$0.28_{-0.12}^{+0.17}$	&	$0.25_{-0.09}^{+0.11}$	&	$0.30_{-0.13}^{+0.10}$	\\
Central Plane	&	$0.22_{-0.10}^{+0.09}$	&	$0.51 \pm 0.09$	&	$0.47_{-0.23}^{+0.16}$	\\
Outer Plane	&	$<0.38$	&	$>-0.31$	&	$0.47_{-0.35}^{+0.19}$	\\
\label{nei_abund}
\end{tabular}
\tablenotetext{1}{Square brackets indicate logarithmic values}
\end{minipage}
\end{table}

\begin{table}[h!]
\centering
\begin{minipage}{160mm}
\caption{Emission properties from the hot gas in the Plane and Halo regions and their subregions}
\begin{tabular}{lccccc}
\hline
Region & EM\tablenotemark{a} & $\mathrm{f}^{\mathrm{therm}}_{0.3-10.0\:\mathrm{keV}}$\tablenotemark{b,c} & ${\mathrm{f}^{\mathrm{pl}}_{0.3-10.0\:\mathrm{keV}}}$\tablenotemark{b} & $\mathrm{L}^{\mathrm{therm}}_{0.3-10.0\:\mathrm{keV}}$\tablenotemark{c,d} & ${\mathrm{L}^{\mathrm{pl}}_{0.3-10.0\:\mathrm{keV}}}$\tablenotemark{d} \\
 & ($10^{61} \;\; \mathrm{cm}^{-3}$) & ($10^{-14} \;\; \mathrm{ergs} \;\; \mathrm{cm}^{-2} \; \mathrm{s}^{-1}$) & ($10^{-14} \;\; \mathrm{ergs} \;\; \mathrm{cm}^{-2} \; \mathrm{s}^{-1}$) & ($10^{39} \;\; \mathrm{ergs} \;\; \mathrm{s}^{-1}$) & ($10^{39} \;\; \mathrm{ergs} \;\; \mathrm{s}^{-1}$) \\
\hline
Plane & $>2.32$ & $8.30$ & $13.13_{-1.19}^{+1.25}$ & $1.17$ & $1.23_{-0.12}^{+0.11}$ \\
Halo & $0.90_{-0.30}^{+0.29}$ & $13.89$ & $5.67_{-1.32}^{+1.36}$ & $1.78$ & $0.49_{-0.12}^{+0.11}$ \\
Outflow & $>2.40$ & $9.61$ & $4.91_{-1.19}^{+1.17}$ & $0.77$ & $0.37 \pm 0.09$ \\
Halo- & $<8.05$ & $4.20$ & $3.13_{-0.74}^{+0.72}$ & $2.29$ & $0.33 \pm 0.07$ \\
Central Plane & $3.57_{-1.09}^{+2.04}$ & $5.42$ & $5.79_{-0.86}^{+0.85}$ & $0.89$ & $0.56 \pm 0.08$ \\
Outer Plane & $1.50_{-0.41}^{+0.69}$ & $3.37$ & $5.63_{-0.85}^{+0.80}$ & $0.53$ & $0.53 \pm 0.08$ \\
\label{em_par}
\end{tabular}
\tablenotetext{1}{Emission measure $EM = n^{2}V$}
\tablenotetext{2}{observed fluxes from the absorbed thermal and power-law components in the range $0.3-10.0$ keV}
\tablenotetext{3}{uncertainties could not be calculated because the Monte Carlo simulation used to calculate the uncertainties in the flux failed}
\tablenotetext{4}{absorption corrected X-ray luminosities in the range $0.3-10.0$ keV from the thermal and power-law components}
\end{minipage}
\end{table}

\eject

\begin{table}[h!]
\centering
\begin{minipage}{160mm}
\caption{Physical properties of the hot gas in the Plane and Halo regions and their subregions}
\begin{tabular}{lccccc}
\hline
Region & $\mathrm{n}_{e}$ & M\tablenotemark{a} & p & $\mathrm{E}_{th}$ & $\tau_{c}$\tablenotemark{b} \\
 & ($\eta^{-1/2} 10^{-3} \; \mathrm{cm}^{-3}$) & ($\eta^{1/2} 10^{7} \; \mathrm{M}_{\odot}$) & ($\eta^{-1/2} 10^{-12} \; \mathrm{dyne} \; \mathrm{cm}^{-2}$) & ($\eta^{1/2} 10^{55} \; \mathrm{ergs}$) & ($\eta^{1/2} 10^{9} \; \mathrm{yr}$) \\
\hline
Plane & $>1.87$ & $>1.1$ & $<5.04$ & $3.73_{-0.64}^{+1.27}$ & $1.0$ \\
Halo & $0.21_{-0.04}^{+0.03}$ & $3.7$ & $0.58_{-0.07}^{+0.22}$ & $18.77_{-2.18}^{+7.02}$ & $3.3$ \\
Outflow & $>3.25$ & $>0.6$ & $4.00_{-1.93}^{+1.03}$ & $1.36_{-0.66}^{+0.32}$ & $0.6$ \\
Halo- & $<0.59$ & $<10.6$ & $0.48_{-0.22}^{+0.15}$ & $11.00_{-1.94}^{+6.21}$ & $1.5$ \\
Central Plane & $7.98_{-1.33}^{+2.08}$ & $0.4$ & $12.68_{-1.93}^{+2.02}$ & $1.07_{-0.16}^{+0.17}$ & $0.4$ \\
Outer Plane & $1.60_{-0.25}^{+0.32}$ & $0.8$ & $<3.92$ & $2.67_{-0.42}^{+0.89}$ & $1.6$ \\
\label{phys_par}
\end{tabular}
\tablenotetext{1}{total mass of hot ISM gas}
\tablenotetext{2}{uncertainties could not be calculated because the Monte Carlo simulation used to calculate the uncertainties in the flux failed}
\end{minipage}
\end{table}

\eject

\eject

\eject

\eject

\eject

\eject

\begin{figure}[h!]
\centering
\mbox{
	\includegraphics[width=75mm]{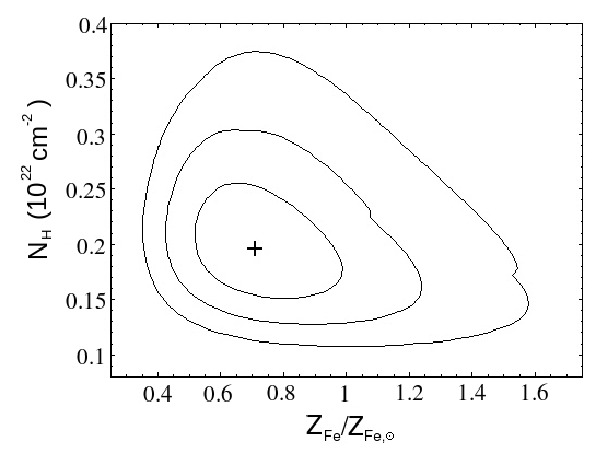}
	\includegraphics[width=75mm]{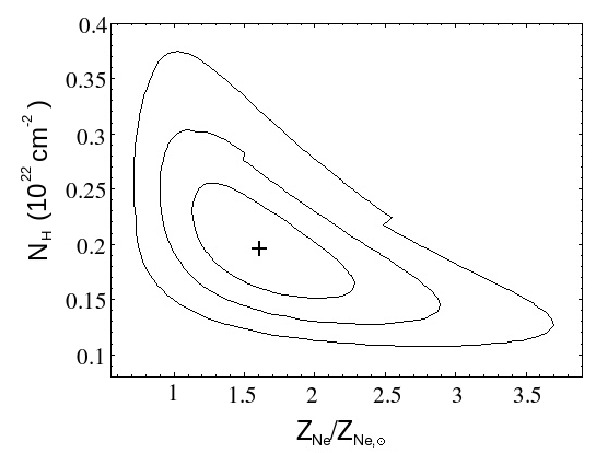}}
\caption{1, 2 and 3$\sigma$ confidence level contours in the $Z_{Fe}$ vs $N_{H}$ plane (left) and the $Z_{Ne}$ vs $N_{H}$ plane (right) for the plane region of NGC 4490.}
\label{nh_z_contours}
\end{figure}

\eject

\eject

\clearpage

\clearpage

\appendix

\section{Point Source Detections in NGC 4490/85}
\label{point_sources}

\begin{table}[h!]
\centering
\begin{minipage}{100mm}
\caption{X-ray sources detected in NGC 4490/85.}
\begin{tabular}{cccc}
\hline
Source Designation & Count Rate\footnote{In the energy range $0.3-6.0$keV.} & Comments\footnote{(R) detected by \citet{roberts}; (F) detected by \citet{fridriksson}; (ULX) ultraluminous X-ray source \citep{gladstone}; (U) shows some evidence for extension beyond the PSF, but uncertain as significance $<2\sigma$.} & Location \\
CXOU J & ($10^{-3} \; \mathrm{counts s}^{-1}$) & & \\
\hline
123023.4+413652	&	$1.61	\pm	0.14$	&	R	&	Background	\\
123023.9+413921	&	$0.12	\pm	0.04$	&		&	Background	\\
123023.9+413841	&	$0.18	\pm	0.05$	&	R	&	Background	\\
123025.2+413924	&	$2.77	\pm	0.17$	&	R, F	&	NGC 4490	\\
123026.1+413639	&	$0.05	\pm	0.03$	&		&	Background	\\
123026.7+413822	&	$0.39	\pm	0.07$	&	F	&	NGC 4490	\\
123027.2+413814	&	$0.92	\pm	0.10$	&	R, F	&	NGC 4490	\\
123027.6+413941	&	$0.27	\pm	0.06$	&	F	&	NGC 4490	\\
123028.2+413958	&	$0.34	\pm	0.07$	&	R, F	&	NGC 4490	\\
123028.5+413926	&	$0.43	\pm	0.07$	&	R, F	&	NGC 4490	\\
123028.7+413757	&	$1.07	\pm	0.11$	&	R, F	&	NGC 4490	\\
123029.0+414046	&	$0.68	\pm	0.09$	&	R, F, U	&	NGC 4485	\\
123029.4+413927	&	$11.06	\pm	0.34$	&	R, F, ULX	&	NGC 4490	\\
123029.4+414058	&	$0.20	\pm	0.05$	&	F	&	NGC 4485	\\
123030.3+413853	&	$5.83	\pm	0.25$	&	R, F	&	NGC 4490	\\
123030.3+414126	&	$0.73	\pm	0.10$	&	F, U	&	NGC 4485	\\
123030.4+413956	&	$0.19	\pm	0.05$	&		&	NGC 4490	\\
123030.4+414142	&	$46.75	\pm	0.72$	&	R, F, ULX	&	NGC 4485	\\
123030.5+413945	&	$0.17	\pm	0.05$	&		&	NGC 4490	\\
123030.7+413911	&	$33.29	\pm	0.59$	&	R, F, ULX	&	NGC 4490	\\
123031.0+413838	&	$0.54	\pm	0.09$	&	R, F	&	NGC 4490	\\
123031.2+413901	&	$0.71	\pm	0.10$	&	R, F	&	NGC 4490	\\
123031.6+414141	&	$8.50	\pm	0.31$	&	F	&	NGC 4485	\\
123032.1+413753	&	$0.15	\pm	0.04$	&		&	NGC 4490	\\
123032.1+413918	&	$34.28	\pm	0.60$	&	R, F, ULX	&	NGC 4490	\\
123032.8+413845	&	$0.29	\pm	0.06$	&	F	&	NGC 4490	\\
\label{source_list}
\end{tabular}
\end{minipage}
\end{table}

\addtocounter{table}{-1}
\begin{table}[H]
\centering
\begin{minipage}{100mm}
\caption{{\it (Continued)} X-ray sources detected in NGC 4490/85.}
\begin{tabular}{cccc}
\hline
Source Designation & Count Rate\footnote{In the energy range $0.3-6.0$keV.} & Comments\footnote{(R) detected by \citet{roberts}; (F) detected by \citet{fridriksson}; (ULX) ultraluminous X-ray source \citep{gladstone}; (U) shows some evidence for extension beyond the PSF, but uncertain as significance $<2\sigma$.} & Location \\
CXOU J & ($10^{-3} \; \mathrm{counts s}^{-1}$) & & \\
\hline
123032.9+414014	&	$0.69	\pm	0.09$	&	F, U	&	NGC 4490	\\
123033.6+414057	&	$0.10	\pm	0.03$	&		&	Background	\\
123034.1+413859	&	$0.52	\pm	0.09$	&	F	&	NGC 4490	\\
123034.1+413819	&	$0.17	\pm	0.05$	&		&	NGC 4490	\\
123034.2+413805	&	$0.65	\pm	0.09$	&	R, F	&	NGC 4490	\\
123034.2+413845	&	$0.18	\pm	0.06$	&		&	NGC 4490	\\
123034.3+413850	&	$4.28	\pm	0.22$	&	R, F	&	NGC 4490	\\
123034.4+413834	&	$0.35	\pm	0.07$	&	F	&	NGC 4490	\\
123035.1+413847	&	$11.19	\pm	0.35$	&	R, F	&	NGC 4490	\\
123035.8+413832	&	$0.55	\pm	0.09$	&	R, F, U	&	NGC 4490	\\
123036.2+413838	&	$23.94	\pm	0.51$	&	R, F, ULX	&	NGC 4490	\\
123036.3+413803	&	$0.08	\pm	0.03$	&		&	NGC 4490	\\
123036.6+413759	&	$0.16	\pm	0.05$	&		&	NGC 4490	\\
123037.5+413803	&	$0.07	\pm	0.03$	&		&	NGC 4490	\\
123037.8+413823	&	$0.27	\pm	0.06$	&	F	&	NGC 4490	\\
123038.3+413831	&	$8.58	\pm	0.30$	&	R, F, ULX	&	NGC 4490	\\
123038.4+413743	&	$1.79	\pm	0.14$	&	R, F	&	NGC 4490	\\
123038.8+413810	&	$3.29	\pm	0.19$	&	R, F	&	NGC 4490	\\
123038.9+413822	&	$0.20	\pm	0.05$	&	R	&	NGC 4490	\\
123039.0+413751	&	$0.09	\pm	0.04$	&	R	&	NGC 4490	\\
123040.3+413813	&	$5.49	\pm	0.24$	&	R, F	&	NGC 4490	\\
123042.9+414030	&	$0.53	\pm	0.08$	&		&	Background	\\
123043.0+413756	&	$0.54	\pm	0.08$	&	R, F	&	NGC 4490	\\
123043.1+413818	&	$43.06	\pm	0.67$	&	R, F, ULX	&	NGC 4490	\\
123045.5+413640	&	$1.08	\pm	0.11$	&	R, F	&	Background	\\
123046.5+414031	&	$0.10	\pm	0.04$	&		&	Background	\\
123047.7+413807	&	$0.37	\pm	0.06$	&	R, F	&	Background	\\
123047.7+413727	&	$0.79	\pm	0.09$	&	R, F	&	Background	\\
123049.4+414056	&	$8.62	\pm	0.31$	&	R	&	Background	\\
\end{tabular}
\end{minipage}
\end{table}

\end{document}